\pdfoutput=1

\documentclass{article}
\usepackage[sort,compress]{natbib} 
\usepackage[utf8]{inputenc}
\usepackage{gensymb}
\usepackage[english]{babel}
\usepackage[pagebackref=true]{hyperref}
\addto\extrasenglish{

}
\hypersetup{colorlinks=true,citecolor=blue,linkcolor=blue,urlcolor=blue}
\usepackage{subcaption} 
\usepackage{amsmath}
\usepackage{units} 
\usepackage[a4paper]{geometry}
\usepackage[labelfont=bf]{caption}
\usepackage{xcolor}
\usepackage{multirow}
\usepackage{bm} 
\usepackage{authblk} 

\usepackage{graphicx}
\usepackage{amssymb,amsfonts}
\usepackage{lineno}
\newcommand{\R}{\texttt{R}\xspace}
\newcommand{\fMRIscrub}{\texttt{fMRIscrub}\xspace}
\newcommand{\ciftiTools}{\texttt{ciftiTools}\xspace}

\usepackage{algorithm}
\usepackage{algorithmic}

\newcommand{\pcatf}{FusedPCA\xspace}
\renewcommand{\hat}{\widehat}
\usepackage{amsthm}

\usepackage{xspace}
\usepackage{mathtools}
\mathtoolsset{showonlyrefs}
\DeclarePairedDelimiterX{\norm}[1]{\lVert}{\rVert}{#1}
\newcommand{\X}{\bfX}
\DeclareMathOperator{\argmin}{argmin}
\DeclareMathOperator{\argmax}{argmax}

\renewcommand{\top}{\mathsf{T}}

\def\bfH{\mathbf H}

\def\bfX{\mathbf X}
\def\bfY{\mathbf Y}

\def\bfv{\mathbf v}

\def\bfx{\mathbf x}

\def\boldfacefake#1{\kern-4pt
    \hbox{ \mathsurround=0pt
    \hbox to 0.4pt{$#1$\hss}\hbox to 0.4pt{$#1$\hss}\hbox {$#1$}}}

\newcommand{\be}{\begin{eqnarray}}
\newcommand{\ee}{\end{eqnarray}}
\newcommand{\ba}{\begin{eqnarray*}}
\newcommand{\ea}{\end{eqnarray*}}
\newcommand{\bc}{\begin{center}}
\newcommand{\ec}{\end{center}}
\newcommand{\btab}[1]{\begin{tabular}{#1}}
\newcommand{\etab}{\end{tabular}}

\newcommand{\reals}{\mbox{\rm I\kern-.20em R}}
\newcommand{\sreals}{\mbox{\small \rm I\kern-.20em R}}

\title{Less is more: balancing noise reduction and data retention in fMRI with data-driven scrubbing}
\author[1]{Damon Pham}
\author[2]{Daniel McDonald}
\author[1]{Lei Ding}
\author[3,4]{Mary Beth Nebel}
\author[1]{Amanda Mejia}
\affil[1]{\small Department of Statistics, Indiana University, Bloomington, IN, USA}
\affil[2]{Department of Statistics, University of British Columbia, Vancouver, BC, Canada}
\affil[3]{{\small Center for Neurodevelopmental and Imaging Research, Kennedy Krieger Institute, Baltimore, MD, USA}}
\affil[4]{{\small Department of Neurology, Johns Hopkins University, Baltimore, MD, USA}}
\date{}

\begin{document}

\maketitle

\begin{abstract}
Functional MRI (fMRI) data may be contaminated by artifacts arising from a myriad of sources, including subject head motion, respiration, heartbeat, scanner drift, and thermal noise. These artifacts cause deviations from common distributional assumptions, introduce spatial and temporal outliers, and reduce the signal-to-noise ratio of the data---all of which can have negative consequences for the accuracy and power of downstream statistical analysis. Scrubbing is a technique for excluding fMRI volumes thought to be contaminated by artifacts and generally comes in two flavors. Motion scrubbing based on subject head motion-derived measures is popular but suffers from a number of drawbacks, among them the need to choose a threshold, a lack of generalizability to multiband acquisitions, and high rates of censoring of individual volumes and entire subjects. Alternatively, data-driven scrubbing methods like DVARS are based on observed noise in the processed fMRI timeseries and may avoid some of these issues. Here we propose ``projection scrubbing'', a novel data-driven scrubbing method based on a statistical outlier detection framework and strategic dimension reduction, including independent component analysis (ICA), to isolate artifactual variation. We undertake a comprehensive comparison of motion scrubbing with data-driven projection scrubbing and DVARS. 
We argue that an appropriate metric for the success of scrubbing is maximal data retention subject to reasonable performance on typical benchmarks such as the validity, reliability, and identifiability of functional connectivity.  We find that stringent motion scrubbing yields worsened validity, worsened reliability, and produced small improvements to fingerprinting. Meanwhile, data-driven scrubbing methods tend to yield greater improvements to fingerprinting while not generally worsening validity or reliability. Importantly, however, data-driven scrubbing excludes a fraction of the number of volumes or entire sessions compared to motion scrubbing. The ability of data-driven fMRI scrubbing to improve data retention without negatively impacting the quality of downstream analysis has major implications for sample sizes in population neuroscience research.

\end{abstract}


\section{Introduction}

Neural activity measured by the blood-oxygen-level-dependent (BOLD) signal makes up only a small proportion of the total fluctuation in fMRI data \citep{lindquistStatisticalAnalysisFMRI2008,bianciardiSourcesFunctionalMagnetic2009}. Artifactual sources of variation cause fMRI data to exhibit low signal-to-noise ratio (SNR) and deviations from common distributional assumptions---e.g., Gaussianity, stationarity, homoskedasticity---due to artifacts' spatiotemporal patterns and variable magnitude. Therefore, prior to statistical analysis, it is vital to remove as much noise from fMRI data as possible without discarding too much signal. This is a challenging task, because the potential sources of noise are diverse in their origin and manifestation. For example, artifacts may be introduced by subject head motion, physiological processes such as respiration, heartbeat, and blood flow fluctuation unrelated to neural activity, thermal noise, magnetic field non-uniformities, scanner drift, and even prior data pre-processing steps \citep{bianciardiSourcesFunctionalMagnetic2009,liuNoiseContributionsFMRI2016,caballero-gaudesMethodsCleaningBOLD2017}.  

While a single gold standard denoising pipeline for fMRI data has yet to be widely accepted, many methods for modeling, detecting, and removing artifacts exist. These methods can be loosely grouped into two categories: regression-based methods and ``scrubbing''.\footnote{"Scrubbing" is also known as ``censoring'' or ``spike regression''. While ``spike regression'' has previously been defined by \cite{satterthwaiteImprovedFrameworkConfound2013} as scrubbing with a DVARS-based measure, here we use it to refer to any scrubbing method.} fMRI data cleaning pipelines commonly employ methods from both categories in various combinations and implementations. A primary goal of this work is to compare the efficacy of motion-based and data-driven scrubbing methods for fMRI. The difficulty is that the impact of scrubbing is only observed after downstream analyses, and any improvement (or lack thereof) may be unique to the specific analysis at hand. For this reason, and in light of recent work showing that very large samples are required to estimate reliable population-level trends in brain-behavior associations \citep{Marek2022}, we suggest that the success of a denoising pipeline should be judged for its ability to simultaneously retain as much of the original data as possible while improving, or at least not degrading, typical measures of data quality such as validity and reliability of functional connectivity (FC). A second objective is to provide a new methodology for scrubbing that achieves this goal. 

\begin{figure}
    \centering
    \includegraphics[width=0.6\textwidth]{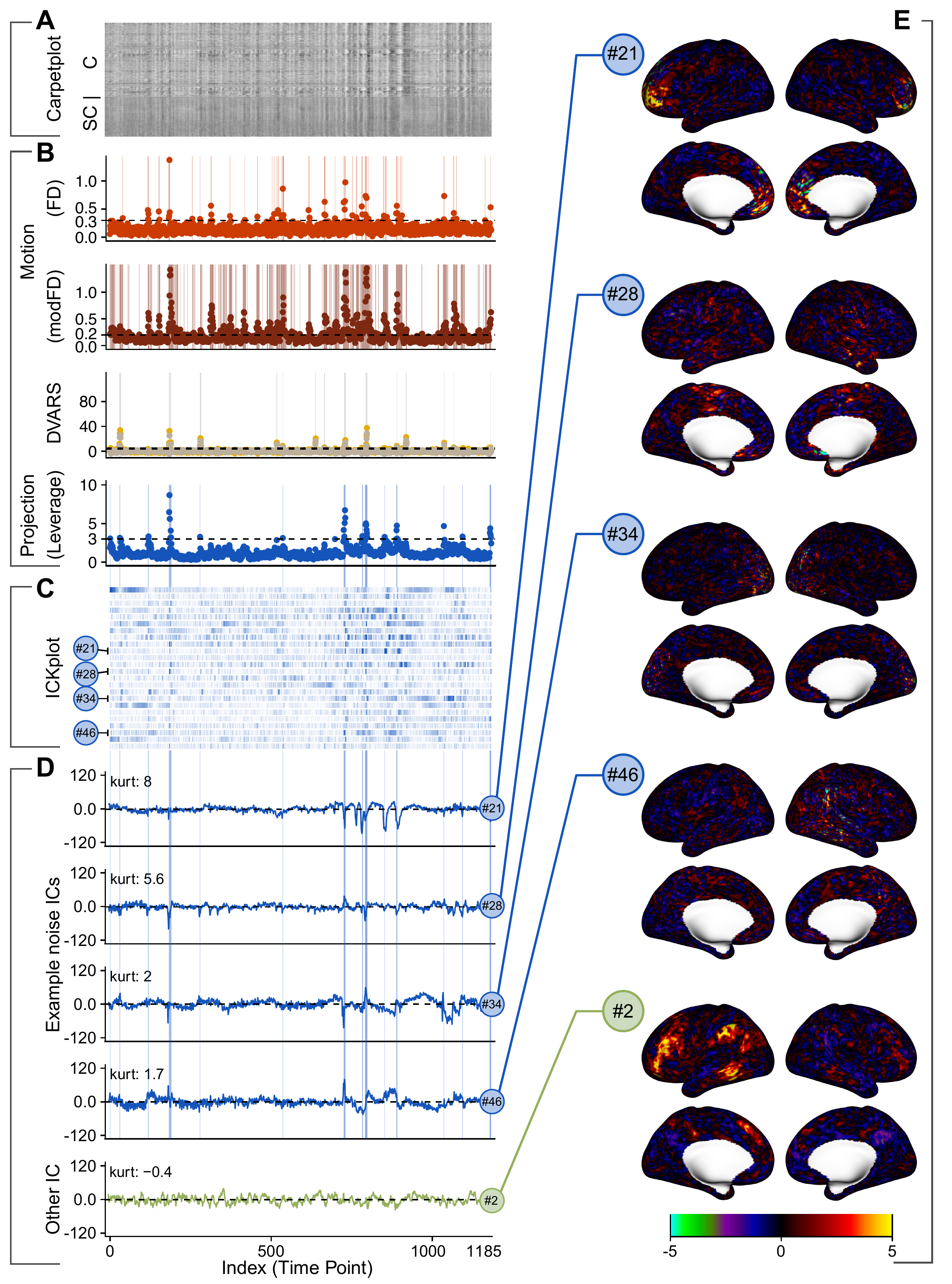}
    \caption{\small \textbf{Illustration of ICA projection scrubbing, which typically removes fewer volumes than motion scrubbing.} The scan shown is HCP subject 111312, visit 2, LR phase encoding, and is a moderate-motion scan (45th quantile of mean FD). Two additional scans (high and low motion) are shown in \autoref{app:additional_scans}. \textbf{(A)} A ``grayplot" or ``carpetplot'': the vectorized fMRI data matrix after regression-based denoising (see \autoref{sec:cleaning_methods}) with time along the x-axis and locations along the y-axis. Lighter colors represent higher BOLD signal. \textbf{(B)} Four different scrubbing measures: FD, modified FD for multiband data, DVARS, and ICA projection scrubbing. Dashed lines indicate selected cutoffs for each method (see \autoref{sec:thresholds}). Projection scrubbing and DVARS retain more volumes than motion scrubbing, and this is generally true for subjects in our study. \textbf{(C-E)} ICA projection scrubbing decomposes the fMRI timeseries into ICs, selects ICs corresponding to transient or ``burst" noise, and then computes a summary measure across those ICs to flag volumes containing burst noise. The ``ICKplot" in \textbf{(C)} shows all noise components; \textbf{(D)} shows the timecourses for a few selected noise ICs (in blue). A non-selected IC is shown in green for comparison. Spatial maps for each IC in \textbf{(E)} illustrate that the selected ICs tend to represent artifacts rather than neural signals. All four selected ICs exhibit moderate or large deviations in their timecourses when leverage surpasses the threshold, suggesting the presence of global abnormalities manifested across multiple artifactual patterns at these timepoints.}
    \label{fig:Example}
\end{figure}

In \autoref{fig:Example}, we provide an illustrative example of common artifactual patterns in a scan from the Human Connectome Project \citep[HCP,][]{vanessenWUMinnHumanConnectome2013} and present metrics for flagging problematic volumes based on two standard techniques, along with our proposed method. The scan shown represents moderate head motion; two additional scans representing high and low motion are shown in \autoref{app:additional_scans}. It is clear that, for these scans, motion scrubbing flags far more volumes than actually exhibit significant abnormalities. As we will show, this is also true more generally. We will also demonstrate that there is little if any loss for downstream tasks associated with retaining more volumes through data-driven scrubbing. Rather, data-driven scrubbing often outperforms motion scrubbing, which is typically much more aggressive. Before describing our contributions in detail, we first review standard denoising procedures, their benefits, and their drawbacks. 

\subsection{Regression-based denoising and scrubbing procedures}

Regression-based denoising methods assume that artifactual fluctuations can be modelled using either external or data-driven sources of information and linearly separated from the data. A number of techniques use exogenous information from recordings related to movement or physiologic processes. As estimates of head motion, rigid body realignment parameters (RPs; also known as the rigid body coordinates) and their extensions can be regressed from the data to minimize artifacts that coincide with motion \citep{powerSpuriousSystematicCorrelations2012,vandijkInfluenceHeadMotion2012,satterthwaiteImpactInScannerHead2012,yanComprehensiveAssessmentRegional2013}; however, spatially variable spin-history effects introduced by motion can persist after the application of such methods \citep{Yancey2011Spin-historyCorrection}. Relatedly, BOLD fluctuations caused by respiration and heartbeat can be adjusted for directly by measuring the physiological cycles with a pulse oximeter, by estimating them from the data 
\citep{leRetrospectiveEstimationCorrection1996, powerSourcesImplicationsWholebrain2017, agrawalModelbasedPhysiologicalNoise2020, salas2021reconstruction}, or by using phase information if available \citep{cheng2010respiratory}.

Rather than using exogenous modeling, one can derive certain signals thought to isolate artifactual variation directly from the BOLD signal. One example is removing tissue-based regressors from brain compartments thought to contain little meaningful neural signals, i.e.\ cerebrospinal fluid (CSF) and white matter \citep{behzadiComponentBasedNoise2007, muschelliReductionMotionrelatedArtifacts2014, caballero-gaudesMethodsCleaningBOLD2017,satterthwaiteMotionArtifactStudies2019}. Another example is removal of the global signal across the brain parenchyma, which remains a popular catch-all denoising technique due to its ability to remove noise for which more targeted techniques may lack sensitivity \citep{powerMethodsDetectCharacterize2014,ciricBenchmarkingParticipantlevelConfound2017,parkesEvaluationEfficacyReliability2018,satterthwaiteMotionArtifactStudies2019}. However, global signal regression (GSR) is controversial due to its potential to worsen existing artifacts, remove neurologically relevant variation between subjects, and induce anti-correlations \citep{liuGlobalSignalFMRI2017,powerSourcesImplicationsWholebrain2017,caballero-gaudesMethodsCleaningBOLD2017,parkesEvaluationEfficacyReliability2018}. Finally, noise-related signals can be identified via independent component analysis (ICA) using ICA-FIX \citep{salimi-khorshidiAutomaticDenoisingFunctional2014} or ICA-AROMA \citep{pruimICAAROMARobustICAbased2015} and regressed from the data.  

Regression-based denoising methods are effective at mitigating nuisance signals that vary across the duration of the scan but are less powerful against noise that is transient and intense---such as that caused by sudden head motion, irregular breaths, or hardware malfunction \citep{powerSpuriousSystematicCorrelations2012,yanComprehensiveAssessmentRegional2013,parkesEvaluationEfficacyReliability2018,satterthwaiteMotionArtifactStudies2019}. This type of transient and intense noise is often called ``burst'' noise. Therefore, a complementary approach for denoising is to identify and directly remove or censor volumes that are highly contaminated, a process known as \textit{scrubbing}. The most common scrubbing techniques are \textit{motion scrubbing} based on framewise displacement (FD), which tracks head movement \citep{powerSpuriousSystematicCorrelations2012,powerMethodsDetectCharacterize2014},  and {DVARS}, which tracks change in image intensity \citep{smyserLongitudinalAnalysisNeural2010,powerSpuriousSystematicCorrelations2012}. Both FD and DVARS are derived relative to the previous volume(s).

While motion scrubbing is widely used, it has a number of disadvantages. First, there is no universally accepted FD threshold above which volumes are considered contaminated.  A range of thresholds between $0.2$ and $0.5$ mm are commonly used \citep{powerMethodsDetectCharacterize2014, ciricBenchmarkingParticipantlevelConfound2017, afyouniInsightInferenceDVARS2018}, but the sensitivity of this threshold to subject and scan acquisition factors, as well as the downstream effects of different choices, is not well-understood. Second, the percentage of volumes flagged for removal is often high, particularly for more stringent (lower) FD thresholds, which can result in 
unintentional removal of relevant signal. It is also common practice to exclude subjects for whom too many volumes are scrubbed, so aggressive removal of volumes can result in the exclusion of a substantial number of study participants. This can lead to loss of power for detecting brain-behavior correlations requiring data from hundreds or thousands of subjects \citep{Marek2022}. It can also introduce selection bias due to exclusion of subjects who tend to move more in the scanner, a marker which may correlate with clinical or sociodemographic variables of interest \citep{nebel2022accounting, cosgrove2022limits}.  Third, head motion may induce artifacts that persist into subsequent time points. \cite{powerSpuriousSystematicCorrelations2012} and others have proposed flagging time points adjacent to high-FD volumes, but this expansion may excessively remove non-contaminated volumes, causing signal loss. Fourth, FD is computed prior to, and independent of, any other denoising strategies, and therefore may flag volumes that actually exhibit low noise levels after regression-based noise reduction strategies have been applied. 

Additionally, several recent studies have highlighted challenges associated with the use of FD in multiband acquisitions. Due to the faster TR of multiband data, respiratory signals are no longer aliased and introduce artifacts into head motion measures, leading to higher and noisier FD that fails to accurately capture head motion \citep{fairCorrectionRespiratoryArtifacts2020, grattonRemovalHighFrequency2020}. The use of FD in multiband data will therefore result in the erroneous removal of volumes not consistent with head motion. Given the accelerating adoption of multiband acquisitions and the growing availability of multiband fMRI data through the HCP and other publicly available datasets adopting HCP-style acquisitions---e.g., the UK Biobank \citep{miller2016multimodal}, the Developing Human Connectome Project \citep{hughes2017dedicated}, the Adolescent Brain and Cognitive Development (ABCD) study \citep{casey2018adolescent}, and the Baby Connectome Project \citep{howell2019unc}---the likelihood of over-aggressive data removal, causing loss of relevant signal, presents a major concern.
\cite{power2019distinctions} proposed modifications to the calculation of FD to accommodate HCP-specific artifacts in FD. However, no generalizable solution was proposed for modifying FD for other multiband acquisitions. Furthermore, since the majority of studies comparing regression-based denoising and scrubbing techniques have focused on single-band fMRI acquisition \citep{ciricBenchmarkingParticipantlevelConfound2017, parkesEvaluationEfficacyReliability2018}, the effects of motion scrubbing in multiband fMRI remain largely understudied.  

DVARS scrubbing, referred to in this manuscript simply as ``DVARS", is a data-driven scrubbing technique based on the variation of the change in BOLD intensities between consecutive volumes \citep{smyserLongitudinalAnalysisNeural2010}. In contrast to FD, DVARS can be computed after regression-based denoising in order to flag only those volumes which still exhibit high levels of contamination, as well as volumes potentially contaminated with burst noise not immediately coincident with head motion, such as those due to lagged effects of motion. Like FD, however, DVARS only compares subsequent volumes. This approach does not leverage information from spatiotemporal patterns which occur throughout the scan and may lead to unnecessary removal of non-contaminated volumes that follow contaminated ones. While its basis in differences makes DVARS sensitive to TR, a standardized form was proposed by \cite{afyouniInsightInferenceDVARS2018}, along with a dual cutoff method that allows for principled thresholding. 

\subsection{Our contributions}

Here we present a data-driven scrubbing approach based on statistical outlier detection known as \textit{projection scrubbing}. This approach is a generalization of the PCA leverage method proposed by \cite{mejiaPCALeverageOutlier2017}. Projection scrubbing consists of two main steps. The data are initially projected onto directions likely to represent artifacts using strategic dimension reduction and selection. Each volume's temporal activity associated with those directions is then summarized and compared to the distribution across all volumes to identify abnormalities representing burst noise. Robust outlier detection methods can be used to enhance sensitivity and specificity in identifying abnormalities. 

We introduce two important improvements to the original projection scrubbing method proposed by \cite{mejiaPCALeverageOutlier2017}. First, while the original framework used principal component analysis (PCA) for the projection step, we consider two alternative techniques that may be more effective at identifying artifactual directions in the data: ICA and a novel technique, FusedPCA. Second, we propose a kurtosis-based selection method to separate and retain components of artifactual origin from those more likely to represent neural sources.  
To examine the effectiveness of projection scrubbing compared with existing scrubbing methods, we evaluate its ability to reduce noise via its effect on several FC quality metrics, including validity, reliability, and identifiability. We also consider the amount of data retained by each method, since over-scrubbing can be detrimental, and optimal data cleaning achieves a balance between removal of noise and retention of true signal. We find that projection scrubbing and DVARS tend to be more beneficial to FC than motion scrubbing, while removing a fraction of the volumes or entire sessions. 



The remainder of this paper proceeds as follows.  In \autoref{sec:methods}, we describe our proposed projection scrubbing framework and the analyses we perform to compare it with existing scrubbing techniques. In \autoref{sec:results}, we compare the performance of projection scrubbing, motion scrubbing and DVARS across our different analyses. Finally, we conclude with a discussion of our findings, limitations and future directions in \autoref{sec:discussion}.

\section{Methods}
\label{sec:methods}

\subsection{Projection scrubbing}

\textit{Projection scrubbing} is a novel statistical framework for identifying and removing fMRI volumes exhibiting significant deviations from expected patterns due to the presence of artifacts. It is a generalization and improvement upon the PCA leverage method proposed by \cite{mejiaPCALeverageOutlier2017}. Projection scrubbing is a data-driven outlier detection method. It involves projecting high-dimensional fMRI volumes onto a small number of directions likely to represent artifacts, then summarizing those directions into a single measure of deviation or outlyingness for each volume. This process consists of three steps, described in turn below: (1) dimension reduction and selection of directions likely to represent artifacts, (2) computing a measure of outlyingness for each volume, and (3) determining a standardized threshold to identify artifactual volumes. All steps are implemented in the open-source \fMRIscrub \R package. 

\subsubsection{Dimension reduction and selection of artifactual directions}

Let $T$ be the duration of the fMRI timeseries, and let $V$ be the number of brain locations included for analysis.  Then $\bfY_{T\times V}$ is the vectorized fMRI BOLD timeseries after basic preprocessing, typically including rigid body realignment to adjust for participant motion, spatial warping of the data to a standard space (distortion correction, co-registration to a high resolution anatomical image), and possibly spatial smoothing.  After robustly centering and scaling the timecourse of each brain location, we aim to project $\bfY$ to $\bfX_{T\times Q}$, representing $Q$ latent directions in the data.  In \cite{mejiaPCALeverageOutlier2017}, PCA was used for this purpose, and the top $Q$ principal components (PCs) explaining 50\% of the data variance were retained. This choice was based on the observation that severe burst noise, such as banding artifacts caused by head motion, results in high-variance fluctuations, so its artifact pattern will manifest in a high-variance PC. Here, we propose several innovations to better distinguish directions representing artifactual variation from those representing neural variation: (1) a principled method to select the embedding dimensionality; (2) the use of alternative projection methods for better separation of components representing signal and noise; and (3) a novel kurtosis-based method to identify those components likely to represent artifacts.

First, for all projection methods, we determine the number of components automatically using penalized semi-integrated likelihood \citep[PESEL,][]{sobczykBayesianDimensionalityReduction2017}, a Bayesian probabilistic PCA approach that generalizes and improves upon the historically popular method of \cite{minkaAutomaticChoiceDimensionality2000}.  

Second, we consider two alternative projection techniques to better identify artifactual directions in the data: ICA and FusedPCA.  Spatial ICA applied to fMRI aims to decompose the data into a set of maximally independent spatial components and a mixing matrix representing the temporal activity associated with each spatial component. The ability of ICA to separate fMRI data into spatial sources of neural and artifactual origin is well-established \citep{mckeown1998analysis}. ICA has been used successfully for noise reduction in fMRI, most notably via ICA-FIX \citep{griffantiICAbasedArtefactRemoval2014} and ICA-AROMA \citep{pruimICAAROMARobustICAbased2015}. 
FusedPCA is a novel PCA technique that employs a fusion penalty on the principal component time series. The fusion penalty has two effects: (1) it encourages a small number of larger jumps in the time series, which may make it easier to detect volumes contaminated with artifacts; and (2) it more easily finds sequential outliers, a common feature of fMRI artifacts. FusedPCA recognizes that volumes contaminated by artifacts, as well as those not contaminated by artifacts, tend to occur in sequential groups. FusedPCA is a PCA version of constant trend filtering \citep{tibshiraniAdaptivePiecewisePolynomial2014, kimEllTrendFiltering2009} and is an instance of the penalized matrix decomposition framework presented in \cite{wittenPenalizedMatrixDecomposition2009}. The optimization problem defining FusedPCA and its estimation procedure are described in \autoref{app:PCATF}. For clarity, and given the historical success of ICA at separating and removing noise in fMRI data and these results, we primarily focus on ICA projection scrubbing. PCA and FusedPCA will be compared in Appendix figures or only in some analyses where convenient. 

Finally, we propose a novel kurtosis-based method for selection of components likely to represent artifacts. Artifacts caused by burst noise tend to be both severe and transient. Therefore, an IC or PC representing a spatial pattern related to burst noise will have associated timeseries---respectively, IC mixing or PC scores---containing spikes or \textit{outliers} at those time points where the artifact exists. The presence of these outliers alters the distribution of temporal activity associated with that component. In particular, \textit{kurtosis}, the standardized fourth moment of a distribution, will be increased. \textit{Excess kurtosis}, hereaftrer referred to as simply \textit{kurtosis}, is computed as 
$$
\text{Kurt}(\bfx) = \frac{1}{N} \sum_{i=1}^N \left(\frac{x_i - \bar{x}}{s} \right)^4 - 3,
$$
for a set of values $\bfx=(x_1,\dots,x_N)$, where $\bar{x}$ and $s$ are respectively the sample mean and standard deviation of $\bfx$. The true kurtosis of Gaussian-distributed data is $0$, but sample kurtosis will exhibit sampling variance.  

Appendix \ref{app:kurtosis} gives further details about our kurtosis-based dimension selection method. Sampling distributions of kurtosis for Gaussian data are shown in \autoref{app:fig:KurtAR}. We select those components whose timeseries have sample kurtosis significantly larger than what would be expected under Gaussianity, defined as exceeding the $0.99$ quantile of the sampling distribution for a particular sample size. This corresponds to a probability of $0.01$ for a false positive, i.e. falsely selecting an outlier-free component.  As illustrated in \autoref{app:fig:KurtAR}, for time series of long duration ($T\geq 1000$), it is reasonable to use the theoretical asymptotic sampling distribution of kurtosis to determine the $0.99$ quantile; for shorter time series, we use Monte Carlo simulation. 

For kurtosis-based component selection, it is important that the data be detrended prior to projection or that the component timeseries be directly detrended. \autoref{app:fig:KurtTrends} illustrates the consequences of failing to detrend properly. For the analyses described below, we detrend the data prior to projection as described in \autoref{sec:data} and therefore forego component detrending.  Both options (data detrending and component mean and variance detrending) are implemented in the \fMRIscrub \R package.  

\subsubsection{Quantifying and thresholding outlyingness}

Let $\bfX_{T\times Q}$ contain the timeseries associated with the $Q$ high-kurtosis components (PCs or ICs) after data or component detrending. Since the selected components are believed to represent artifacts, outliers or spikes in their associated timeseries can be used to locate the temporal occurrence of these artifacts.  We compute an aggregate measure of the temporal outlyingness of each volume, \textit{leverage}, which can be used to identify spikes occurring across any of the components or in multiple components. Leverage is defined as the diagonal entries of the orthogonal projector onto the column space of $\bfX$, $\bfH=\bfX(\bfX^\top \bfX)^{-1}\bfX^\top$, and is related to Mahalanobis distance \citep{kutnerAppliedLinearStatistical2005, weinerHandbookPsychologyResearch2012}.  In linear regression, $\bfH$ is known as the ``hat matrix'', and leverage is related to the influence of each observation on the coefficient estimates.  In our context, leverage is an approximation of the influence of each volume on the selected components \citep{mejiaPCALeverageOutlier2017}. Since these components are likely to represent artifacts, leverage can be interpreted as a {summary measure of the strength of artifactual signals} in each volume.

The final step in projection scrubbing is to threshold leverage to identify volumes contaminated by artifacts.  To determine an appropriate threshold, it is important to first note that the total leverage over all $T$ volumes is mathematically fixed at $Q$, and the existence of each outlier will decrease the leverage of all other volumes. The mean leverage will always be $Q/T$, but the leverage of a typical (non-outlying) volume may be much lower.  Therefore, we use the median leverage across all volumes as a reference value and consider volumes with leverage greater than a factor of the median to be outliers. \cite{mejiaPCALeverageOutlier2017} considered a range of thresholds between $3$ and $8$ times the median, and found $4\times$ to be near optimal. Here, because we are using new and modified methods, different evaluation metrics, and multiband instead of single-band data, we reexamine a range of thresholds up to $8\times$ to determine an appropriate threshold. Note that because leverage is not based on temporal differences like FD and DVARS, it is not necessarily sensitive to change in TR, so a similar threshold may be appropriate for both single-band and multiband data.

\subsection{Other scrubbing methods}\label{sec:other_scrubbing}

For comparison, we consider two established scrubbing methods: motion scrubbing using framewise displacement \citep[FD,][]{powerSpuriousSystematicCorrelations2012, powerMethodsDetectCharacterize2014} and DVARS \citep{smyserLongitudinalAnalysisNeural2010, afyouniInsightInferenceDVARS2018}. While FD is based on head motion and DVARS is based on BOLD signal change, both measure the change from the previous volume. Because FD has been shown to be inflated and noisy in multiband data, we consider a lagged and filtered version of FD based on changes suggested by \cite{power2019distinctions} and \cite{fairCorrectionRespiratoryArtifacts2020}.

\paragraph{FD} Head motion is estimated by assuming the brain is a rigid body and realigning all volumes to a chosen reference (e.g., the mean across all volumes). FD is calculated from the rigid body RPs according to the formula 
$$
\text{FD}(t) = |\Delta d_{xt}| + |\Delta d_{yt}| + |\Delta d_{zt}| + |\Delta \alpha_{t}| + |\Delta \beta_{t}| + |\Delta \gamma_{t}|
$$ 
where $\Delta q_t = q_t - q_{t-1}$ \citep{powerSpuriousSystematicCorrelations2012}. At volume $t$, $d_{xt}$, $d_{yt}$ and $d_{zt}$ represent the change in the brain's estimated position along the $x$, $y$ and $z$ coordinates, respectively relative to the reference; $\alpha_{t}$, $\beta_{t}$ and $\gamma_{t}$ represent the estimated rotation of the brain along the surface of a 50-mm-radius sphere (an approximation of the size of the cerebral cortex) relative to the same reference. Thus, $\text{FD}(t)$ is equal to the sum of absolute changes in the realignment parameters between volume $t$ and the previous volume. By convention $\text{FD}(1)=0$. FD thresholds reported in the literature vary; thus, we examine a range of FD thresholds between 0.2 mm and 0.8 mm. 

\paragraph{Modified FD for multiband data}\label{sec:modFD} \cite{power2019distinctions} and \cite{fairCorrectionRespiratoryArtifacts2020} demonstrate that the higher sampling rate of multiband data introduces two issues which make it more difficult to distinguish high-motion volumes based on FD. First, large movements lasting several seconds are subdivided across volumes, which leads to shorter peaks in FD for detecting volumes to scrub. Second, the constant, moderate head movements due to respiration are no longer aliased, making all volumes have higher FD values: at the respiratory frequency, the RPs now include both true head movements caused by respiratory motion and artifactual variance, especially in the translational RP along the phase encoding direction \citep{fairCorrectionRespiratoryArtifacts2020}. To address the first issue, \cite{power2019distinctions} proposed a lagged version of FD for the HCP, which takes lag-4 differences instead of lag-1 differences when calculating motion from the RPs: $\Delta q_t = q_t - q_{t-4}$. Since the HCP uses a TR of $0.72$ seconds, measuring motion with lag-4 differences yields a sampling rate more comparable to that of the original studies used in the development of FD.  To address the second issue and avoid including head movements which accompany normal respiration in the FD measure, both \cite{power2019distinctions} and \cite{fairCorrectionRespiratoryArtifacts2020} proposed applying a notch filter to the RPs at the respiratory frequency, resulting in a filtered version of FD. Since the specific filters proposed differ, in \autoref{app:filteredFD} we perform a comparison of the proposed filters in conjunction with lag-1 or lag-4 differences based on their effect on FC reliability (see \autoref{sec:reliability_methods}). Based on this analysis, we adopt lag-4 FD with a Chebyshev filter for comparison with other scrubbing approaches. We refer to this combination of lagged and filtered FD as \textit{modified FD} (modFD).

\paragraph{DVARS} 
DVARS measures the standard deviation of the change in the BOLD signal across space and was first used for scrubbing by \cite{smyserLongitudinalAnalysisNeural2010}. Unlike FD, DVARS is a unitless measure and is affected by prior pre-processing of the BOLD data, so appropriate thresholds can vary widely between studies. \cite{afyouniInsightInferenceDVARS2018} proposed a dual DVARS-based method which applies two variations of DVARS that can be thresholded in a principled and standardized manner: ZDVARS measures statistical significance and $\Delta$\%DVARS measures ``practical" significance. Defining
$
\text{A}(t) = \frac{1}{V} \sum_{v=1}^V [\textbf{Y}_{t,v}]^2$ and
$\text{D}(t) = \frac{1}{V} \sum_{v=1}^V [\frac{1}{2} (\textbf{Y}_{t,v} - \textbf{Y}_{t-1,v})]^2
$,
the two variations are

$$
\textrm{ZDVARS}(t) = \Phi^{-1} \left[ 1-P_{\chi^2}\left(\frac{\widetilde{\text{D}}\times2\text{D}(t)}{\text{s}_{\text{hIQR}}^2(\text{D})},\frac{\widetilde{\text{D}}^2\times2}{\text{s}_{\text{hIQR}}^2(\text{D})}\right) \right]
\quad\text{and}\quad
\Delta \%\textrm{DVARS}(t) = \frac{\text{D}(t) - \widetilde{\text{D}}}{\text{mean}(A)} \times 100\%,
$$

where $\widetilde{\text{D}}$ is the median of $\text{D}(t)$ across $t$, $\text{s}_{\text{hIQR}}$ is a robust estimate of  standard deviation, $P_{\chi^2}(q,n)$ is the probability density function for a chi-squared random variable with $n$ degrees of freedom, and $\Phi^{-1}$ is the inverse cumulative distribution function for a standard Normal random variable. 
ZDVARS represents a z-score for signal change, whereas $\Delta$\%DVARS measures the excess variance in the signal change as a percentage of the mean signal change. Following \cite{afyouniInsightInferenceDVARS2018}, we use the same formulation of $\text{s}_{hIQR}$, an upper-tail 5\% Bonferonni FWER significance level as the cutoff for ZDVARS, and 5\% as the cutoff for $\Delta$\%DVARS. Only volumes exceeding both cutoffs are flagged by the dual DVARS method. And by convention, $\text{ZDVARS}(1) = \Delta \%\textrm{DVARS}(1) = 0$

\subsection{Data collection and processing}\label{sec:data}

\subsubsection{Human Connectome Project resting-state data}

To compare the effects of different scrubbing methods, we use resting-state fMRI (rs-fMRI) from the Human Connectome Project (HCP) 1200 Subjects Data Release and Retest datasets \citep{vanessenWUMinnHumanConnectome2013}. 45 young adults participated in both studies, each of whom contributed a total of eight resting-state fMRI runs collected across four visits and two acquisition sequences at each visit (LR and RL phase encoding). We use the 42 participants who have both MPP and ICA-FIX versions of all rs-fMRI scans in full duration.

Each rs-fMRI run includes 1200 volumes over approximately 14 and a half minutes (TR = 0.72 s). We remove the first $15$ volumes of each run to account for magnetization stabilization. We analyze the data in grayordinates space resulting from the HCP surface processing pipeline of cortical data combined with the HCP volume processing pipeline for subcortical and cerebellar structures. This data consists of 91,282 gray matter brain locations across the left and right cortical surfaces, subcortical structures, and cerebellum. The grayordinates data are stored as CIFTI files, and we use the \ciftiTools \R package \citep{pham2022ciftitools} to read, process and analyze them. We use the minimally preprocessed (MPP) rs-fMRI data \citep{glasserMinimalPreprocessingPipelines2013} for our analyses, but we include the HCP's rs-fMRI data denoised with ICA-FIX \citep{griffantiICAbasedArtefactRemoval2014} for comparison with other regression-based denoising strategies.

\subsubsection{Functional connectivity}

\begin{figure}
    \centering
    \includegraphics[width=1\textwidth]{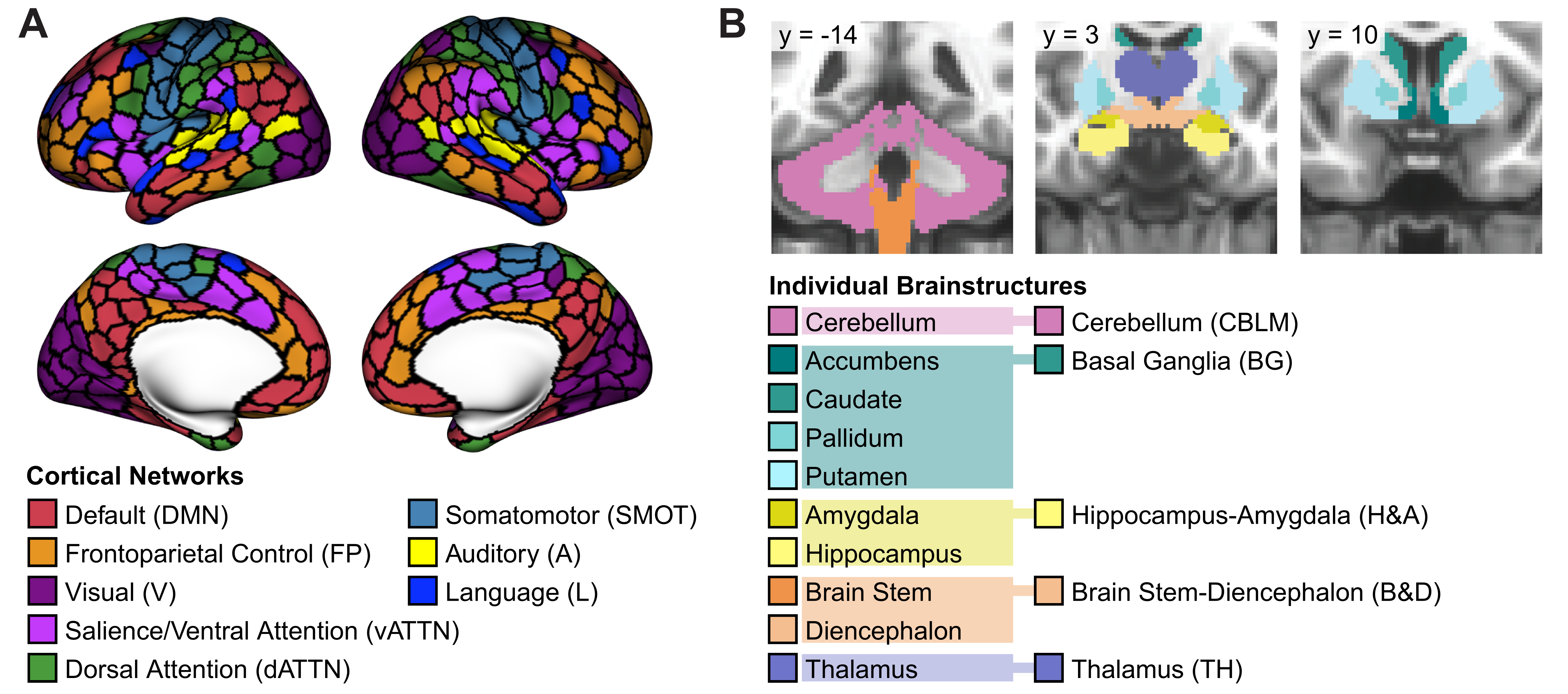}
    \caption{\small \textbf{Schaefer cortical parcellation and Freesurfer cerebellar and subcortical parcellation.} The 400 cortical parcels are outlined in black and grouped into 8 cortical networks.  Groupings of cerebellar and subcortical structures are indicated using color gradients. Note that all cerebellar and subcortical regions except the brainstem are separated into left and right hemisphere in our analysis.}
    \label{fig:Parcels}
\end{figure}

In the series of analyses described below, we assess the impact of scrubbing on the validity, reliability, and identifiability of functional connectivity. To compute FC, we calculate the average timeseries within each of $400$ cortical regions \citep{schaeferLocalGlobalParcellationHuman2018, kong2021individual} and within each of $19$ Freesurfer parcels representing subcortical and cerebellar regions (\autoref{fig:Parcels}).  The $419\times419$ matrix of FC is computed as the Pearson correlation between each parcel timeseries. Prior to further analyses, a Fisher z-transformation is applied to map the FC values to the real line.  Letting $r$ represent Pearson's correlation, the Fisher z-transformation is computed as $z = \frac{1}{2}\ln\left\{{(1+r)}/{(1-r)}\right\}$. We assign the 400 cortical functional areas to eight networks by grouping the 17 smaller networks of \cite{kong2021individual} by name (for example, Visual A--C are grouped into a single visual network, \autoref{fig:Parcels}A). We also define five subcortical groupings based on location, function, and classical separation between midbrain and hindbrain structures (\autoref{fig:Parcels}B). 

\subsubsection{Scrubbing and denoising}\label{sec:cleaning_methods}

\begin{figure}
    \centering
    \fbox{\includegraphics[width=0.95\textwidth]{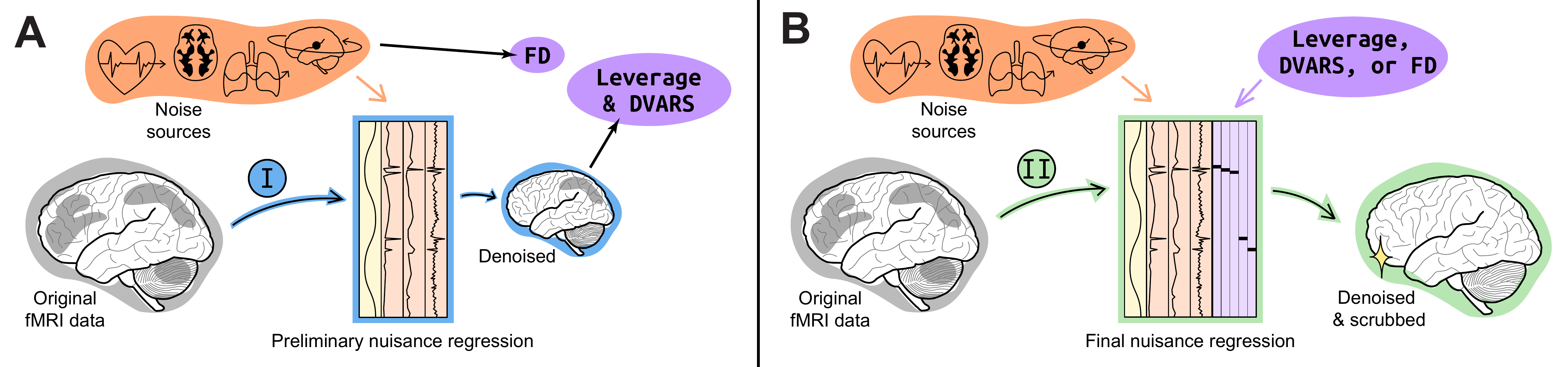}}
    \caption{\small \textbf{Simultaneous regression framework for nuisance regression and scrubbing.} \textbf{(A) Calculation of scrubbing measures.} FD is calculated from the RPs, while projection scrubbing and DVARS are calculated from the fMRI data after a preliminary nuisance regression. The design matrix for the preliminary nuisance regression includes DCT bases (yellow) which remove low-frequency trends, as well as a combination of noise components (orange) such as aCompCor bases, the RPs, noise ICs, and/or the global signal, depending on the selected denoising method. Data-driven scrubbing measures are computed from the nuisance-regressed data (blue brain) rather than the minimally preprocessed data (gray brain) in order to provide a lower noise floor for easier identification of severe, transient artifacts not already eliminated by nuisance regression. \textbf{(B) Inclusion of scrubbing in the regression model.}  The final cleaned data are obtained with a single nuisance regression that includes spike regressors for each volume flagged by the selected scrubbing method, in addition to the original nuisance regressors. The final nuisance regression is performed on the original data (gray brain), rather than the data after preliminary denoising (blue brain), to avoid issues associated with modular or sequential preprocessing. Thus our processing framework performs detrending, denoising, and scrubbing simultaneously. Note that this is equivalent to censoring flagged volumes from the data and design matrix prior to nuisance regression, but is not equivalent to censoring the residuals after an initial nuisance regression.}
    \label{fig:Flowchart}
\end{figure}

There are many choices of denoising and scrubbing procedures to include in an fMRI data cleaning pipeline. Moreover, even a single, fixed set of procedures can be merged into a unified pipeline in many different ways. In this work we use the processing framework illustrated in \autoref{fig:Flowchart}. This pipeline cleans an fMRI run by detrending, denoising, and scrubbing the minimally pre-processed data with a multivariate nuisance regression. Performing these operations simultaneously with a single design matrix, rather than as sequential operations, avoids problems associated with modular preprocessing \citep{lindquistModularPreprocessingPipelines2019}, and has been proposed previously by \cite{jo2013effective}. Further details on our processing framework, as well as comparisons to alternative pipelines, is given in \autoref{app:nreg_framework}. 

As illustrated in \autoref{fig:Flowchart}A, we perform a preliminary nuisance regression for detrending and denoising the data before computing the data-driven scrubbing measures. This attenuates trends and certain noise patterns, providing a lower noise floor and allowing for greater sensitivity of data-driven scrubbing to burst noise. Once these scrubbing measures are obtained, we modify the original nuisance regression by adding a spike regressor for each flagged volume to the design matrix of nuisance regressors. This is illustrated in \autoref{fig:Flowchart}B. This regression is also performed on the original data, so that the cleaned data are produced from a single simultaneous regression. Flagged volumes are removed from the data before computing FC.

\autoref{fig:Flowchart} also illustrates how the marginal impact of scrubbing can be measured: differences between the data obtained from the blue path versus that obtained from the green path will reflect the improvements due to scrubbing, above and beyond the impact of denoising. In this work we obtain baseline FC estimates using denoising alone (\autoref{fig:Flowchart}A) and then estimate FC again using the denoised and scrubbed data (\autoref{fig:Flowchart}B). The change in validity, reliability, or identifiability of the FC estimates compared to the baseline is the improvement attributable to a particular scrubbing method.


\paragraph{Regression-based denoising} 
The benefit of scrubbing depends on which baseline denoising method is used. For example, if regression-based denoising is highly effective at removing artifacts, less scrubbing may be required to eliminate residual noise. We therefore first consider the following candidate strategies, described in \autoref{tab:denoising}: no additional denoising (MPP), anatomical CompCor (CC$x$), 2 parameter (P), 9P, and 36P 
\citep{behzadiComponentBasedNoise2007, satterthwaiteImprovedFrameworkConfound2013, muschelliReductionMotionrelatedArtifacts2014, satterthwaiteMotionArtifactStudies2019, parkesEvaluationEfficacyReliability2018}. We also investigate including motion parameters (MPs) in addition to white matter and CSF-derived signals in aCompCor. Specifically, we consider the inclusion of the 6 RPs (CC$x$+6P) or those parameters along with their one-back differences, squares, and squared one-back differences (CC$x$+24P).  For each denoising strategy, we include four discrete cosine transform (DCT) bases for detrending and high-pass filtering. We adopt high-pass rather than band-pass filtering because removing high-frequency components from rs-fMRI has been shown to worsen signal-noise separation and reliability of FC estimates \citep{shirerOptimizationRsfMRIPreprocessing2015}.  

\begin{table}
\small
    \centering
    \begin{tabular}{@{}p{0.13\linewidth}|p{0.82\linewidth}@{}}
    \textbf{Short Name} & \textbf{Description} \\
    \hline
         MPP & No additional denoising \\    \hline

         CC$x$ & Anatomical CompCor based on the top $x\in\{1,\dots,10\}$ principal components within masks of cerebral white matter, cerebral spinal fluid (CSF), and cerebellar white matter in volume space. Two voxel layers are eroded from the white matter ROIs, and one voxel layer is eroded from the CSF ROI. \\    \hline

         2P & Mean signals from white matter (cerebral and cerebellar) and CSF. Erosion as in aCompCor. \\    \hline

         9P & 2P signals along with the six motion RPs and the global signal.  Global signal is calculated across in-mask voxels in the volumetric MPP data. \\    \hline

         36P & 9P signals along with their one-back differences, squares, and squared one-back differences \\     \hline
         
         
    \end{tabular}
    \caption{\small \textbf{Candidate denoising strategies.} All denoising strategies are applied to the minimally preprocessed (MPP) data via a regression framework using a design matrix including the noise regressors, an intercept, and (possibly) four DCT bases for detrending and high-pass filtering. The strategy resulting in the most reliable FC estimates will be adopted as the baseline denoising strategy, to be applied simultaneously with scrubbing.}
    \label{tab:denoising}
\end{table}

We also include the HCP implementation of ICA-FIX for comparison. 
Briefly, the volumetric and surface MPP BOLD data and 24 motion parameters are all initially detrended with a high-pass filter. ICA is performed on the volumetric BOLD data, and noise ICs are identified based on a trained classifier. The noise ICs along with the 24 MPs are then ``softly'' regressed from the surface BOLD data, meaning that any shared variance with the signal ICs is removed from these nuisance signals prior to the regression \citep{griffantiICAbasedArtefactRemoval2014}. Because ICA-FIX causes attenuation of spikes in the fMRI timeseries (similar to the ``DVARS dips'' observed by \citep{glasser2018using}), it is not a viable choice as a baseline denoising method prior to data-driven scrubbing in our framework. 

\paragraph{Scrubbing} For several of the analyses described below, we must determine a scrubbing threshold. We consider a range of thresholds for projection scrubbing ($2\times$ to $8\times$, in multiples of $1$) and both FD and modFD ($0.2$ to $0.8$ mm, in multiples of $0.1$) .
We use the dual DVARS cutoff method as described above and therefore do not need to consider different cutoffs for DVARS. 

\subsection{Quantifying the effects of scrubbing on downstream analysis}

We perform an extensive comparison of projection, motion, and DVARS scrubbing in terms of their effects on downstream analysis. Here we focus on functional connectivity (FC), which is often the measure of interest in resting-state fMRI-based analyses. We use FC quality to gauge the effects of scrubbing on signal-to-noise in fMRI, and we expect our observations on FC to be fairly generalizable to other fMRI-derived measures of brain function and organization.

When large amounts of data are available for individuals as in the HCP, patterns in FC have been shown to be dominated by organizing principles that are common across participants, as well as individual-specific elements of brain networks that are stable across time \citep{Gratton2018FunctionalVariation}. Effective noise removal strategies will therefore produce FC estimates that are more stable or reliable across multiple sessions collected from the same subject. On the other hand, overly aggressive denoising or scrubbing can result in the loss of meaningful signal to the detriment of subsequent analyses. The goal for data cleaning is to achieve a balance between removal of noise and retention of true signal.

We examine the effect of scrubbing on the following measures of FC estimation quality: 
validity, mean absolute change (MAC), reliability, and fingerprinting.  In addition, we assess the agreement of projection scrubbing with motion scrubbing and DVARS in terms of the volumes they flag, and the effect of each scrubbing method on the exclusion of sessions due to a lack of sufficient data remaining.

\subsubsection{Validity of FC}\label{sec:validity_methods}

We quantify the accuracy or validity of FC estimated from short (10 min.) resting-state fMRI scans, relative to a ``ground truth'' value for FC based on approximately 1.7 hours of data across all four sessions (visits) in the HCP main and retest datasets. More specifically, to compute the FC estimate based on the held-out 10 minutes of data, we combine  five minutes from the middle of the LR run with five minutes from the middle of the RL run for a given session. This strategy balances the proportions of LR- and RL-acquisition data used for each FC estimate. Any subjects for which less than $5$ minutes of the original $10$ minutes remain after scrubbing, for at least one session, are removed from analysis to ensure results for all methods are based on sufficient data. The ground truth for a given subject is then calculated from the remaining volumes of both runs, along with all $14.4 \times 2 \times 3$ minutes of data from the other three sessions, totaling approximately 1.7 hours. We scrub the data for the ground truth by censoring any volumes with modFD over 0.2mm (in addition to baseline denoising). Given the long duration and aggressive motion scrubbing, this can be assumed to produce a nearly noise-free FC matrix for each subject to serve as ground truth. This process is repeated for all four sessions (visit 1, visit 2, retest visit 1, retest visit 2) for each subject. 

Effective scrubbing should yield more accurate FC estimates by reducing noise and retaining as much signal as possible. Scrubbing that achieves noise removal but discards too much signal can be detrimental to FC, as well as other fMRI-based measures, by increasing sampling variability, thereby worsening accuracy. We quantify the accuracy of the 10-minute FC estimates using root mean squared error (RMSE), defined as
$$
\textrm{RMSE}=
\frac{1}{AS}\sum^{A}_{a=1}\sum^{S}_{s=1}\sqrt{\frac{1}{P}\sum^{P}_{p=1}{\left(\hat{z}_{asp}-z^*_{asp}\right)^2}},
$$
where $A=4$ is the number of sessions, $S=42$ is the number of subjects, and  $P$ is the number of unique FC pairs. The 6-minute FC estimate for pair $p$, session $a$, and subject $s$ is denoted $\hat{z}_{asp}$ and ${z^*_{asp}}$ is the corresponding ground truth value based on the held-out data. Note that the ground-truth for a given subject $s$ and pair $p$ is session-specific, because the held-out data is slightly different for each session.

\subsubsection{Mean absolute change (MAC)}

Mean absolute change (MAC) has been proposed as an alternative to QC-FC as an evaluation metric for scrubbing efficacy \citep{williams2022advancing} and is related to a measure proposed earlier by \cite{powerMethodsDetectCharacterize2014}.  
QC-FC is defined as the correlation across subjects between mean FD and the estimated strength of each FC pair, and it measures scrubbing's effectiveness for eliminating the influence of head motion on FC estimates \citep{powerSpuriousSystematicCorrelations2012, satterthwaiteImpactInScannerHead2012}. If head motion phenotype truly has no relationship with any functional connection, then QC-FC could be used a measure of validity, with lower-magnitude QC-FC indicating more effective data cleaning. Although commonly used, QC-FC has been criticized because in-scanner movement may relate to neural patterns of interest and because it does not consider signal loss \citep{williams2022advancing,raval2022pitfalls}. \cite{williams2022advancing} observed that QC-FC behaved in unexpected ways when the FD cutoff for motion scrubbing was varied. In some cases, motion scrubbing had minimum QC-FC despite the cutoff being so high that virtually nothing was censored. In other cases, QC-FC values continued to decrease despite motion scrubbing resulting in the removal of more than half of the scan. They proposed an alternative, mean absolute change (MAC), defined as the absolute value of the change in FC estimates relative to random scrubbing of the same number of volumes:
$$
\textrm{MAC}=\frac{1}{SP}\sum^{S}_{s=1}{\sum^{P}_{p=1}{\bigl\lvert\frac{1}{R}\sum^{R}_{r=1}\Delta z_{rsp}\bigr\rvert}},
$$
where $S$ and $P$ are as defined above, and $R=8$ is the number of runs per subject. Based on $Q$ random scrubbing permutations, $\Delta z_{rsp} = \frac{1}{Q}\sum^{Q}_{q=1}\left(z_{rsp} - z_{{RAND}_{rsp}^q}\right)$ is the change in FC values for a given scan between scrubbing and random scrubbing across $Q$ permutations, where $z_{{RAND}_{rsp}^q}$ is $z_{{RAND}_{rsp}}$ for permutation $q$. {MAC} is based on the premise that if scrubbing measures do not positively correlate with true neural signals, then any systematic change in FC represents a reduction in noise. Thus, if two scrubbing methods flag volumes at a similar rate, the one with the higher MAC is deemed to have achieved greater signal-to-noise ratio improvement. Under this premise, MAC can be used as an indicator for validity. 

Note that MAC is based on the \textit{absolute} change in FC for each unique pair. Therefore, if removal of motion artifacts through scrubbing tends to increase the FC magnitude of long-range connections and decrease the decrease the magnitude of short-range connections, as observed in prior work \citep{powerSpuriousSystematicCorrelations2012}, these changes will not cancel out but will instead combine to increase MAC.

It is important to note that 
using increasingly strict cutoffs tends to yield higher MAC values, regardless of the cutoff's effectiveness at noise removal versus signal retention. The fact that MAC increases with greater censoring does not imply better performance. It is therefore inappropriate to use MAC to determine a threshold for a particular scrubbing method. Rather, MAC should only be used to compare across different scrubbing methods at cutoffs for which they remove similar amounts of data \citep{williams2022advancing}. 


\subsubsection{FC Reliability}\label{sec:reliability_methods}

The intra-class correlation coefficient \citep[ICC,][]{shroutIntraclassCorrelationsUses1979} has been used in a number of previous studies \citep[e.g.,][]{shehzad2009resting, zuo2010reliable, thomason2011resting, guo2012one, shirerOptimizationRsfMRIPreprocessing2015, noble2019decade} as a measure of FC reliability. ICC represents a ratio of variances, with $\text{ICC}=1$ indicating that all of the variance in the estimates is attributable to between-subject differences, with no noise variance across repeated measures of the same individual.  At the other extreme, $\text{ICC}=0$ indicates that all of the variance is attributable to within-subject noise variance across repeated measures. ICC therefore reflects the presence of unique individual signal and reduced noise and measures how well estimated FC reflects unique subject-specific traits.

We adopt the ICC$(3,1)$ because it is appropriate for measuring test-retest reliability when the reliability of a single measurement is of interest \citep{kooGuidelineSelectingReporting2016}, and to facilitate comparison with previous studies of FC reliability \citep{caceresMeasuringFMRIReliability2009, parkesEvaluationEfficacyReliability2018}. ICC$(3,1)$ is based on the relationship between mean sum-of-squares between (MSB) and mean sum-of-squares within (MSW). For a given FC pair, let $z_{sr}$ represent the FC estimate for subject $s$ based on run $r$. Let $M_s = \frac{1}{R} \sum_{r=1}^R z_{sr}$ be the average FC estimate across the $R$ runs for that subject, and let $\bar{M} = \frac{1}{S} \sum_{s=1}^S \sum_{r=1}^R M_s$ be the average across subjects. Then the ICC$(3,1)$ is given by
$$
\text{ICC}(3,1) = \frac{\text{MSB} - \text{MSW}}{\text{MSB} + (R-1)\text{MSW}},
$$
where
$$
\text{MSB} = \frac{R}{S-1} \sum_{s=1}^S (M_s - \bar{M})^2\quad\text{and}\quad
\text{MSW} = \frac{1}{S(R-1)} \sum_{s=1}^S \left( \sum_{r=1}^R (z_{sr} - M_r)^2 \right).
$$

We must note that some patterns of motion-induced noise in fMRI data may be reliable across multiple runs from the same subject. Removing such sources of noise could reduce ICC. Reductions in ICC are therefore difficult to interpret, since they can result from loss of reliable signal, removal of reliable noise patterns, or both.  However, we assert that \textit{increases} in ICC that occur due to scrubbing are most likely due to removal of noise while retaining reliable signal. That is, scrubbing cannot \textit{introduce} reliable noise or any other type of signal or noise, only remove.  Therefore, we interpret increases in ICC due to scrubbing as beneficial, representing an improved balance between signal and noise. 


\subsubsection{Fingerprinting}\label{sec:fingerprinting_methods}
Previous work has shown FC of individual subjects to be identifiable across multiple sessions via ``fingerprinting'' \citep{finnFunctionalConnectomeFingerprinting2015}. We therefore examine the effect of scrubbing on the fingerprinting match rate. We divide the eight runs available from each participant into four sets paired by acquisition and dataset: the two test LR scans, the two test RL scans, the two retest LR scans, and the two retest RL scans. For each set of paired runs, we perform fingerprinting using the first run as the database set and the second as the query set. For each scan in the query set, we calculate Pearson correlations between its FC and that of each scan in the database set. If the database scan with the highest correlation is from the same subject, the fingerprint is a match. This procedure is repeated, swapping the roles of database set and query set within each pair, to yield eight rounds of matching. The overall match rate is computed as the rate of successful matches across $8 \times 42$ total queries.
Because previous studies have shown that using connections only within individual brain networks can yield more accurate fingerprinting than using all connections across the brain \citep{finnFunctionalConnectomeFingerprinting2015}, we repeat this procedure using only the connections within each network or the subcortex. 

As with ICC above, we note that scrubbing could potentially diminish the fingerprint match rate if noise patterns being removed are reliable and identifiable across multiple runs from the same subject.  It is therefore important to interpret reductions in fingerprint match rate due to scrubbing with caution. However, we interpret increases in match rate due to scrubbing as beneficial, since exclusion of volumes via scrubbing can only remove signal or noise; scrubbing cannot introduce reliable noise patterns that would enhance the success of fingerprinting.




\section{Results}
\label{sec:results}

\subsection{Selection of baseline denoising strategy}
\label{sec:baseline-denoising}

\begin{figure}
    \centering
    \includegraphics[width=1\textwidth]{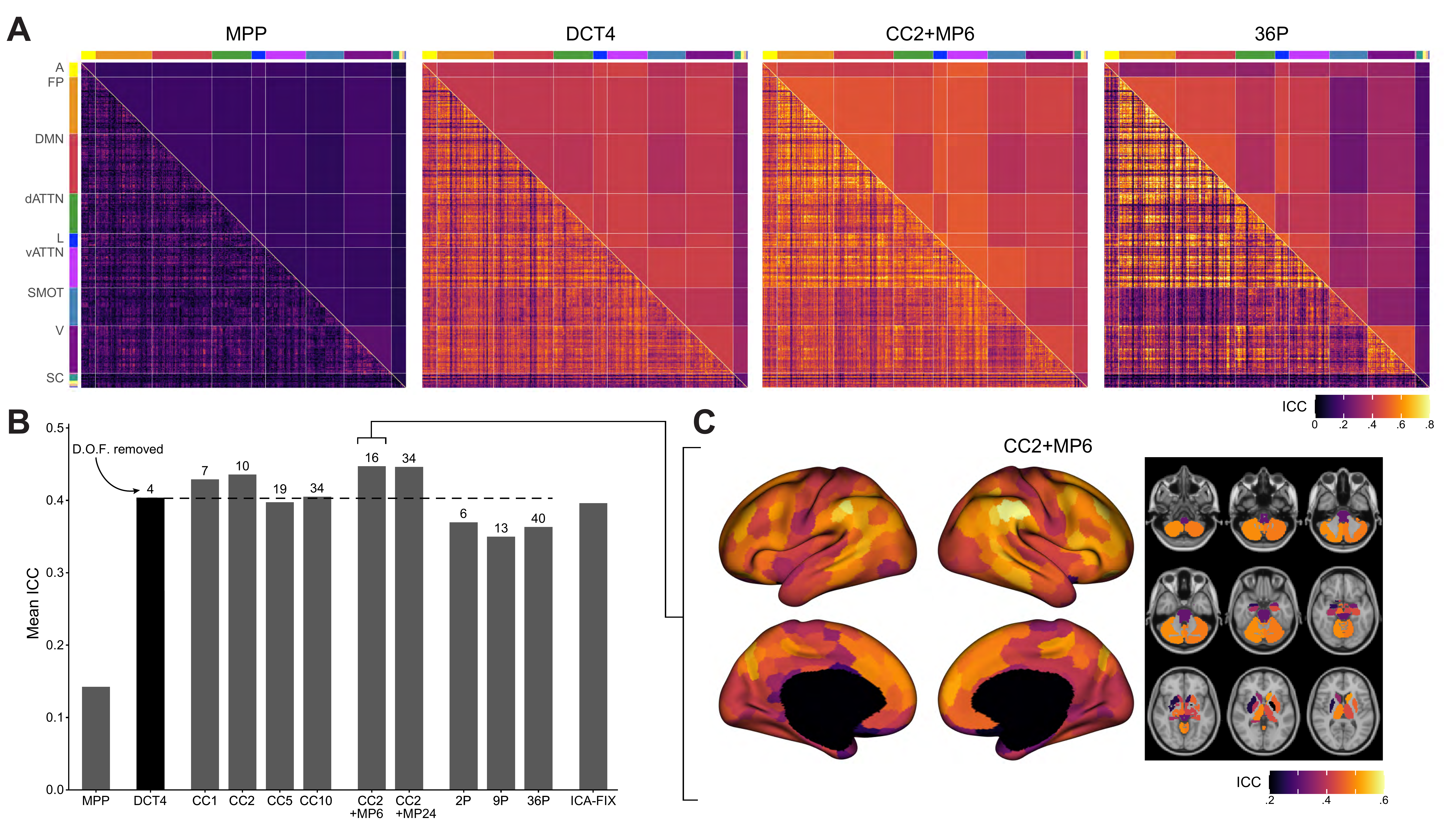}
    \caption{\small \textbf{CC2+MP6 yields the highest reliability across the connectome, while preserving temporal degrees of freedom.} \textbf{(A)} Average ICC of functional connectivity (FC) estimates across subjects and sessions for different denoising strategies, from left to right: minimally preprocessed (MPP), four DCT bases (DCT4), aCompCor with two components per noise ROI (CC2) plus six RPs (CC2+6MP), and the 36 parameter model (36P). CC2+6MP and 36P both include DCT detrending. Higher ICC values indicate more reliable FC estimates. The top halves of the matrices represent the mean ICC values for each network pair, i.e. for each corresponding region in the lower triangles. \textbf{(B)} The mean ICC across all functional connections for each denoising method. All methods include DCT detrending except MPP and ICA-FIX. CC2+MP6 and CC2+MP24 both achieve the highest reliability, while CC2+MP6 maintains many more degrees of freedom, and so we adopt it as the baseline denoising strategy for subsequent analyses. \textbf{(C)} The mean ICC of FC estimates across all connections involving each parcel or subcortical structure, for the CC2+MP6 denoising method.}
    \label{fig:baselineICC}
\end{figure}

\begin{figure}
{
    \centering
    \includegraphics[width=.95\textwidth]{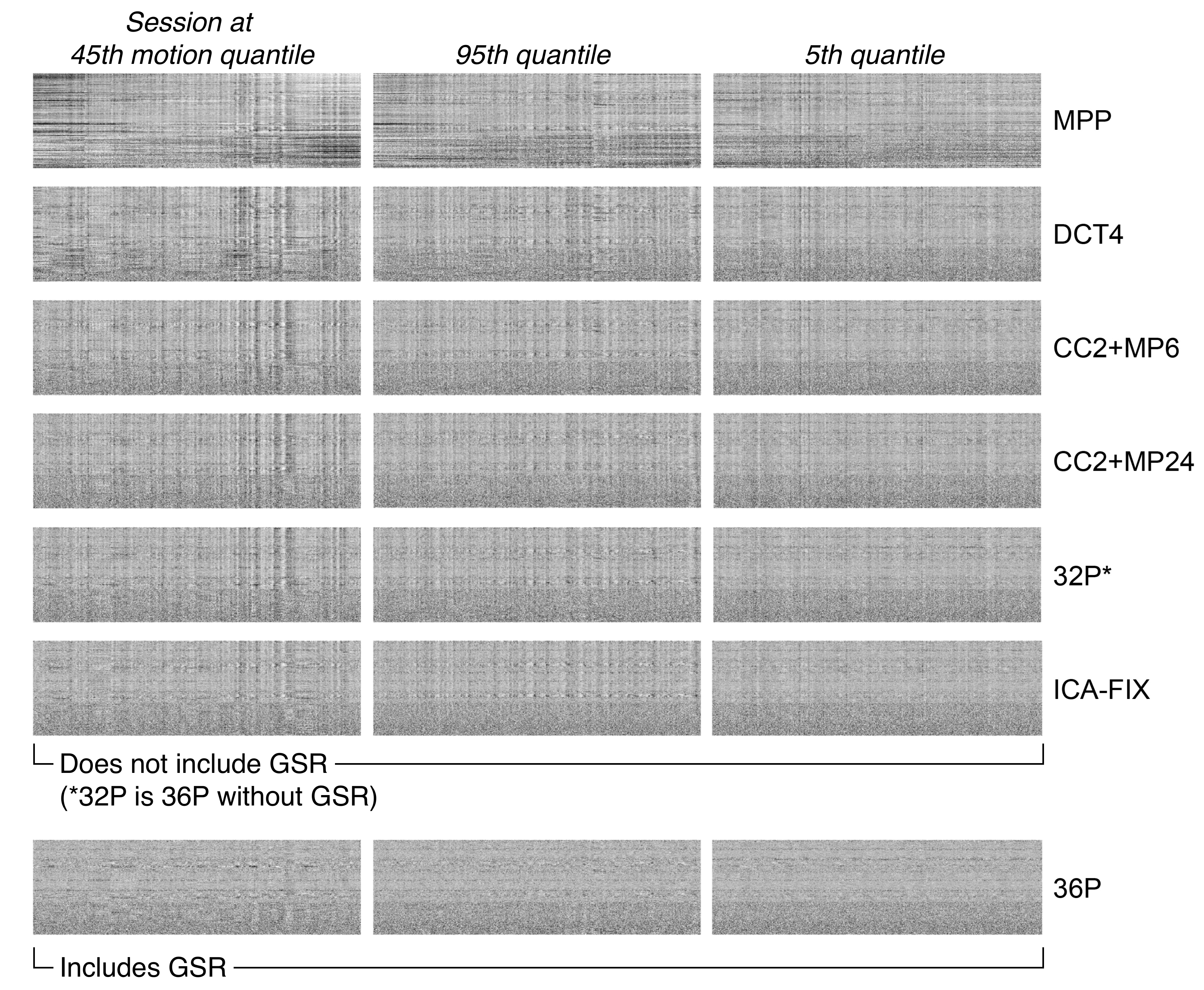}
    }
    \caption{\small
    \textbf{Grayplots for different denoising strategies for three example sessions.} The example sessions represent a range of head motion levels and are the same as those shown in \autoref{fig:Example} and Appendix Figures \ref{app:fig:Example2} and \ref{app:fig:Example3}. Horizontal bands on the grayplots indicate local low-frequency variation, while vertical bands indicate global high-frequency variation. 36P clearly eliminates the latter more than other methods due to inclusion of global signal components. 32P is a version of 36P without any global signal components. Among the methods that do not remove the global signal, CC2+MP6, CC2+MP24 and 32P all appear to achieve similar reductions of both low- and high-frequency noise based on visual inspection.}
    \label{fig:grayplots}
\end{figure}

\autoref{fig:baselineICC} examines the effect of different regression-based denoising strategies on the reliability of FC. Panel A displays the connection-level ICC for MPP data and three selected denoising strategies. All strategies generally improve reliability over MPP. Among those shown in Panel A, CC2+MP6 achieves the highest reliability, followed by DCT4, followed by 36P. For 36P, there is substantial variability in reliability across connections. Connections involving some cortical parcels are commonly reliable while inter-hemispheric connections in the somatomotor network are typically very unreliable, and subcortical connections are almost always unreliable. By comparison, CC2+MP6 achieves more uniform reliability across the connectome. Many subcortical connections, which were less reliable with either MPP or DCT4 denoising, improve with CC2+MP6 denoising. 

Panel B compares all denoising strategies considered in our study by showing the mean ICC across all connections. 
First, DCT4 detrending alone dramatically improves reliability compared with the MPP data. Of all the denoising methods we consider, only aCompCor improves upon DCT4, with CC2 resulting in the highest reliability. The inclusion of 6 or 24 MPs in CC2 slightly improves reliability over CC2 alone. Including 24 MPs rather than 6 achieves a similar improvement but removes many more degrees of freedom. 


Panel C displays the average reliability of connections involving each cortical parcel and subcortical/cerebellar region after CC2+MP6 denoising. Although CC2+MP6 results in more uniform FC reliability compared to other denoising methods, baseline differences in reliability remain across the brain prior to scrubbing. The most reliable connections are those involving the fronto-parietal, attention and default mode networks. Somewhat surprisingly, the cerebellum is fairly reliable, possibly because we include cerebellar white matter signals in our aCompCor implementation. The least reliable connections include those involving the somatomotor network and subcortical regions. 

Appendix \autoref{app:fig:baselineVarDecomp} expands on \autoref{fig:baselineICC} by decomposing ICC into within-subject and between-subject variability.  Within-subject variation (MSR) is residual variance after accounting for differences between subjects and any visit effects. ICC increases with higher MSB and/or lower MSR. We see that all denoising methods reduce both within-subject variance (MSR) and between-subject variance (MSB). Within a given denoising strategy (e.g., CC$x$), both measures of variance decrease when more nuisance regressors are included. The highest reliability is achieved when there is an optimal trade-off. In contrast, 36P is highly effective at reducing within-subject variance but also reduces between-subject variance dramatically, resulting in lower ICC. CC2+MP6 maintains a high level of between-subject variance while reducing within-subject variance, ultimately achieving the most reliable FC estimates. 


Recall that because removal of reliable noise patterns can actually reduce ICC, higher ICC does not uniformly indicate more accurate functional connectivity. To further assess the efficacy of CC2+MP6 compared with methods that more explicitly target head-motion induced noise, we conduct two additional analyses. First, \autoref{app:fig:baselineFC} and \autoref{app:fig:baselineFC_diff} display the strength of functional connectivity with different denoising methods, including the difference in FC strength between inter- and intra-hemispheric connections and between inter- and intra-network connections.  Effective denoising is expected to increase inter-hemispheric connectivity relative to intra-hemispheric connectivity (attenuation of inflated connectivity for close-range connections), and increase intra-network connectivity relative to inter-network connectivity (clearer resolution of network signals). \autoref{app:fig:baselineFC} Panel A shows mean FC strength across subjects for each FC pair. Some patterns are apparent, including strengthened within-network FC along the diagonal blocks, but visual comparison is difficult because 36P shifts the distribution due to the inclusion of global signal components. To facilitate a direct comparison of inter/intra network FC strength, \autoref{app:fig:baselineFC_diff} shows histograms for the FC pairs, aggregated separately for inter-network connections and intra-network connections. Compared with DCT4 detrending only, all denoising methods strengthen intra-network FC relative to inter-network FC. As a summary measure, we report the difference between median intra-network FC and median inter-network FC.  Besides ICA-FIX, the largest median difference is achieved by CC10 and 9P, followed closely by CC5, 36P and CC2+MP6. \autoref{app:fig:baselineFC} Panel B shows that CC2MP6 also increases intra-hemispheric connectivity relative to inter-hemispheric connectivity, consistent with the idea of mitigating motion-induced inflation of proximal connections.

We also examine grayplots \citep{powerSimpleUsefulWay2017} from the same example sessions as those shown in \autoref{fig:Example} and Appendix Figures \ref{app:fig:Example2} and \ref{app:fig:Example3}, which represent the 45th, 95th and 5th quantiles of head motion, respectively. \autoref{fig:grayplots} displays the grayplots for these three sessions using minimal preprocessing, DCT4 detrending, CC2+MP6, CC2+MP24, 36P, and ICA-FIX. With MPP, horizontal bands indicative of low-frequency variation are clearly seen in all three sessions, as well as vertical bands indicative of transient changes. DCT4 detrending yields obvious reductions in low-frequency variation. Note that 36P quite obviously yields the greatest reductions in high-frequency variation, since it is the only method here that includes global signal components. For a more fair visual comparison with the other methods, we display a version of 36P without global signal components, which we denote 32P.   The methods that do not include GSR, namely CC2+MP6, CC2+MP24 and 32P appear to do a similar job of mitigating both low-frequency and high-frequency noise. We adopt CC2+MP6 as the baseline denoising strategy for all subsequent analyses, as it achieves a balance between noise reduction and degrees of freedom retained and was shown to improve reliability (see \autoref{fig:baselineICC}). Recall that all scrubbing is applied on top of baseline denoising, using the simultaneous regression framework illustrated in \autoref{fig:Flowchart}.

\subsection{Effect of different scrubbing methods on MAC}

\begin{figure}
    \centering
    \includegraphics[width=0.7\textwidth]{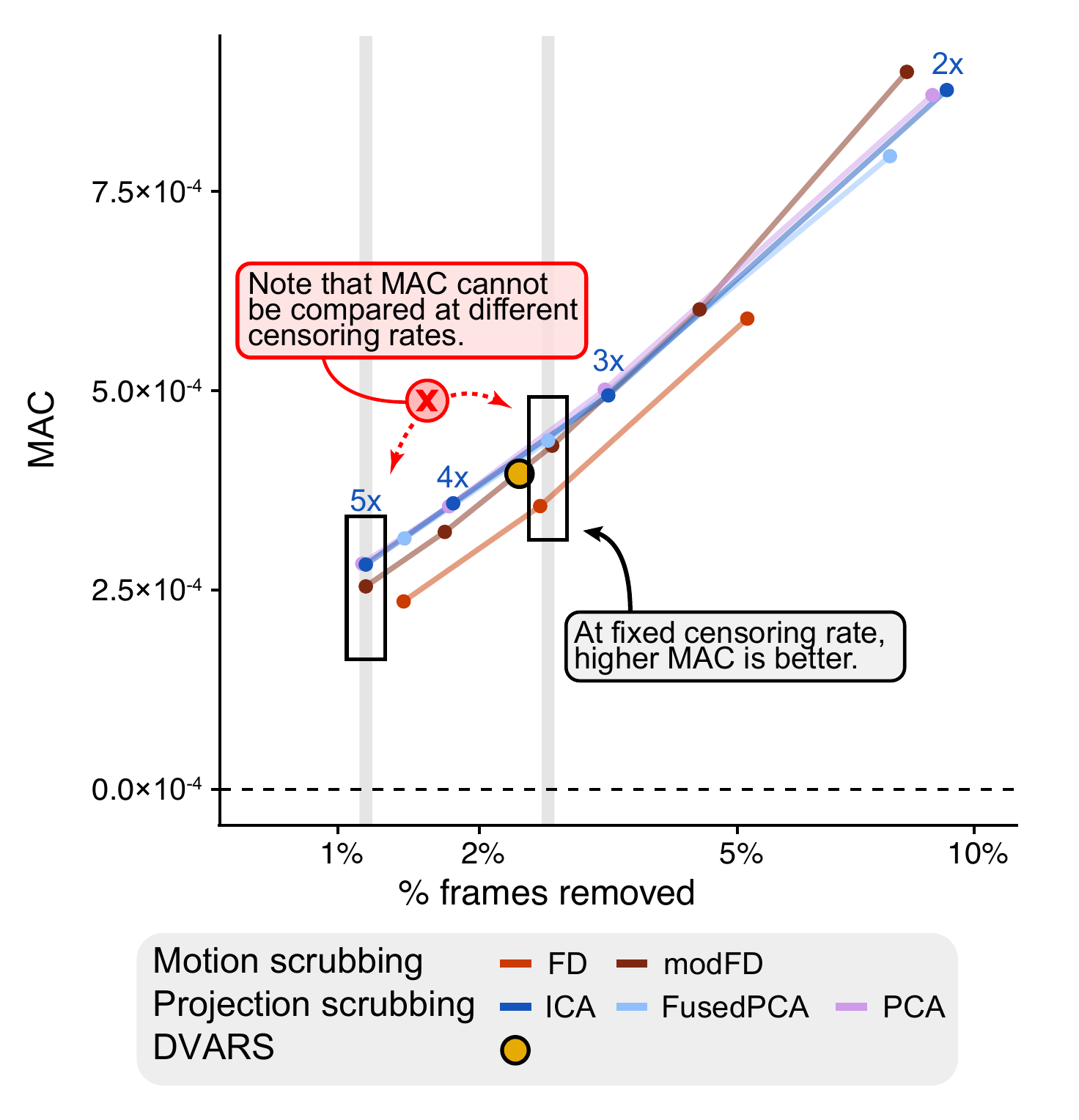}
    \caption{\small 
    \textbf{Effect of different scrubbing methods on MAC.} We compare the effect of projection scrubbing, motion scrubbing, and DVARS on mean absolute change (MAC) in functional connectivity. MAC is measured relative to random removal of the same number of volumes. At a fixed censoring rate, higher MAC is thought to indicate more effective scrubbing because it suggests more efficient removal of the influence of noise. Since MAC increases monotonically with higher censoring rates, it should not be used to compare across different censoring rates or to choose a scrubbing threshold. To facilitate method comparisons at the same censoring rate, we show results for all candidate thresholds that remove $1\%$ to $10\%$ of volumes, on average. The x-axis is on a log scale. Because projection scrubbing tends to retain more volumes than motion scrubbing, this means that here, in order to compare across methods at higher censoring rates (near 10\%), we include a threshold of $2\times$ for projection scrubbing, more stringent than is generally recommended. At these higher censoring rates, modFD has higher MAC than projection scrubbing, suggesting that more aggressive censoring with projection scrubbing may not be beneficial over motion scrubbing. At the lower censoring rates typically seen with data-driven scrubbing (below $5\%$), however, projection scrubbing with any projection method has higher MAC than motion scrubbing or DVARS, suggesting it best improves the signal-to-noise ratio.}
    \label{fig:MAC}
\end{figure}


We next compare scrubbing methods in terms of mean absolute change (MAC) in functional connectivity, a measure of scrubbing efficacy at a given volume removal rate \citep{williams2022advancing}. As described in \autoref{sec:validity_methods}, MAC measures the change in FC estimates relative to randomly scrubbing at the same censoring rate. It is thought to reflect the effect of scrubbing on the validity of FC. That is, higher MAC is indicative of improved validity \textit{at a given rate of removal}. MAC is not comparable across different censoring rates, and can therefore not be used to determine an appropriate threshold for scrubbing. However, it is useful for comparing the efficacy of different scrubbing methods at a fixed rate of removal. 

\autoref{fig:MAC} shows MAC versus censoring rate for projection scrubbing and motion scrubbing. In order to facilitate comparisons between motion scrubbing and projection scrubbing, we show results for all candidate thresholds that remove $1\%$ to $10\%$ of volumes. Thus, we include more stringent projection scrubbing thresholds (e.g., $2\times$) and more lenient motion scrubbing thresholds (e.g, 0.6mm) than are typically recommended. If we consider a threshold of $2\times$ the median, which is more aggressive than recommended in previous work, the projection-scrubbing censoring rate is more comparable to motion scrubbing. At this higher censoring rate, motion scrubbing with modFD has higher MAC. This suggests that a $2\times$ threshold for projection scrubbing is overly stringent and likely results in more removal of true signal and/or less effective noise removal, compared with motion scrubbing. Yet at the lower censoring rates typical of data-driven scrubbing ($<5\%$ of volumes), projection scrubbing outperforms motion scrubbing and DVARS, suggesting that it yields the most improvement to signal-to-noise ratio.

Motion scrubbing with FD results in substantially less effective noise removal than other methods, based on its consistently lower MAC across all censoring rates. This underscores that standard FD is not an appropriate scrubbing metric for motion scrubbing in multiband data, and lagged and/or filtered versions like modFD should be used.

The three projection scrubbing methods perform very similarly. For simplicity, we focus primarily on ICA as the projection method in projection scrubbing in the remainder of analyses.

\subsection{Selection of scrubbing thresholds}\label{sec:thresholds}

\begin{figure}
    \centering
    \includegraphics[width=1\textwidth]{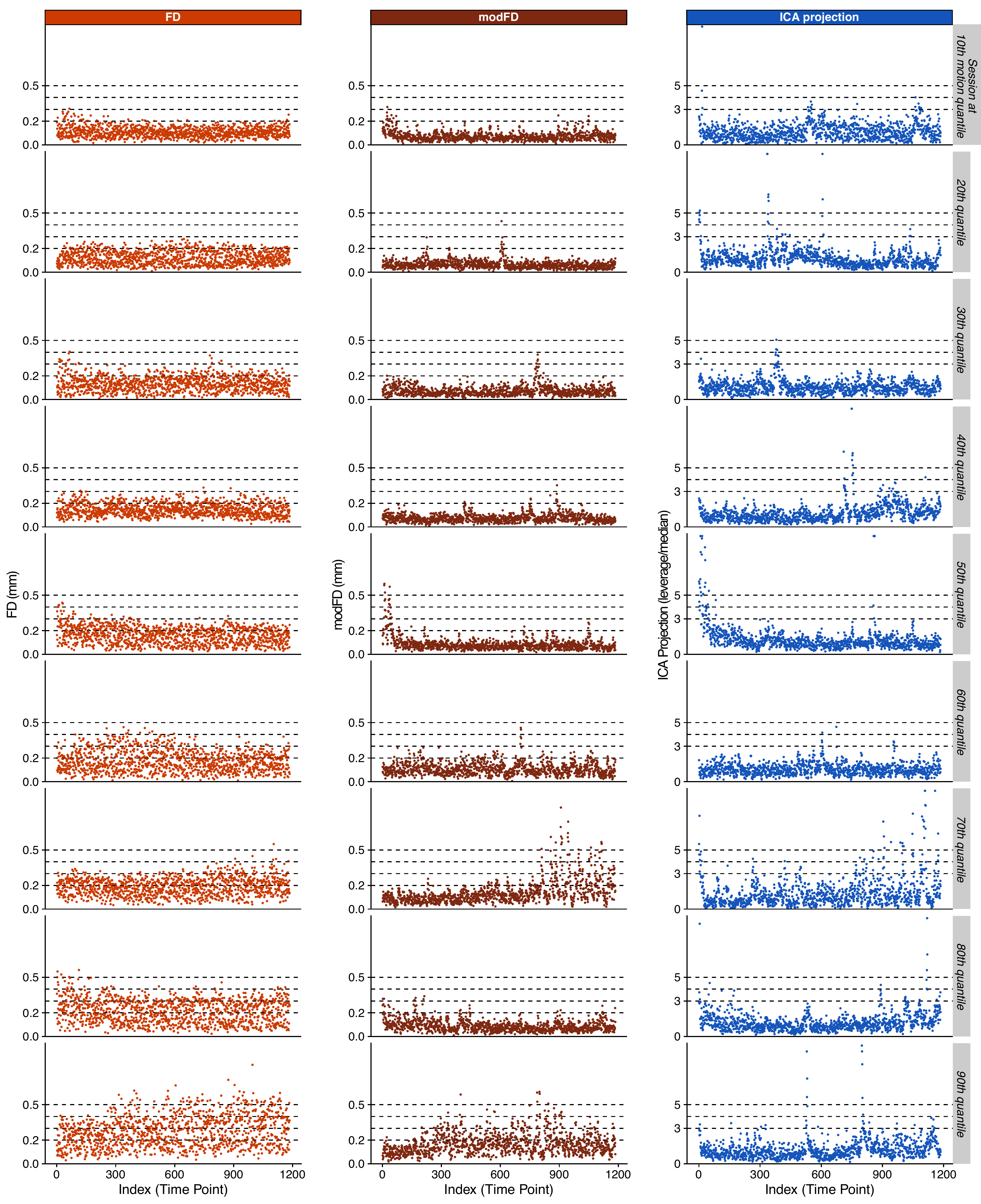}
    \caption{\small \textbf{Comparison of scrubbing thresholds for several example sessions.} Nine example sessions are shown, representing the 10th to 90th quantiles of head motion. For FD and modFD, thresholds of 0.2mm, 0.3mm, 0.4mm and 0.5mm are shown. For projection scrubbing, thresholds of 3, 4 and 5 times the median leverage are shown. Based on these results, in subsequent analyses we adopt thresholds of 0.2mm for modFD, 0.3mm for FD, and 3$\times$ for projection scrubbing. These results show that standard FD is noisy and inflated and largely fails to capture spikes in head motion that are apparent with lagged and filtered FD, in agreement with prior work. These results also show common discrepancies between the modFD timeseries and actual abnormalities in the data occurring near head motion, seen as spikes in projection scrubbing leverage.}
    \label{fig:NineSubjects}
\end{figure}

Before proceeding to subsequent analyses, we first consider the choice of an appropriate threshold for motion scrubbing and projection scrubbing. For motion scrubbing, FD of 0.2mm is often recommended in single-band data, but there is not yet a consensus around an appropriate choice of threshold for lagged and/or filtered variants of FD in multiband data. Here, we consider thresholds between 0.2mm and 0.5mm, ranging from the ``stringent" 0.2mm cutoff to the ``lenient" 0.5mm cutoff as characterized by \citet{powerMethodsDetectCharacterize2014}. For projection scrubbing, previous work based on single-band data looked at thresholds of 3 to 8 times the median leverage across the scan and found $4\times$ to be near-optimal \citep{mejiaPCALeverageOutlier2017}. Leverage is in theory less sensitive to TR than measures like FD and DVARS based on temporal differencing, so a similar threshold may work equally well in single-band and multiband data. However, here we revisit the choice of threshold to ensure it is appropriate for the data at hand. For DVARS, we adopt the dual thresholding method proposed by \cite{afyouniInsightInferenceDVARS2018} and therefore do not consider different thresholds. 

\autoref{fig:NineSubjects} displays standard FD, modFD, and ICA projection scrubbing leverage for 9 sessions representing the 10th to 90th quantiles of head motion, based on mean FD. For FD and modFD, candidate thresholds of 0.2mm, 0.3mm, 0.4mm and 0.5mm are indicated; for projection scrubbing, thresholds of 3-, 4-, and 5-times the median are indicated. In agreement with previous work \citep{fairCorrectionRespiratoryArtifacts2020,power2019distinctions} standard FD appears inflated, noisy, and largely fails to capture spikes in head motion apparent in the modFD timeseries. Though we do not recommend adopting FD in multiband data, we evaluate it in some of the subsequent analyses for completeness and therefore must choose a threshold. Given that FD is artifactually inflated in multiband data, the standard 0.2mm threshold appears to be overly aggressive, and we therefore consider a threshold of 0.3mm. For modFD, 0.2mm does not appear to be overly aggressive, while higher thresholds may be too lenient. Similarly, for projection scrubbing, the more stringent threshold of 3-times the median leverage ($3\times$) does not appear overly aggressive, while less stringent thresholds may be too lenient. For subsequent analyses, we therefore adopt thresholds of 0.2mm for modFD, 0.3mm for FD, and $3\times$ for projection scrubbing. In some analyses we display multiple thresholds while highlighting the adopted threshold, to provide a more complete picture. We note that the selected modFD cutoff of 0.2mm is consistent with prior work in multiband fMRI \citep{fairCorrectionRespiratoryArtifacts2020, kaplan2022filtering}; also, the selected projection scrubbing cutoff is close to the $4\times$ found to be near optimal by \cite{mejiaPCALeverageOutlier2017}.

\subsection{Effect of scrubbing on validity of FC}

\begin{figure}
    \centering
    \hspace{-1in}\includegraphics[page=1, width=.7\textwidth]{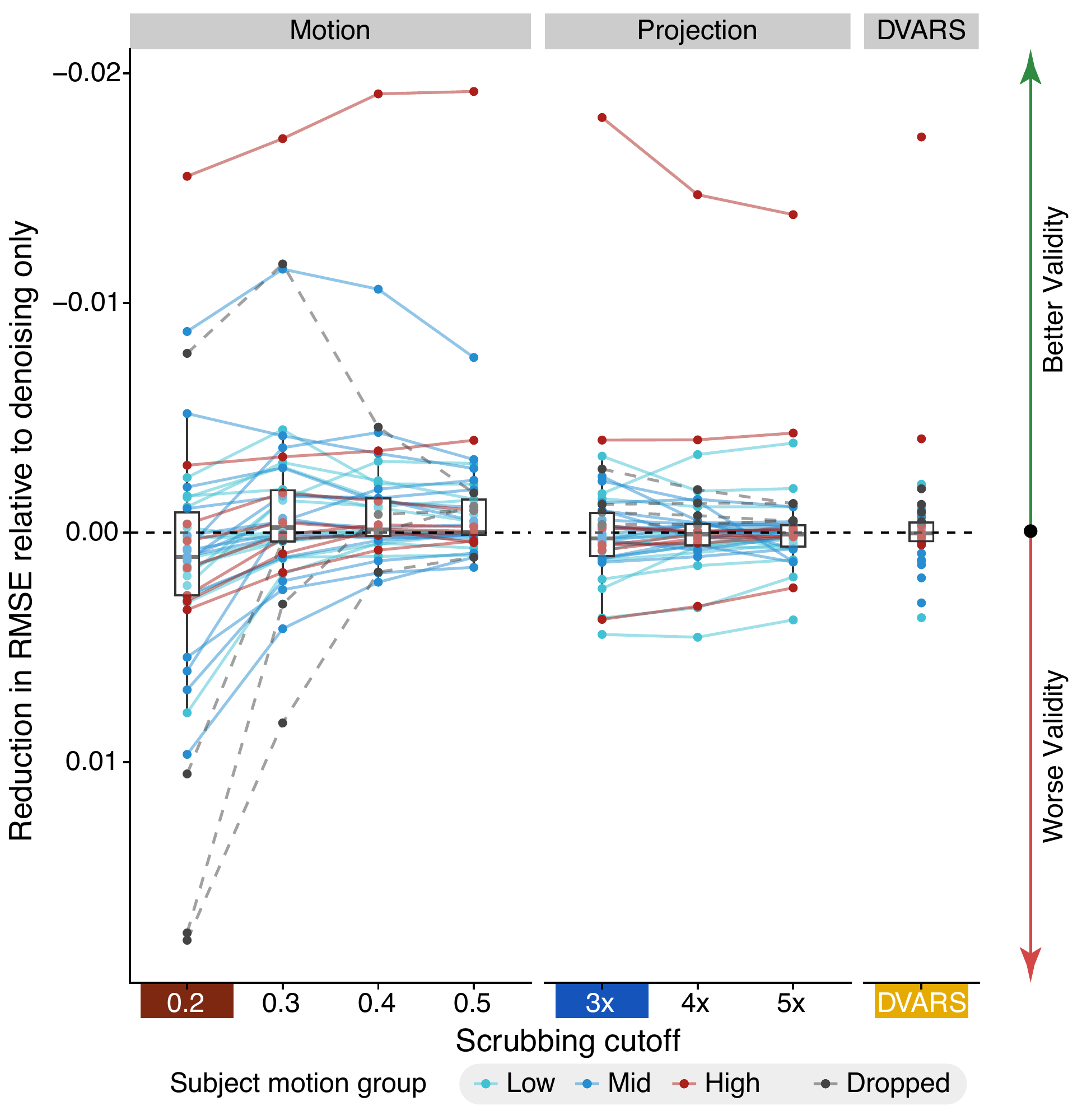}
    \caption{\small \textbf{Effect of scrubbing on validity of FC.} The reduction in root mean squared error (RMSE) of FC after motion scrubbing (based on modFD), ICA projection scrubbing, and DVARS. The selected thresholds based on Figure \ref{fig:NineSubjects} are highlighted on the x-axis. Positive values indicate greater reductions in RMSE and better validity, i.e. lower sampling variability due to removal of noise and retention of signal. Negative values indicate increased RMSE and worse validity, i.e. higher sampling variability of FC estimates due to removal of signal. FC estimates for each subject are based on 10 minutes of resting-state data, and RMSE is computed relative to ``ground truth'' FC based on 1.7 hours of data and stringent motion scrubbing. Each line indicates a subject, colored based on their mean FD: low ($<0.15$mm), mid ($0.15$-$0.2$mm), and high ($>0.2$mm). Four subjects (gray dashed lines) are excluded from the boxplots due to insufficient data remaining after motion scrubbing. 
    Note that this slightly biases the boxplots in favor of motion scrubbing, since data-driven scrubbing actually slightly improves validity for those subjects while motion scrubbing tended to worsen them.
    Stringent motion scrubbing (modFD 0.2mm) worsens validity of FC for most subjects, sometimes dramatically, more lenient motion scrubbing (e.g. modFD 0.5mm) is slightly beneficial to validity on average, and data-driven scrubbing does not tend to dramatically change validity.}
    \label{fig:RMSE}
\end{figure}

\autoref{fig:RMSE} summarizes the effect of scrubbing on the validity of FC estimates for low, medium, and high-motion subjects. To quantify validity, we establish a ``ground truth'' value of FC for each subject using 1.7 hours of rest data, which we censor using rigorous motion scrubbing (modFD over 0.2mm). This should produce highly accurate estimates of FC that are nearly free of motion-induced noise. As described in \autoref{sec:validity_methods}, we then apply each scrubbing technique to the held-out 10 minutes of resting-state data before estimating FC. The root mean squared error (RMSE) relative to the ground truth for each subject is a measure of validity of the FC estimates, where higher RMSE indicates worse validity and lower RMSE indicates better validity. An effective scrubbing method will improve validity (lower RMSE) by removing noise while retaining as much signal as possible. Figure \autoref{fig:RMSE} shows the reduction in RMSE after scrubbing relative to baseline, where positive values indicate reduced RMSE (better validity) and negative values indicate increased RMSE (worse validity). Boxplots summarize across subjects at each cutoff, without including the subjects having too little data after scrubbing for any method and cutoff ("dropped" subjects in gray and with dashed lines). Those four subjects are not negatively impacted by and actually slightly benefit from data-driven scrubbing; thus, the boxplots are slightly biased toward motion scrubbing. Stringent motion scrubbing actually tends to worsen validity on average, while more lenient motion scrubbing tends to be beneficial to validity, likely due to greater signal retention. Data-driven projection scrubbing and DVARS tend to maintain baseline validity on average, while not dramatically worsening validity for any subjects. While more lenient projection scrubbing is slightly better in terms of validity, projection scrubbing appears to be much more robust to the choice of threshold than motion scrubbing.


\subsection{Effect of scrubbing on reliability}

\begin{figure}
    \centering
    \includegraphics[width=1\textwidth]{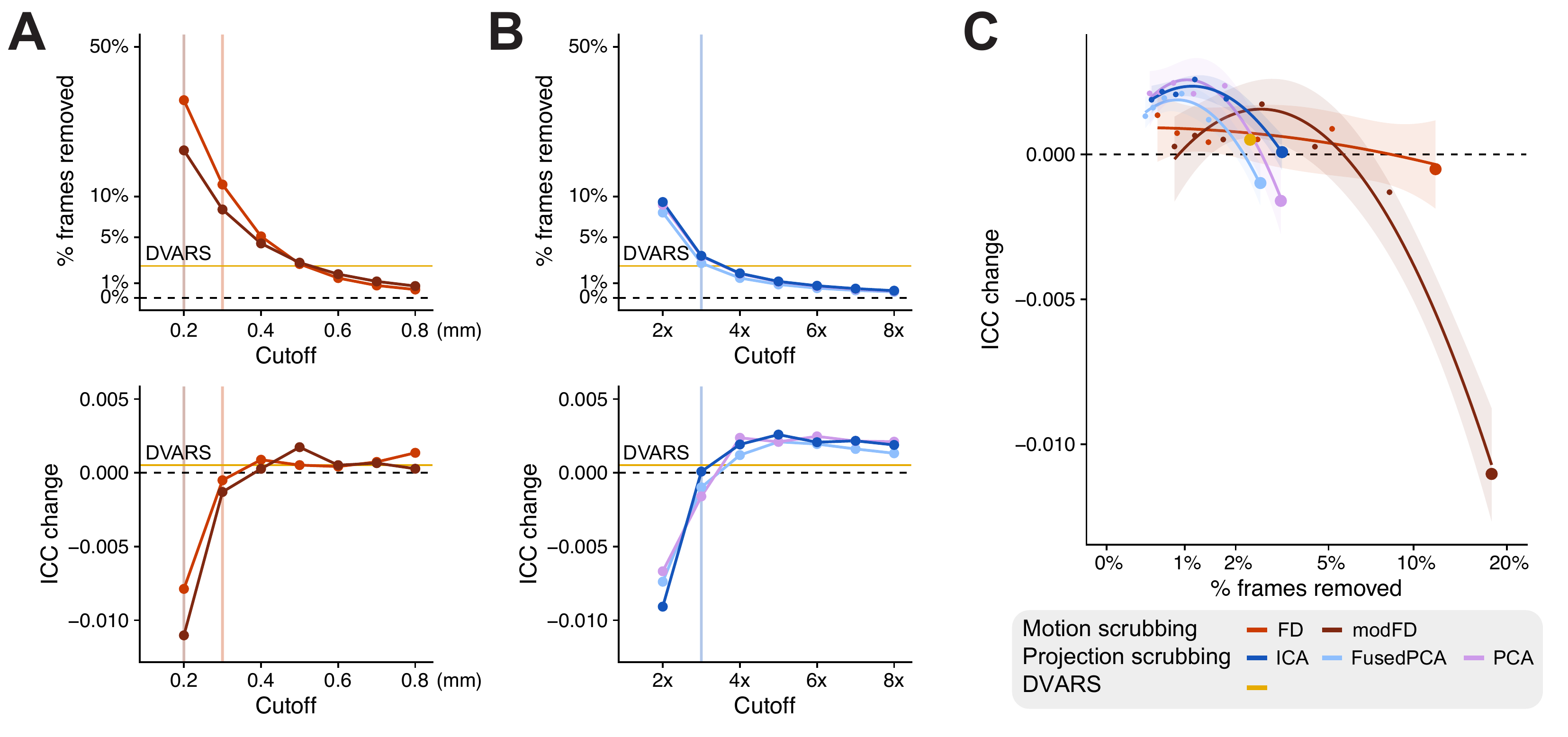}
    \caption{\small 
    \textbf{Effect of different scrubbing thresholds on censoring rates and reliability of FC.} 
    \textbf{(A)} \textbf{Motion scrubbing:} FD and modFD cutoffs between 0.2 and 0.8 mm are shown, with the selected thresholds of modFD = 0.2mm and FD = 0.3mm highlighted. At the selected thresholds, FD censors approximately $12\%$ 
    of volumes on average, while modFD censors approximately $18\%$ 
    of volumes; FD decreases ICC slightly, while modFD causes a much larger decrease in ICC. \textbf{(B)} \textbf{Projection scrubbing:} cutoffs between 2 and 8 times the median leverage are shown, with the selected threshold of $3\times$ highlighted.  Projection scrubbing retains many more volumes than motion scrubbing, censoring only $2.6$ to $3.3\%$ of volumes on average, depending on the projection method. This is similar to DVARS, which censors $2.4\%$ of volumes on average. At the selected thresholds, ICA projection scrubbing and DVARS both increase ICC slightly. We consider increases in ICC due to scrubbing as uniformly beneficial, while the reductions in ICC seen with motion scrubbing may be due to loss of reliable signal, the removal of reliable noise, or both.  \textbf{(C)} \textbf{Relationship between censoring rate and ICC.} Note that censoring rate is displayed on the log scale. For each method, we show all cutoffs at or above the selected one, indicated with larger dots. Quadratic fits and their 95\% confidence interval are overlaid. At lower censoring rates ($2\%$ or less), both motion scrubbing and projection scrubbing increase ICC. At the selected thresholds, projection scrubbing and FD have a similar effect on ICC, but FD results in much higher censoring rates. modFD has the highest censoring rates and drastically reduces ICC.}
    \label{fig:scrubICC}
\end{figure}

\autoref{fig:scrubICC} examines the effect of different scrubbing thresholds on FC reliability and on the censoring rate. Panels (A) and (B) show ICC change and the censoring rate as a function of scrubbing threshold for motion scrubbing and projection scrubbing, respectively. The effect of DVARS is indicated in yellow on each plot, based on the dual cutoff proposed by \citet{afyouniInsightInferenceDVARS2018}. The $18\%$ average censoring rate of motion scrubbing using a $0.2mm$ modFD cutoff is markedly higher than that of projection scrubbing, which censors only $2.6\%$ to $3.3\%$ of volumes, or DVARS, which censors $2.4\%$ of volumes. At the chosen thresholds, ICA projection scrubbing and DVARS slightly increase ICC, while other projections slightly decrease ICC, and motion scrubbing with modFD sharply decreases ICC. We note again that decreased ICC is difficult to interpret since it may be driven by removal of reliable noise in addition to removal of reliable signal, and whether one factor or the other dominates cannot be determined.  However, we assert that \textit{increased} ICC due to scrubbing is likely beneficial, since it is not possible to \textit{introduce} reliable noise (or indeed any type of signal or noise) by removing volumes. 
Therefore, we interpret increases in ICC that occur due to scrubbing as representing an improved balance between signal and noise. While the large decrease in ICC with modFD at the threshold of 0.2mm cannot unquestionably be attributed to an excessive loss of signal versus removal of reliable noise, it does indicate a substantially higher risk of signal loss associated with motion scrubbing, especially given that nearly 20\% of volumes are censored on average.

An expanded comparison of FD cutoffs showing additional forms of motion scrubbing based on lagged and/or filtered FD is provided in \autoref{app:fig:MotionExpanded}. Similarly, an expanded comparison of projection scrubbing cutoffs with and without kurtosis-based component selection is provided in \autoref{app:fig:ProjectionExpanded}. These plots also explore the effect of scrubbing on the reliability of FC estimated using a shorter scan duration. \autoref{app:fig:ICCexpanded} disaggregates ICC changes by FC connection type, showing that data-driven scrubbing methods are especially beneficial for subcortical connections. Finally, a comparison of mean ICC improvement due to scrubbing in conjunction with different baselines is provided in \autoref{app:fig:AltBases}. We generally see that for projection scrubbing, the use of ICA tends to yield slightly higher ICC than other projections. ICA also tends to flag slightly more volumes than PCA or FusedPCA, as seen in \autoref{fig:scrubICC}B, and therefore may be slightly more sensitive than other projections. 



\subsection{Effect of scrubbing on fingerprinting}

\begin{figure}
    \centering
    \includegraphics[width=.95\textwidth]{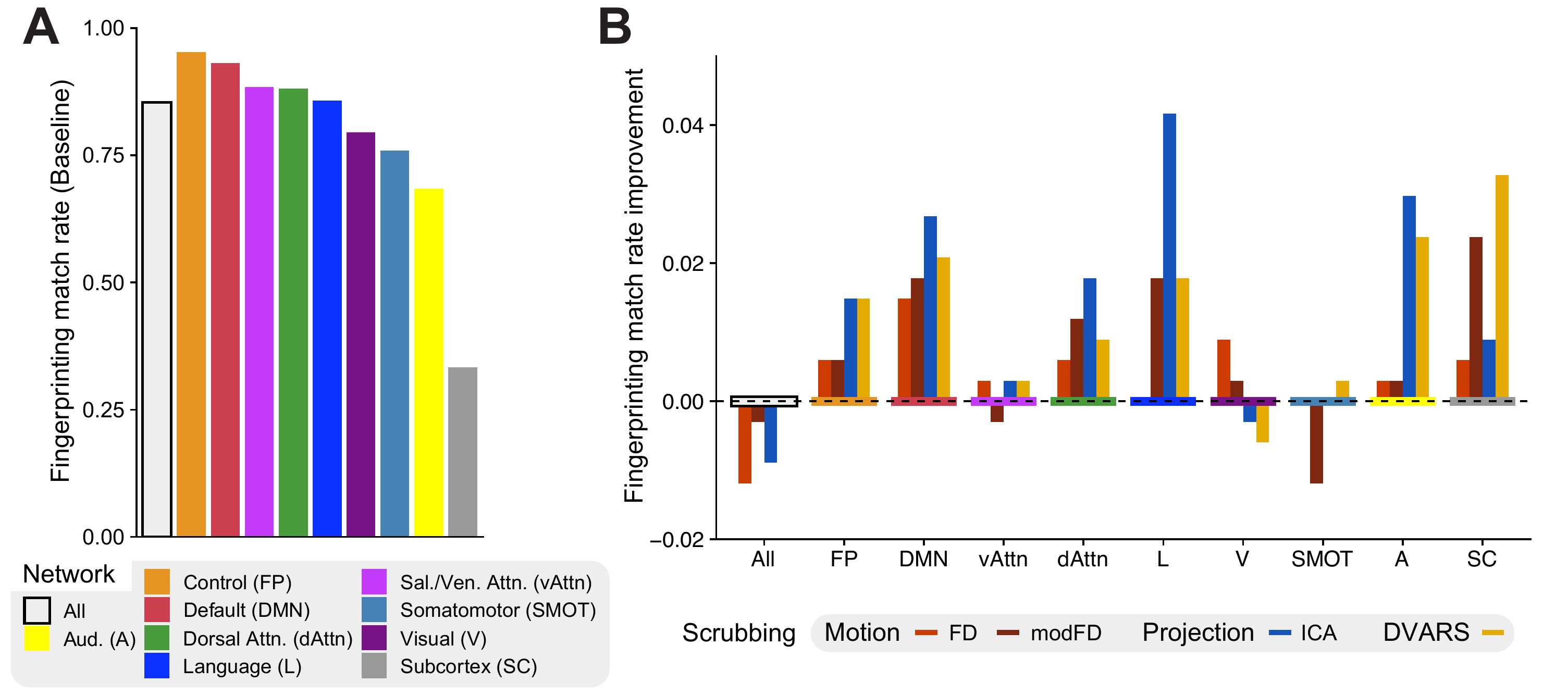}
    \caption{\small \textbf{Scrubbing slightly improves most fingerprinting match rates.} Fingerprinting was performed using all connections or only the connections within each network. \textbf{(A)} \textbf{Baseline fingerprinting match rates}, with networks sorted from most successful fingerprinting to least successful. 
    All match rates are much higher than the success rate of random guesses (1 in 42, or less than 3\%). \textbf{(B)} \textbf{Change in fingerprinting match rates due to scrubbing}, sorted by the baseline rates of fingerprint success. The effects of scrubbing are generally small in magnitude, perhaps because changes in FC tend to be subtle and may not cause a change in subject-to-subject matching. Scrubbing generally benefits network-wise fingerprinting, with projection scrubbing being most beneficial for cortical networks and DVARS being most beneficial for subcortical regions. Scrubbing generally worsens fingerprinting based on all connections, suggesting a differential effect of scrubbing on within-network and between-network connections. Note that worse fingerprinting match rate may not always indicate loss of true signal, since removal of reliable noise can also reduce reliability. However, increases in fingerprinting match rate due to scrubbing can safely be interpreted as more effective noise removal and signal retention.}
    \label{fig:Fingerprint}
\end{figure}

We next examine the effects of scrubbing on fingerprinting, using the thresholds determined above ($3\times$ for projection scrubbing, 0.2mm for modFD, and 0.3mm for FD). \autoref{fig:Fingerprint}A shows baseline fingerprint match rates by network and the changes due to scrubbing. Networks are shown in decreasing order of baseline fingerprint match rates. Panel A shows that the highest fingerprint match rates are achieved using connections within the control, default mode, dorsal attention, ventral attention, or language networks. Using only connections within the visual, somatomotor, or auditory networks yields somewhat worse match rates than using all connections, while using only subcortical connections results in much lower match rates. These network-level differences are consistent with the between-network differences in ICC shown in
\autoref{fig:baselineICC}---networks with higher ICC are generally more
useful for distinguishing individual subjects. This makes sense, as both
fingerprinting and ICC are measures of individuality. 

Panel B shows the change in fingerprinting match rates after scrubbing. As in the case of ICC, it is important to take care when interpreting scrubbing-induced \textit{reductions} in fingerprinting match rate, since it may be possible to worsen fingerprinting by removing reliable noise. For example, using only motor connections, motion scrubbing with modFD decreases fingerprinting match rate. This may indicate removal of reliable noise, loss of reliable signal, or both. However, \textit{improvements} in fingerprinting match rate due to scrubbing are easier to interpret, since scrubbing cannot introduce reliable noise, only change the relative contributions of signal and noise. Therefore, we interpret increased fingerprinting match rate due to scrubbing as indicative of a better balance between signal and noise. 

Based on the results shown in \autoref{fig:Fingerprint}B, scrubbing is generally beneficial for network-based fingerprinting. ICA projection scrubbing generally produces the most improvement for cortical networks, followed by DVARS and motion scrubbing with modFD, while DVARS generally benefits subcortical connections the most. Results for other projections are shown in \autoref{app:fig:scrubFingerprintExpanded}: ICA tends to be more beneficial than PCA or FusedPCA for the most identifiable networks, while PCA or FusedPCA are somewhat more beneficial for the less identifiable ones.  When all connections are used, scrubbing generally reduces the fingerprinting match rate, in contrast to network-wise fingerprinting. This may indicate a differential effect of scrubbing on within-network and between-network connections, with within-network connections clearly becoming more reliable with scrubbing and between-network connections sometimes becoming less reliable, due to removal of reliable noise, reliable signal, or both.

\subsection{Agreement between scrubbing methods}

\begin{figure}
    \centering
    \includegraphics[width=0.75\textwidth]{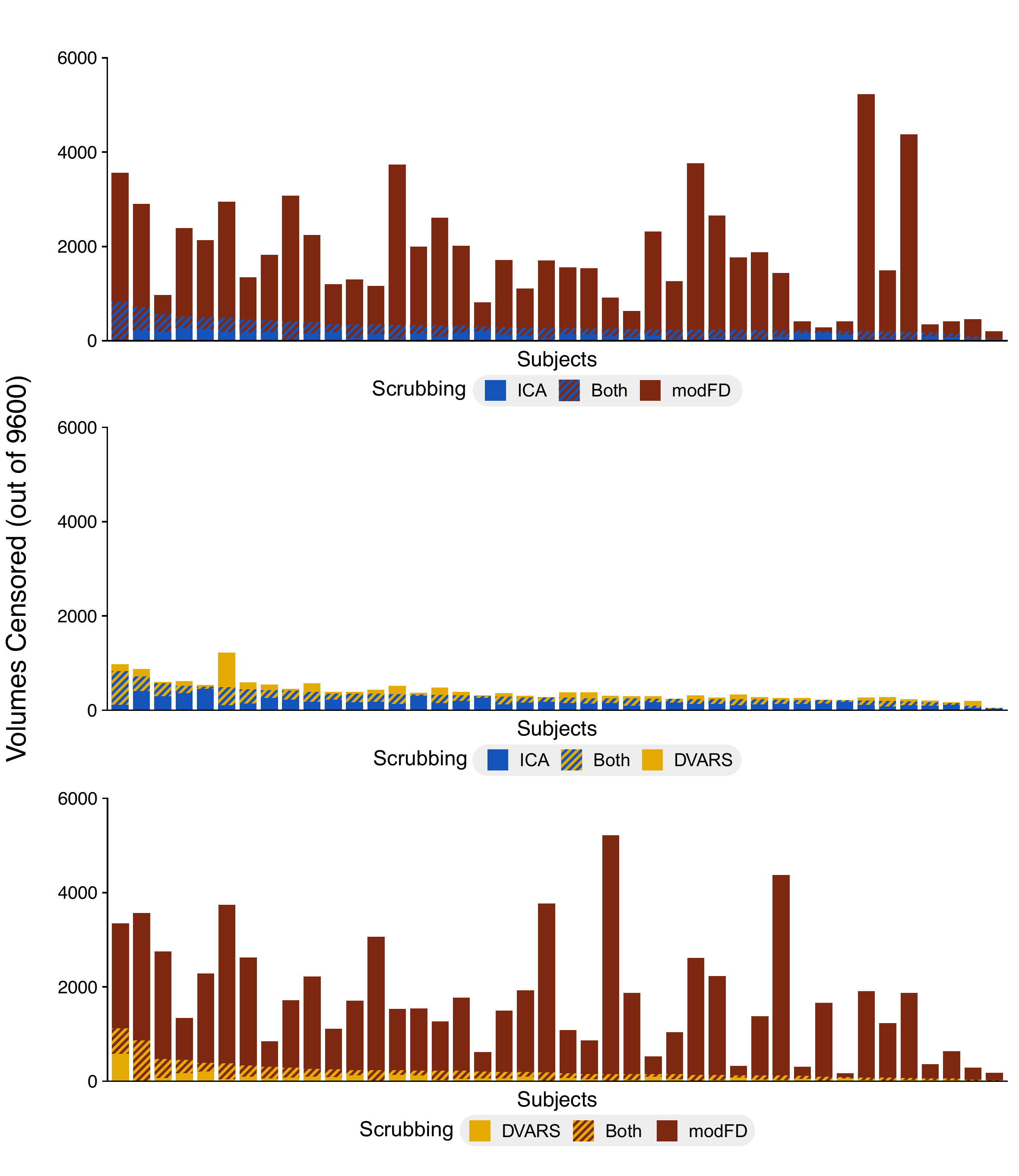}
    \caption{\small \textbf{Agreement between scrubbing methods.} The number of volumes flagged by ICA projection scrubbing, motion scrubbing and DVARS for each of 42 subjects. Cross-hatched areas indicate volumes flagged by both methods. Subjects are ordered by the number of volumes flagged by the method listed first in each panel's legend. Motion scrubbing tends to flag many more volumes compared with projection scrubbing or DVARS. Both data-driven scrubbing methods identify additional volumes where modFD is not elevated, which may indicate sensitivity to artifacts associated with lagged effects of head motion or non-motion-related sources. Agreement between ICA projection scrubbing and DVARS is moderate but not perfect, which suggests spatial differences in their sensitivity to artifacts.}
    \label{fig:overlaps}
\end{figure}

Here, we investigate the agreement or disagreement between ICA projection scrubbing, motion scrubbing, and DVARS to provide some intuition around the drivers of the differences in their downstream effects on FC validity, reliability and identifiability.  First, \autoref{fig:overlaps} displays the overlap between volumes flagged with ICA projection scrubbing versus motion scrubbing or DVARS (top two panels), as well the overlap between motion scrubbing and DVARS (bottom panel). 
Motion scrubbing clearly removes many more volumes than projection scrubbing or DVARS. Indeed, most of the volumes flagged by motion scrubbing do not exhibit statistical abnormalities according to the data-driven metrics. In addition, both projection scrubbing and DVARS identify volumes not flagged by motion scrubbing. This suggests that data-driven scrubbing detects abnormalities that motion scrubbing may miss. There are a couple of reasons this could occur: data-driven scrubbing may better identify volumes actually exhibiting motion-induced noise that are not immediately coincident with increases in modFD; or, data-driven scrubbing may detect noise induced by other sources besides head motion. The bottom panel shows a similar relationship between motion scrubbing and DVARS, suggesting a more general disagreement between motion-based and data-driven measures of abnormality. Motion scrubbing often flags a large number of volumes that neither projection scrubbing nor DVARS identifies as problematic. The number of excess volumes flagged by motion scrubbing suggests that head motion is sometimes only weakly related to abnormal patterns still present after effective regression-based denoising. That is, head motion-induced artifacts may already have been effectively attenuated or eliminated by regression-based methods. 

\begin{figure}
    \centering
    \hspace{-1in}\includegraphics[page=1, width=.8\textwidth]{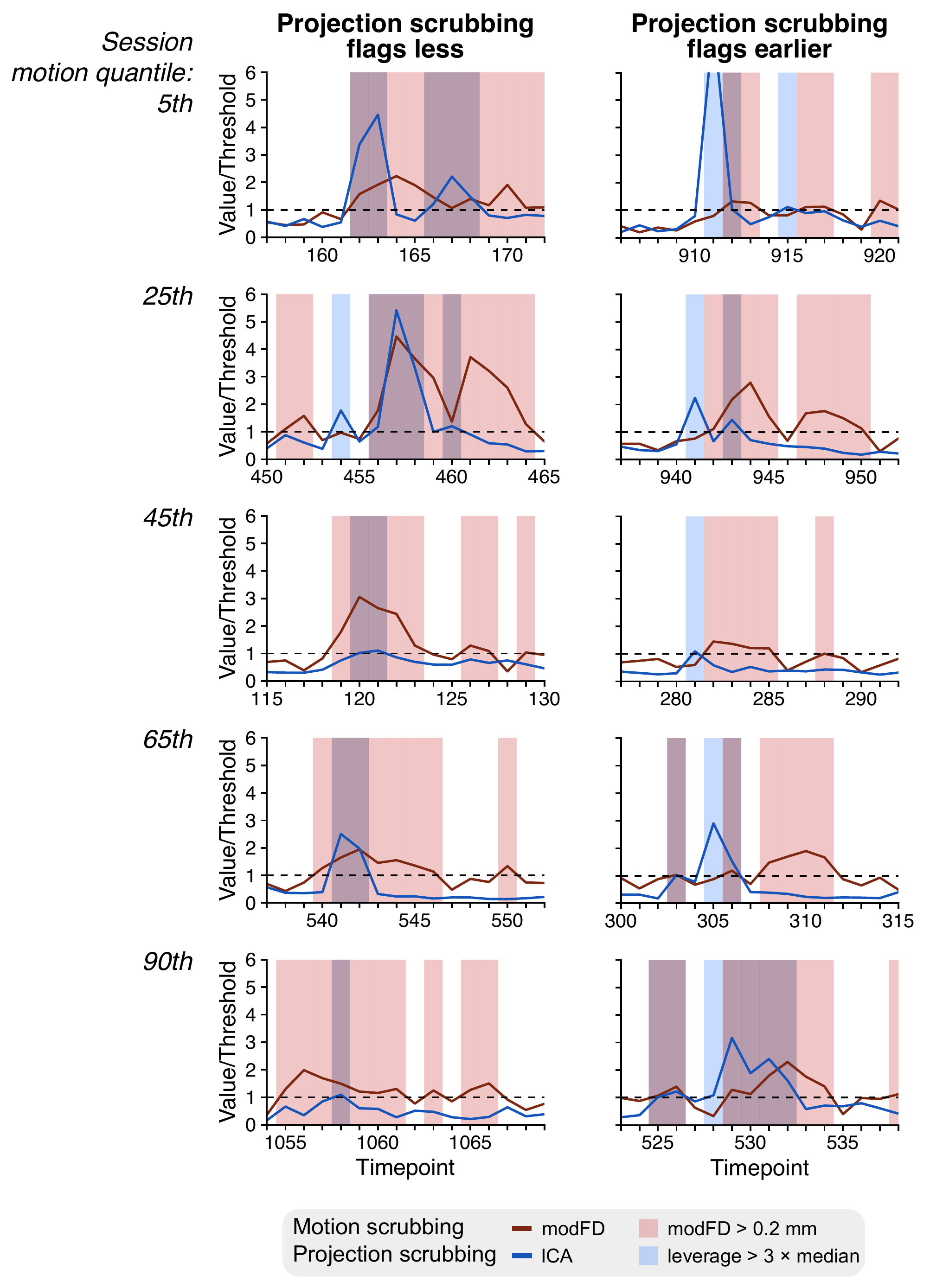}
    \caption{\small \textbf{Examples of projection scrubbing and motion scrubbing.} Five sessions representing a range of head motion (5th to 90th quantiles of mean FD) are shown. The left column shows instances where projection scrubbing flags volumes coincident with head motion but identifies fewer volumes than modFD. These examples illustrate that projection scrubbing often identifies artifacts induced by head motion but may do so with greater \textit{specificity} than modFD. The right column shows examples where leverage flags volumes just before a spike in modFD. Recall that modFD is based on 4-back temporal differences, which may cause lagged sensitivity to motion artifact in some cases. These examples suggest that projection scrubbing may also be more \textit{sensitive} to certain effects of motion than modFD.} 
    \label{fig:Lag}
\end{figure}

\autoref{fig:Lag} takes a deeper look at how projection scrubbing identifies motion-related artifacts in five example sessions representing a range of head motion levels. For each session, the first column shows an example where projection scrubbing tends to flag a \textit{subset} of volumes with high modFD. In these cases, projection scrubbing seems to identify a  motion artifact without flagging all volumes with high modFD. Since projection scrubbing is based on statistical abnormalities in the data after regression-based denoising, this may happen if motion artifacts only manifest in a subset of motion-concurrent volumes after nuisance regression.  The second column shows an example for each run where projection scrubbing identifies a statistical abnormality just \textit{before} a spike in modFD. Recall that modFD is a lagged measure of FD, so this phenomenon may indicate a lag in the sensitivity of modFD to motion artifacts. Note, however, that the lag is not consistent across all sessions, so this effect may not be necessarily remedied by a simple modification to the modFD calculation. Instead, what these examples illustrate is that projection scrubbing detects motion-related abnormalities based on actual patterns in the data, allowing it to be more specific (avoid flagging non-contaminated volumes) and in some cases more sensitive (detect statistical abnormalities even if they do not coincide perfectly with a spike in modFD) than explicit motion scrubbing.

\subsection{Effects of scrubbing on scan exclusion rates}

\begin{figure}
    \centering
    \includegraphics[width=.95\textwidth]{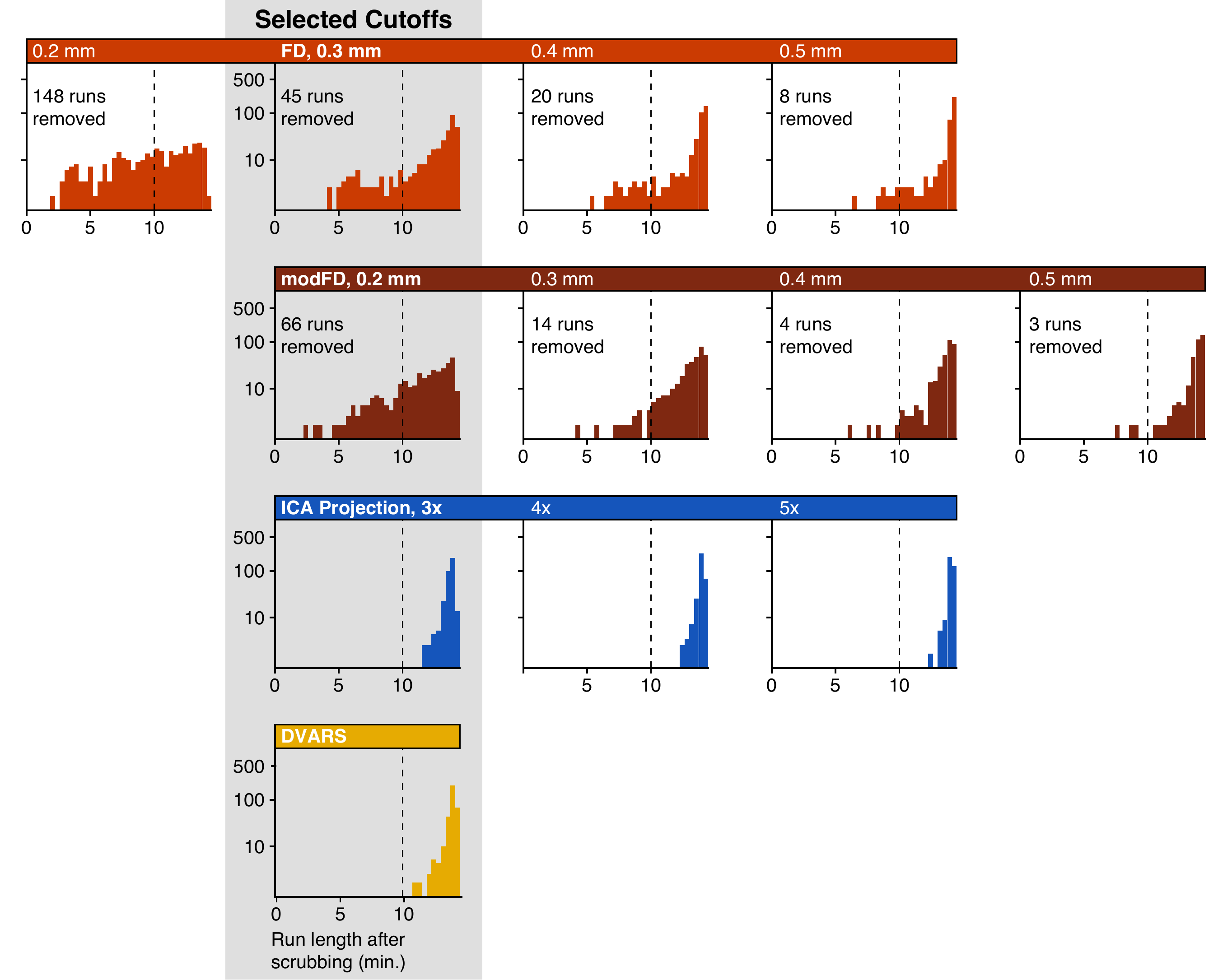}
    \caption{\small \textbf{Motion scrubbing results in high rates of run exclusion.} Histograms show the distribution of minutes remaining in each run after scrubbing.  Note the square-root scale on the y-axis.  Based on a rule that all runs must retain at least $10$ minutes of data after scrubbing (dashed line), motion scrubbing with modFD at the 0.2mm threshold would exclude 66 runs or 20\% of the 336 total runs. Both projection scrubbing and DVARS would leave every run with at least $10$ minutes of data, resulting in no exclusions.}
    \label{fig:Exclusion}
\end{figure}

Aggressive motion scrubbing often leaves many sessions without sufficient data to be included in analysis. It is not uncommon to see one-third to one-half of all sessions discarded due to motion scrubbing. Data-driven scrubbing has the potential to preserve more volumes and avoid such drastic exclusion rates. \autoref{fig:Exclusion} shows histograms of the minutes of data \textit{remaining} per run after scrubbing using motion scrubbing, projection scrubbing or DVARS. Note that the y-axis is shown on the square root scale. We quantified the proportion of the 336 total runs (42 subjects with 8 runs each) that would be excluded based on a minimum of $10$ minutes per run. Projection scrubbing and DVARS would retain at least $10$ minutes of data for all runs, resulting in no exclusion. By contrast, motion scrubbing with modFD at the 0.2mm threshold would result in exclusion of 66 runs, or $20\%$ of all runs. Note that naively using standard FD with a strict 0.2mm threshold would cause the removal of nearly half of all runs. Since the HCP includes only data from healthy young adults, motion scrubbing exclusion rates are likely to be much higher in more diverse datasets or those including participants from higher-motion populations, such as children, the elderly, and clinical populations.

\section{Discussion}
\label{sec:discussion}


In this work, we evaluate the downstream effects of different fMRI scrubbing techniques including projection scrubbing, a new data-driven, outlier detection-based approach of identifying fMRI volumes contaminated by artifacts. In the projection scrubbing framework, the fMRI data is projected onto various directions representing artifacts, and a summary measure of artifactual intensity, \textit{leverage}, is computed for each volume and thresholded to flag problematic volumes. We consider several projection methods: ICA, PCA, and a novel FusedPCA method. We compare the performance of projection scrubbing with motion scrubbing and DVARS at reducing noise and improving the validity, reliability, and identifiability of FC, three indirect reflections of the signal-to-noise ratio in the data. Our results show that data-driven  scrubbing is generally more beneficial than motion scrubbing across these evaluation metrics. Among projection methods, we find ICA to have the best performance, building on established methods that use ICA to separate and remove noise from fMRI data. Below, we discuss these results in more detail, describe some potential reasons for the effectiveness of projection scrubbing, and note some considerations for scrubbing more generally. 

\subsection{Data-driven scrubbing is less aggressive but generally more beneficial than motion scrubbing}

One of the most striking findings in our analysis is that projection scrubbing tends to be more beneficial to validity, reliability, and identifiability of FC compared with stringent motion scrubbing, while removing less than a third as many volumes. Motion scrubbing was detrimental to FC validity and reliability for most subjects, while projection scrubbing and DVARS were generally beneficial or at least not harmful. This provides evidence that motion scrubbing using common stringent thresholds likely results in substantial loss of true signal along with motion-induced noise. Using less stringent modFD thresholds is not necessarily the answer, since these appeared too lenient across several example fMRI runs shown in \autoref{fig:NineSubjects}. Indeed, the quality metric MAC showed that when motion scrubbing was made less stringent by increasing the modFD threshold, projection scrubbing performed better.

Examining the overlap between ICA projection scrubbing, motion scrubbing and DVARS, we observed that data-driven scrubbing typically identifies a fraction of those volumes flagged by modFD and often identifies additional volumes not flagged by modFD. Given the generally strong performance of data-driven scrubbing, this suggests a number of possible phenomena at play. First, many motion-related artifacts may have been alleviated by regression-based denoising and therefore do not manifest as burst noise in the nuisance regressed data. Second, projection scrubbing may identify burst noise caused by other sources, such as scanner artifacts, spillover effects of previous head motion, or artifacts introduced through preprocessing. Third, given the challenges associated with tracking head motion in multiband data (see \autoref{sec:modFD}), projection scrubbing may identify volumes affected by head motion that motion scrubbing fails identify. Notably, this is true even using the modified version of FD adapted specifically for HCP-style data; motion scrubbing with standard FD had very poor performance, underscoring the idea that standard FD should not be used to identify motion artifacts in multiband data. Overall, the benefits of data-driven scrubbing over motion scrubbing can be attributed to being much more \textit{specific} by avoiding unnecessary removal of clean volumes and, in some cases, more \textit{sensitive} to the manifestations of head motion in the data.

The less aggressive but more effective censoring achieved by data-driven scrubbing has significant implications for data retention in fMRI studies. Because even submillimeter head movements impact functional connectivity \citep{powerSpuriousSystematicCorrelations2012, vandijkInfluenceHeadMotion2012, satterthwaiteImpactInScannerHead2012}, it is common for researchers to eliminate entire scans from an analysis based on FD restrictions \citep[see, e.g.,][]{ciricBenchmarkingParticipantlevelConfound2017,parkesEvaluationEfficacyReliability2018}. This practice can lead to drastic reductions in sample size, particularly for certain populations such as children and the elderly. For instance, \cite{Nielsen2019EvaluatingDenoising} excluded 75\% of participants due to excessive head motion before attempting to evaluate the impact of motion scrubbing and global signal regression on predictions of brain maturity from functional connectivity. This not only results in the loss of many subjects but also may introduce selection bias, since the retained subjects may differ systematically from the study population of interest \citep{nebel2022accounting, cosgrove2022limits}. Similarly, \cite{Marek2019IdentifyingStudy} and \cite{Marek2022} excluded 40\% and 60\% of ABCD participants respectively. Projection scrubbing and DVARS provide more targeted removal of contaminated volumes, allowing for more data retention and an increase in the number of subjects ultimately used for analysis. Given that thousands of subjects are needed to reliably identify many brain-wide associations with behavior and the high cost associated with imaging so many individuals, the ability to retain more data without sacrificing the quality of downstream analysis could have major implications for such investigations. 

\subsection{Projection scrubbing versus DVARS}

We observed moderate, though not perfect, agreement between projection scrubbing and DVARS. While DVARS and projection scrubbing are both data-driven approaches, a fundamental distinction between them is that DVARS is computed based on the difference between subsequent volumes, while projection scrubbing compares each volume to the distribution across all volumes. As such, projection scrubbing utilizes information from the ``population'' of volumes in quantifying the abnormality of individual volumes and, therefore, may be more accurate at identifying temporal abnormalities. A second distinction between the two methods is that projection scrubbing is based on first extracting directions (e.g., independent components) in the data that are likely to represent transient noise. This may have the effect of making projection scrubbing a more specific classifier for artifact-contaminated volumes. 

Interestingly, DVARS was generally more beneficial for connections between subcortical/cerebellar regions, based on our fingerprinting analysis. This may be because DVARS weights each  voxel or vertex equally, while projection scrubbing is based on deviations from the typical \textit{distribution} of intensities. If subcortical voxels tend to exhibit greater baseline variation than cortical vertices, projection scrubbing will require a more substantial deviation in those voxels to flag a volume as problematic. Therefore, DVARS may be more sensitive to artifacts primarily affecting subcortical areas (or in general, areas of the brain exhibiting greater-than-average baseline variation) and less sensitive to artifacts in other areas of the brain. It would be possible to modify both projection scrubbing and DVARS to operate separately on cortical vertices and subcortical voxels in grayordinates fMRI data, given their very different SNR profiles and sizes.  This is one advantage of data-driven scrubbing: it can be targeted to certain regions of the brain in order to maximize sensitivity and specificity in specific areas of interest. The spatial differences in sensitivity of projection scrubbing and DVARS suggests a possible area of future work: combining both measures by flagging volumes that are identified as abnormal based on either one. Such a data-driven approach would likely be highly sensitive to noise while still being much less aggressive than motion scrubbing: Figure \ref{fig:overlaps} shows that it would censor well below $10\%$ of volumes per run in the HCP data we analyzed, compared with an average of $18\%$ for stringent motion scrubbing. Future work could evaluate the ability of such an approach to effectively reduce noise while enhancing data retention.


\subsection{Relationship of ICA projection scrubbing to ICA-based denoising}

ICA projection scrubbing builds upon established methods using ICA to identify and remove artifacts in fMRI data, most notably ICA-FIX \citep{salimi-khorshidiAutomaticDenoisingFunctional2014} and ICA-AROMA \citep{pruimICAAROMARobustICAbased2015}.  Like those approaches, ICA projection scrubbing takes advantage of the ability of ICA to separate neural sources from those of artifactual origin in fMRI.  However, ICA projection scrubbing has several differences from these other ICA-based denoising methods.  We will focus on comparing with ICA-FIX since it is more general than ICA-AROMA, which is primarily designed to remove motion-related artifacts. First, ICA projection scrubbing is not designed for general denoising like ICA-FIX. However, it can be effectively combined with established regression-based denoising techniques, such as aCompCor or motion regression, as illustrated here. ICA projection scrubbing cannot be combined with ICA-FIX due to the latter's tendency to induce DVARS dips \citep{glasser2018using} in locations with transient artifacts---effectively, ICA-FIX may behave like a spike regressor, and therefore may actually present an alternative to scrubbing. Future work should evaluate the efficacy of ICA-FIX compared with data-driven scrubbing.

Second, classifying each IC as neural or artifactual is simpler for ICA projection scrubbing than it is for ICA-FIX, because we assume the data has been partially denoised using regression-based methods prior to performing ICA, and because we are only interested in identifying burst noise artifacts amenable to scrubbing.  Therefore, the ICs of interest in ICA projection scrubbing can be identified based on the presence of spikes in the IC's timecourse, while the artifactual ICs being classified by ICA-FIX are likely to exhibit a wider range of spatiotemporal properties. Indeed, the classifier used in ICA-FIX is based on over 180 spatial and temporal features, while ICA projection scrubbing is based on a single temporal metric, kurtosis, that is indicative of spikes or outliers. Third, ICA projection scrubbing is fully automated, while ICA-FIX is based on a trained classifier that requires manual re-labelling in each new dataset for high accuracy. Projection scrubbing instead employs a formal hypothesis testing approach to identify those ICs with high kurtosis, avoiding the need for model training. 
Here, we have not directly examined the accuracy of our approach at separating signal and noise ICs. This would be a useful area of future work.  However, we note that perfect IC classification is not necessary in the projection scrubbing framework, since all noise ICs are pooled to identify volumes exhibiting temporal abnormalities representing burst noise. Thus, errors in classification of a small number of individual ICs do not necessarily give rise to errors in classification of volumes.

\subsection{Scrubbing and denoising represent trade-offs between noise removal and signal retention}

Our analyses show that reliability and identifiability of FC generally benefit from data-driven scrubbing, but for some types of functional connections they actually worsen. Similarly, our validity analysis showed that for some subjects, FC estimates become slightly less accurate after scrubbing. This tradeoff illustrates an important consequence of scrubbing: the inadvertent removal of meaningful signal along with noise. Some signal loss is almost inevitable with scrubbing, but over-aggressive discarding of volumes in the pursuit of noise elimination has important downstream consequences that should not be ignored. 

The optimal balance between noise removal and signal retention in a given situation depends the SNR properties of the data generating process, which varies across the connectome and depends on many acquisition and processing factors. Connections with high baseline SNR may benefit more from greater signal retention than from more aggressive noise removal. Therefore, the degree of scrubbing, and whether scrubbing is warranted at all, should depend on the scientific goals, intended downstream analyses, and properties of the data. In particular, multiband acquisitions allow for higher temporal resolution but at the cost of spatially variable noise amplification \citep{risk2021multiband}. In cerebellar and subcortical regions, noise amplification can be moderate to high respectively, so scrubbing is likely to be highly beneficial---as we have indeed observed.  In regions where this noise amplifcation is low (e.g, the visual cortex), on the other hand, a more cautious and measured approach to scrubbing is warranted, as its benefits may outweigh the costs in terms of the power to detect fMRI signal changes \citep{chen2015evaluation, todd2017functional, risk2018impacts}. 

The effect of scrubbing is also somewhat dependent on the choice of a baseline denoising technique.  Generally, more effective regression-based denoising results in cleaner data, leaving less noise for scrubbing to remove. Indeed, previous studies have noted that scrubbing may be unnecessary or even detrimental when applied on top of effective regression-based denoising, such as aCompCor or ICA-FIX \citep{muschelliReductionMotionrelatedArtifacts2014, Cho2021ImpactConnectomics}. In our analysis, we selected a baseline denoising method based in part on its ability to maximally improve FC reliability prior to scrubbing. More effective baseline denoising may increase the challenge of scrubbing to improve reliability and other FC quality metrics. Despite this, we generally saw further, albeit modest, improvements to FC reliability with data-driven scrubbing. Moreover, Appendix \autoref{app:fig:AltBases} shows that these improvements occur with other common baseline denoising methods too, and indeed, the mean improvements to reliability due to scrubbing are greater when used in conjunction with these less effective baseline methods. 

\subsection{Limitations and future directions}

Our proposed projection scrubbing framework has several limitations that could be improved with future work. First, the metric we employ, leverage, is related to Mahalanobis distance, a classic multivariate distance metric. As with most traditional distance measures, it is not robust to outliers, which could influence the parameters underlying the distance metric. This could lead to the phenomenon of \textit{masking}, wherein truly outlying observations appear to fall within the normal range in the multivariate distribution, due to the influence of even larger outliers \citep{rousseeuw1990unmasking}. The use of a robust distance measure in place of leverage is therefore an important future direction and could ultimately improve the sensitivity and specificity of projection scrubbing at identifying artifactual volumes. 

Second, here we employ an ad-hoc threshold for leverage.  While our analysis, considered alongside the earlier findings of \cite{mejiaPCALeverageOutlier2017}, shows that a threshold of 3 times the median leverage is reasonable and likely near-optimal, a more principled threshold with controllable statistical properties is desirable. \cite{mejiaPCALeverageOutlier2017} considered a robust distance metric with a known null distribution \citep{hardin2005distribution} in the case of independent Gaussian data. This null distribution could in theory be used to determine an appropriate threshold for identifying outlying volumes. However, the assumptions underlying the theory are restrictive and unlikely to hold in the case of fMRI data. Further work is needed to determine the null distribution of robust distance metrics for data that deviate from common assumptions such as Gaussianity and independence.

A limitation to our analysis of projection scrubbing is that we have not directly examined its efficacy at identifying artifactual directions in the data.  Our method employs a single temporal metric, kurtosis, to select ICs (or other types of components) representing burst noise artifacts. We contend that this simpler approach relative to ICA-FIX is possible because the data has already been largely denoised through regression-based methods such as aCompCor, and because we only aim to identify burst noise artifacts. The overall strong performance of ICA projection scrubbing suggests that this approach is sufficiently accurate at distinguishing artifactual ICs to form the basis for an effective data-driven scrubbing technique. 

Further work may consider alternative projection methods as the basis for projection scrubbing. Here we considered ICA, FusedPCA, and PCA in conjunction with a subsequent kurtosis cutoff to identify directions likely to represent artifacts. These projections attempt to estimate components that exhibit certain properties, e.g., independence in the case of ICA; however, what is of greater relevance in projection scrubbing is that the components represent patterns of burst noise in the data.  Other projection or rotation methods may be better suited to this task, including methods that are designed to maximize kurtosis, such as projection pursuit \citep{hou2011fast} and varimax rotation \citep{rohe2020vintage}, or dimension reduction methods designed specifically for the purpose of outlier detection \citep{kandanaarachchi2021dimension}. 

In this work, we compared projection scrubbing with popular existing scrubbing techniques. However, an alternative data-driven approach is imputation of voxel or component timeseries using ``despike'' methods \citep{allen2011baseline}, such as AFNI's 3dDespike (\url{http://afni.nimh.nih.gov}; NIMH Scientific and Statistical Computing Core, Bethesda, Maryland) or Wavelet Despike \citep{patelWaveletMethodModeling2014}.\footnote{Note that ``despiking" is distinct from ``spike regression", an implementation of scrubbing.} Since these methods target the timeseries of specific brain locations or components, they are less heavy-handed than scrubbing, which targets entire volumes.  They may therefore be able to retain more signal while reducing the effects of burst noise. A comparison of data-driven scrubbing and despiking methods would be a worthwhile direction for future research. 

Although the HCP remains a highly utilized dataset and many other studies have adopted the HCP protocol, there are several other multiband fMRI data repositories utilizing different scan acquisition techniques. Since our analysis was based on the HCP, it is not clear how generalizable our findings are to other multiband acquisitions, such as those not based on LR/RL phase encoding. Determining the effects of denoising and scrubbing for other multiband acquisition techniques is an important area of future research.

An emerging technique for mitigating the impact of motion on functional connectivity  focuses on using multi-echo fMRI sequences, which involve the acquisition of multiple images at different echo times for each frame/volume. Recent work suggests that multi-echo fMRI, when used in combination with a breath-holding task and ICA processing, can leverage the echo-time dependent decay of signals to more directly separate neural and artifactual components and improve the reliability of functional connectivity \citep{lynch2020rapid}. Future work should assess the utility of projection scrubbing with multi-echo denoising approaches as they become more widely used.

Finally, here we have focused on resting-state fMRI analyses. However, projection scrubbing may also be applied in task fMRI analysis by computing leverage based on the residuals of the task general linear model including nuisance regressors, then re-fitting the GLM with spike regressors (or censoring the original data and design matrix) as in Figure \ref{fig:Flowchart}. This approach has been previous adopted by \cite{mejia2022longitudinal} and is analogous to how the data-driven scrubbing metrics are computed after a preliminary nuisance regression in our denoising pipeline, followed by a simultaneous regression to incorporate scrubbing. Motion confounds in task data are a known challenge, particularly when working with motor tasks and other stimuli where head motion tends to be correlated with the task. Indeed, effective elimination of motion artifact may in some cases be even more salient for task activation measures than for functional connectivity and other resting-state measures, given the often overlapping nature of the task-induced signal and motion-induced noise in task fMRI. Examining the efficacy of projection scrubbing in producing more accurate measures of task activation is an important area of future work.

\section{Conclusion}

We propose projection scrubbing, a novel data-driven, statistically principled scrubbing technique for functional MRI data. This technique combines statistical outlier detection with the ability of dimension reductions like ICA to separate signal from noise, in order to identify the presence of statistically significant abnormalities in the data. We comprehensively evaluate the impact of projection scrubbing, motion scrubbing, and DVARS on the quality of functional connectivity estimates and on the amount of data retained. We find that data-driven projection scrubbing and DVARS are more targeted than stringent motion scrubbing, yielding generally more valid, reliable and identifiable estimates of functional connectivity, while retaining substantially more volumes and avoiding high rates of session exclusion. These findings have major implications for population neuroscience, as data-driven scrubbing can empower researchers to retain more data, thus increasing sample size and scan duration, without sacrificing the quality of downstream analysis.  

\section*{Acknowledgements}

This work was supported in part by National Institute of Biomedical Imaging and Bioengineering (NIBIB) grant R01EB027119 to A.F.M. and M.B.N.; National Science Foundation (NSF) CAREER grant DMS–1753171 and the National Sciences and Engineering Research Council of Canada (NSERC) grant RGPIN-2021-02618 to D.J.M.; and National Institute of Mental Health (NIMH) grant K01MH109766 to M.B.N.	 

\bibliographystyle{apalike}
\bibliography{Scrubbing.bib}

\clearpage

\appendix

\renewcommand\thefigure{\thesection.\arabic{figure}}    
\setcounter{figure}{0} 
\setcounter{page}{1}


\section{Additional example scans}
\label{app:additional_scans}

\begin{figure}
    \centering
    \includegraphics[page=1, width=.95\textwidth]{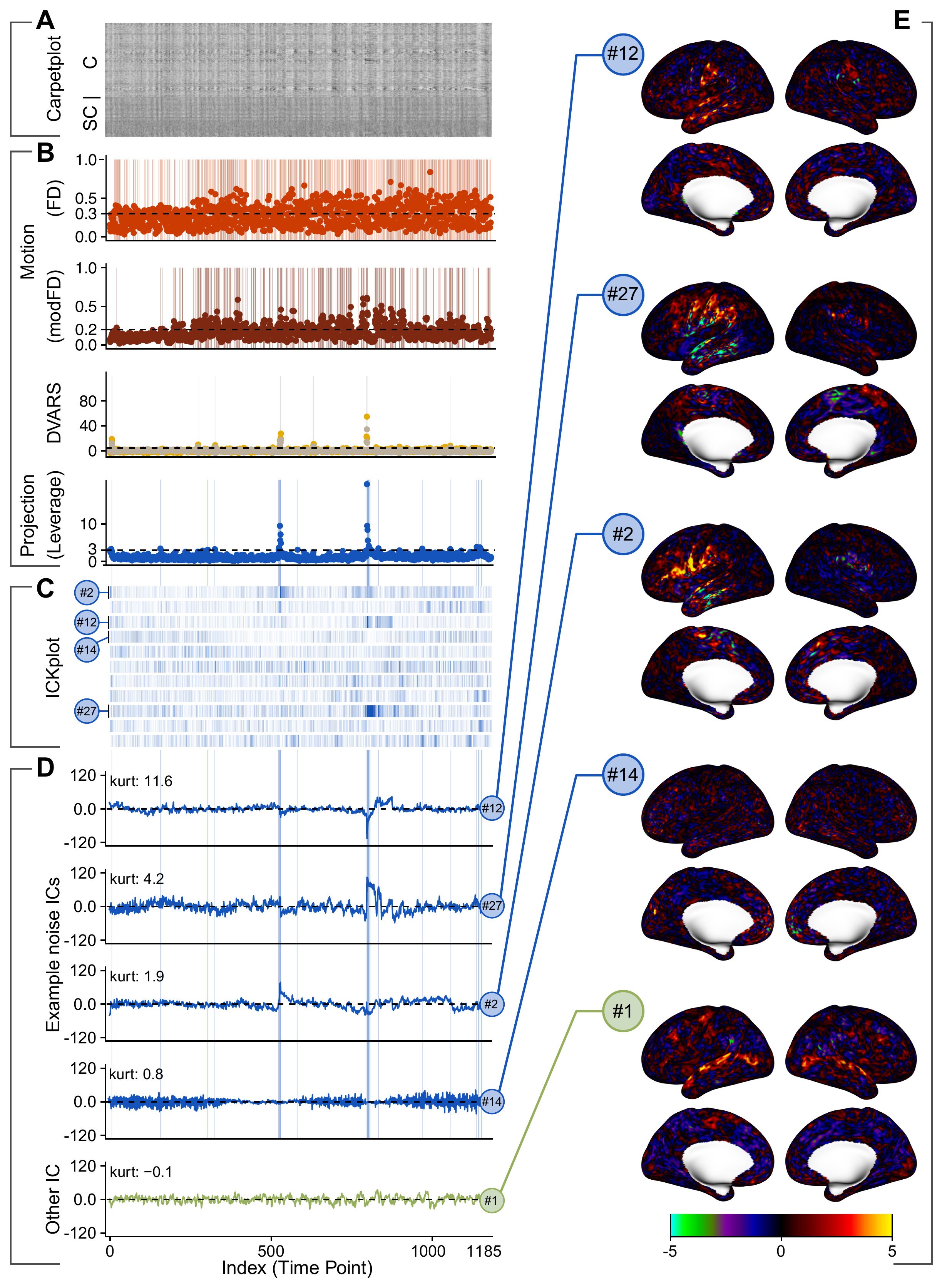}
    \caption{\small \textbf{Example scan with a high amount of subject head motion.} The scan shown is HCP subject 250427, visit 2, LR phase encoding. It represents the 95th percentile of mean FD. As in \autoref{fig:Example}, data-driven scrubbing flags fewer volumes than motion scrubbing. Refer to \autoref{fig:Example} for detailed descriptions of each panel.}
    \label{app:fig:Example2}
\end{figure}

\begin{figure}
    \centering
    \includegraphics[page=1, width=.95\textwidth]{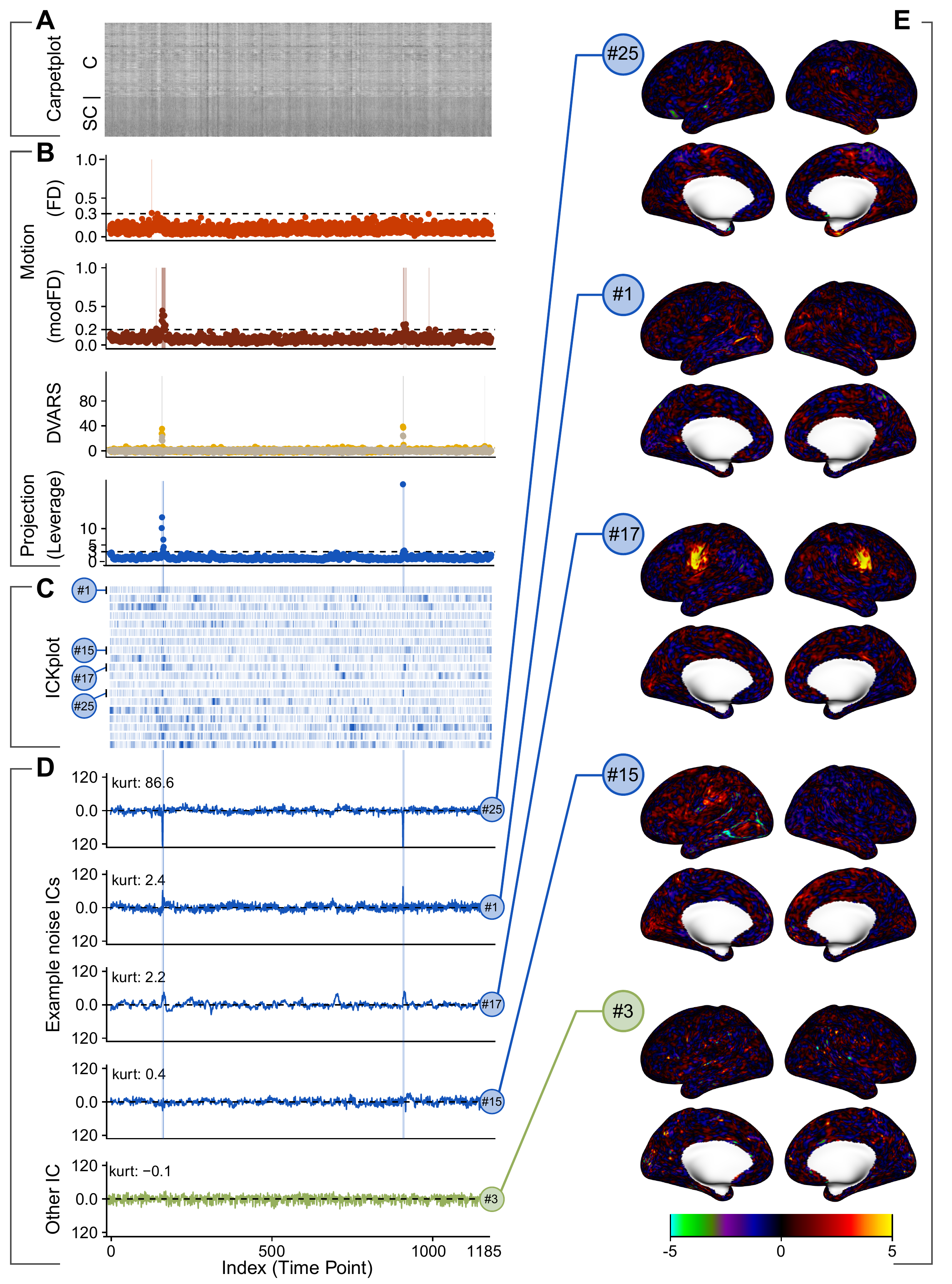}
    \caption{\small \textbf{Example scan with a low amount of subject head motion.} The scan shown is HCP subject 177746, visit 2, RL phase encoding. It represents the 5th percentile of mean FD. The third example noise IC actually resembles somatomotor network activation: transient head motion may be induced or compensated by directed movements. Also, the other example IC (green) actually appears to represent noise, but since its timecourse indicates continuous, non-transient fluctuations, it is not selected by projection scrubbing for computing leverage because constant noise is not amenable to scrubbing. Refer to \autoref{fig:Example} for detailed descriptions of each panel.}
    \label{app:fig:Example3}
\end{figure}

To better understand the differences in behaviors of each scrubbing method, we inspected the timeseries and volumes flagged by each method for 21 scans. The sample was obtained by computing the mean FD for each of the 336 scans in the HCP Retest set, and then inspecting the scan representing the 0th quantile of mean FD, the 5th quantile, the 10th, and so on, such that both low- and high-motion scans would be included. We display the scan at the 45th quantile in \autoref{fig:Example} because as a medium-motion scan it is representative of typical relative flagging rates between the different scrubbing measures. In this Appendix section we include two additional scans: \autoref{app:fig:Example2} shows the scan at the 95th percentile of mean FD, and \autoref{app:fig:Example3} shows the scan at the 5th percentile. 

\begin{figure}
    \centering
    \includegraphics[page=1, width=1\textwidth]{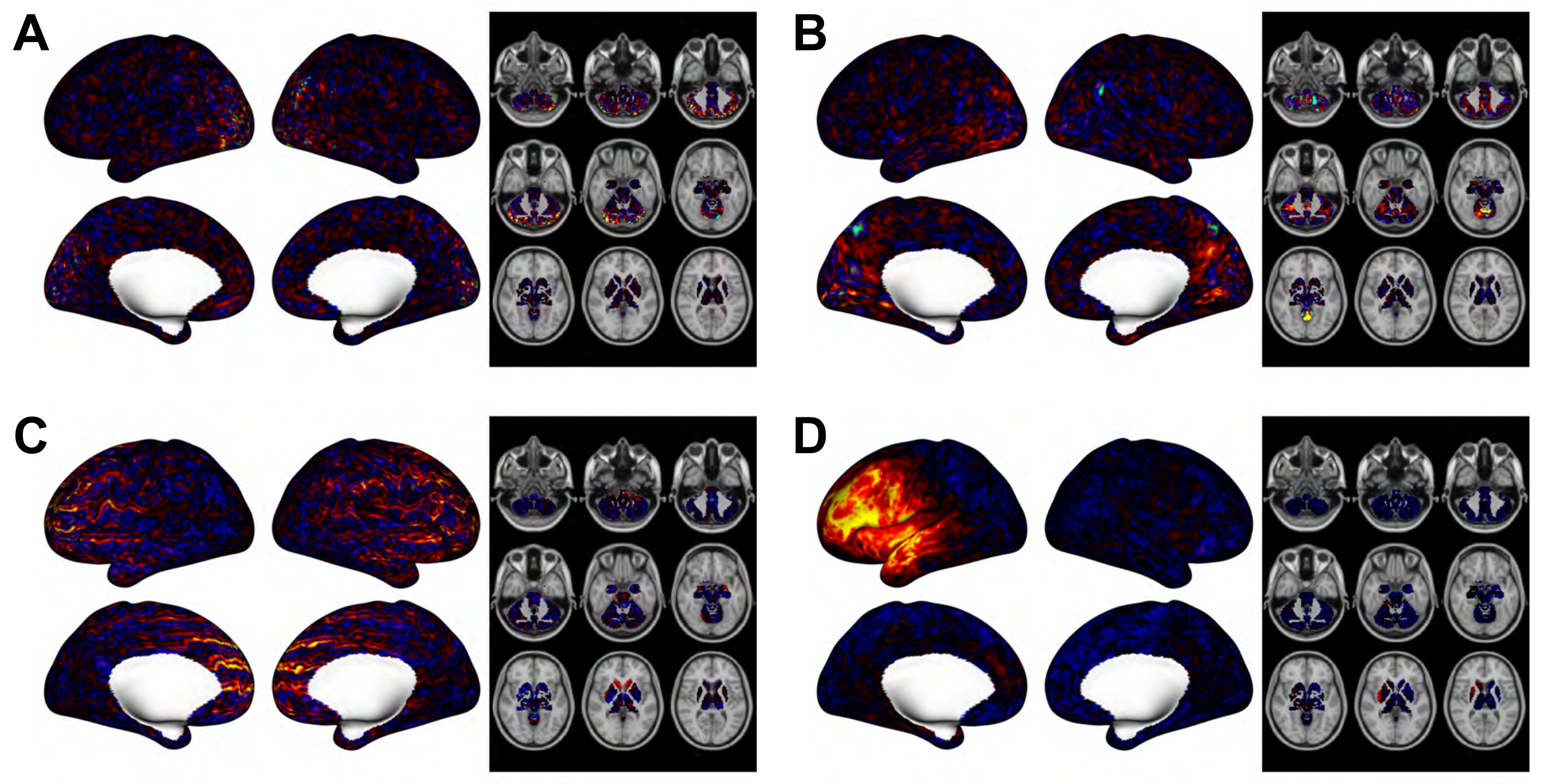}
    \caption{\small \textbf{Full images of four noise components selected by projection scrubbing.} The cortex and subcortex are shown for four selected components. \textbf{(A)} The 34th IC from HCP subject 111312, visit 2, LR phase encoding. This IC is shown in \autoref{fig:Example}, but here we also display the subcortex. Speckles of high intensity are seen on both the posterior cortical surface and the posterior edge of the cerebellum. \textbf{(B)} The 13th IC from HCP subject 917255, visit 2, RL phase encoding. This is a lower motion scan. Spots of high activation appear on the posterior cortical surface, cerebellum, and upper brainstem. \textbf{(C)} The 17th IC from HCP subject 151526, visit 2, RL phase encoding. This scan has the most motion of all scans we used from the HCP Retest set. Many components including this one show a dramatic banding artifact across the cortical surface. \textbf{(D)} The 6th IC from HCP subject 204521, visit 4, LR phase encoding. This high-motion scan has a noise IC with a diffuse high-intensity region on the lateral side of the left hemisphere.}
    \label{app:fig:ExampleSub}
\end{figure}

For visual clarity we have omitted the subcortical portions of the ICA spatial maps in these Figures. Complete images of a few additional spatial maps are provided in \autoref{app:fig:ExampleSub}. Across all inspected scans, we observed a few patterns in the noise IC maps. On the cortex, high-intensity values are often either spatially localized to the front, back, or side of the head, or formed in a speckled pattern across most of the cortical surface. High-intensity values adjacent to the medial wall are also common. When high-intensity values occur in the subcortex, a common pattern is clustering along an edge of the cerebellum. 

\newpage
\section{FusedPCA}
\label{app:PCATF}

The original PCA leverage method \citep{mejiaPCALeverageOutlier2017} calculates the scores by solving the
following optimization problem
\begin{align}
\label{eq:PCAleverage_opt}
\begin{split} \min_{\mathbf{U},\mathbf{V}} & \quad \frac{1}{2}
\norm{\X - \mathbf{U}\mathbf{\Lambda}\mathbf{V}^\top}^2_F \\ \text{subject to} &
\quad \mathbf{U}^\top \mathbf{U} = \mathbf{I}_Q, \mathbf{V}^\top \mathbf{V} =
\mathbf{I}_V.
\end{split}
\end{align}
After calculating $\mathbf{U}$, the PCA leverage is given by the diagonal of
$\mathbf{U} \mathbf{U}^\top$. 
When $Q = 1$, the optimization problem in Equation~\eqref{eq:PCAleverage_opt} is equivalent to
\begin{align}
\label{eq:PCAleverage_opt2}
\begin{split} \max_{\mathbf{u}, \mathbf{v}} & \quad \mathbf{u}^\top \X
\mathbf{v} \\ \text{subject to} & \quad \norm{\mathbf{u}}^2_2 = 1,
\norm{\mathbf{v}}^2_2 = 1.
\end{split}
\end{align} 
If  $(\hat{\mathbf{u}},\hat\bfv) = \argmax_{\mathbf{u},\bfv} \mathbf{u}^\top \X\bfv$  are such that $\norm{\mathbf{u}}_2,\ \norm{\bfv}_2 \geq 1$, then Equation~\eqref{eq:PCAleverage_opt2} has a convex relaxation
\begin{align}
\label{eq:PCAleverage_opt3}
\begin{split} \max_{\mathbf{u}, \mathbf{v}} & \quad \mathbf{u}^\top \X
\mathbf{v} \\ \text{subject to} & \quad \norm{\mathbf{u}}^2_2 \le 1,
\norm{\mathbf{v}}^2_2 \le 1.
\end{split}
\end{align}
To enforce the desired piecewise-constant behavior for $\mathbf{u}$, \pcatf modifies Equation~\eqref{eq:PCAleverage_opt3} by adding
a penalty for the first order differences, that is
\begin{align}
\label{eq:our_opt}
\begin{split} \max_{\mathbf{u}, \mathbf{v}} & \quad \mathbf{u}^\top \X
\mathbf{v} \\ \text{subject to} & \quad \norm{\mathbf{u}}^2_2 \le 1, \norm{\mathbf{D}
\mathbf{u}}_1 \le c, \norm{\mathbf{v}}^2_2 \le 1,
\end{split}
\end{align} where $c$ is a constant and $\mathbf{D}$ is the first order difference
operator
\begin{equation}\mathbf{ D} =
\begin{pmatrix} 1 & -1 & & &\\ & 1 & -1 & & \\ & & \ddots & \ddots & \\ & & & 1
& -1
\end{pmatrix}.
\end{equation} By adding the first order difference penalty, the resulting
$\mathbf{u}$ will be piecewise constant when $c$ is appropriately small. For computational purposes, we re-express the constraints on $\mathbf{u}$ in
Lagrange form 
\begin{align}
\label{eq:our_opt2}
\begin{split} \min_{\mathbf{u}, \mathbf{v}} & \quad - \mathbf{u}^\top \X
\mathbf{v} + \frac{1}{2} \mathbf{u}^\top \mathbf{u} + \kappa\norm{\mathbf{D}
\mathbf{u}}_1 \\ \text{subject to} & \quad \norm{\mathbf{v}}^2_2 \le 1.
\end{split}
\end{align} 
This problem is separately convex in $\mathbf{u}$ and $\bfv$, and can be solved iteratively with $\hat\lambda_1 = \max\mathbf{u}^\top\X\bfv$.
To extend to higher dimensions with $Q > 1$, we 
remove the part in $\X$ that has already been explained by the previous $k$
estimated components. 
\autoref{alg:PCAleverage_trend} makes this procedure explicit.
After estimating the leverage scores, 
we use the same thresholding method to identify outliers as detailed above.

\begin{algorithm}
  \begin{algorithmic}[1] \STATE {\bfseries Input:} centered and scaled $\X$
    \STATE Set $\X^{(1)} = \X$
    \FOR{$k = 1, \ldots, Q$} 
        \STATE Initialize $\mathbf{v}_k$ as the first right singular vector of $\X^{(k)}$ 
        \STATE Iterate until convergence: 
        \STATE \quad\quad (1) $\mathbf{u}_k \leftarrow \argmin_{\mathbf{u}}  \frac{1}{2} \norm{\mathbf{u} - \X^{(k)} \mathbf{v}_k }^2_2 + \kappa \norm{\mathbf{D} \mathbf{u}}_1 $ 
        \STATE \quad\quad (2) $\mathbf{v}_k \leftarrow \frac{\X^{k\top} \mathbf{u}_k}{ \norm{
        \X^{k\top} \mathbf{u}_k}_2}$ 
        \STATE $\lambda_k \leftarrow \mathbf{u}^\top_k \X^k \mathbf{v}_k$ 
        \STATE $\X^{k+1} = \X^k - \lambda_k \mathbf{u}_k \mathbf{v}_k$
    \ENDFOR
    \STATE {\bf Return:} the scores $\mathbf{u}_k$ for $k = 1,
\ldots, Q$
    \end{algorithmic}
  \caption{\pcatf}
  \label{alg:PCAleverage_trend}
\end{algorithm}






\newpage
\section{Kurtosis-based selection of artifactual directions}\label{app:kurtosis}

\begin{figure}[H]
    \centering
    \includegraphics[width=0.65\textwidth]{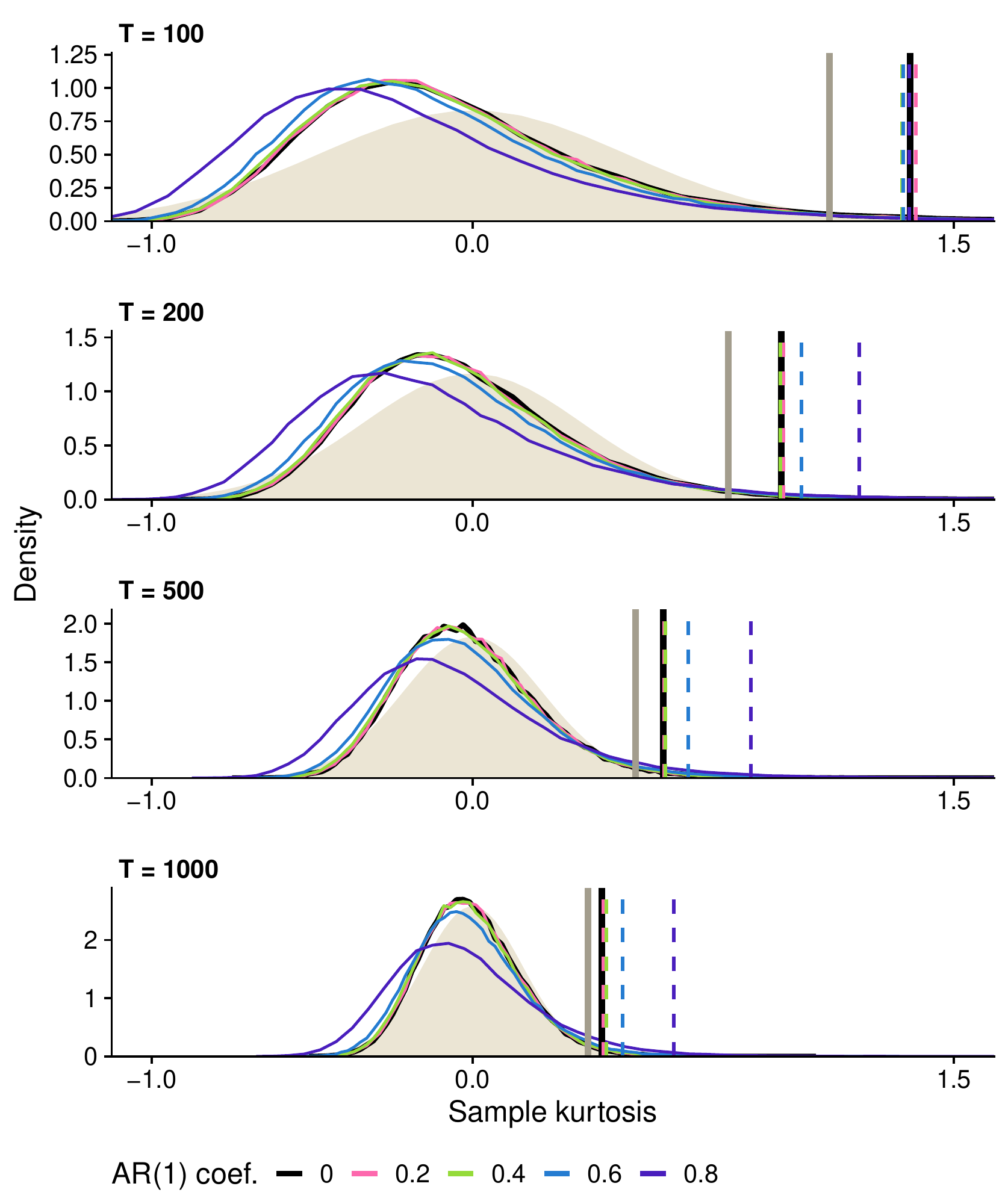}
    \caption{\small \textbf{Effect of sample size and autocorrelation on the sampling distribution of kurtosis in normally distributed data with no outliers.} Each density curve is based on 100,000 Monte Carlo samples of a given sample size and autocorrelation level. The theoretical asymptotic distribution of kurtosis for non-autocorrelated data at each sample size is shown in light brown \citep{fisherMomentsDistributionNormal1930}. Vertical lines indicate each distribution's 0.99 quantile, the threshold used to identify component timeseries likely to contain outliers. For smaller sample sizes, the sampling distribution of kurtosis is right-skewed, becoming more symmetric and converging to a normal distribution as sample size increases. For sample sizes below $1000$, the asymptotic distribution is not appropriate. Therefore, for scan duration $T<1000$ we use simulation to determine the $0.99$ quantile; for longer durations, we use the theoretical $0.99$ quantile based on the asymptotic distribution. The presence of weak to moderate autocorrelation (e.g., AR(1) coefficient $\phi\leq 0.6$) has little effect on the sampling distribution of kurtosis.}
    \label{app:fig:KurtAR}
\end{figure}

\begin{figure}[H]
    \centering
    \includegraphics[width=1\textwidth, trim=0 0 0 1cm, clip]{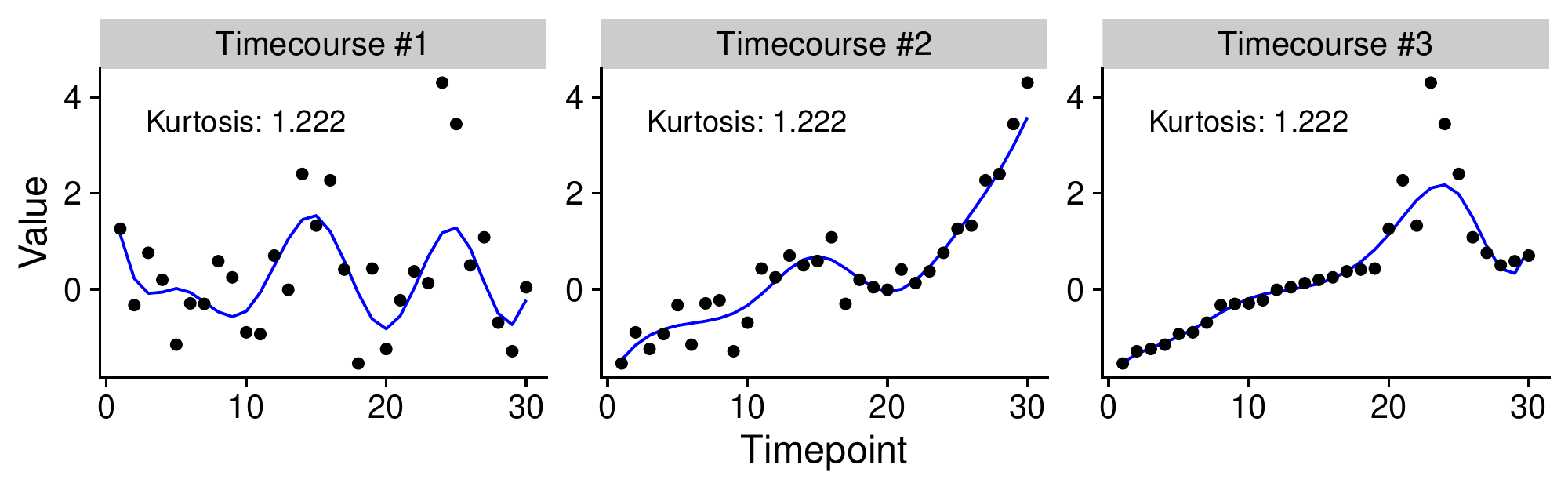}
    \includegraphics[page=2, width=1\textwidth, trim=0 0 0 1cm, clip]{A_kurt_detrending.pdf}
    \caption{\small \textbf{Effects of timeseries trends on kurtosis.} The presence of trends can invalidate kurtosis as a measure of outlier presence. In the top panel, all three timeseries have the same kurtosis value of 1.222, below the high-kursosis threshold. The first has a weak trend but no outliers, the second has a strong trend, and the third has both a trend and outliers. The bottom panel shows the same timeseries after detrending. Detrending has little effect on the kurtosis of the first timeseries, decreases the kurtosis of the second timeseries, and strongly increases kurtosis for the third dataset (the only one containing outliers). This illustrates the importance of detrending before using kurtosis to detect the presence of outliers.}
    \label{app:fig:KurtTrends}
\end{figure}


\newpage
\section{Comparison of filtering methods for FD in multiband data}
\label{app:filteredFD}

\begin{figure}
    \centering
    \includegraphics[page=1, width=.98\textwidth]{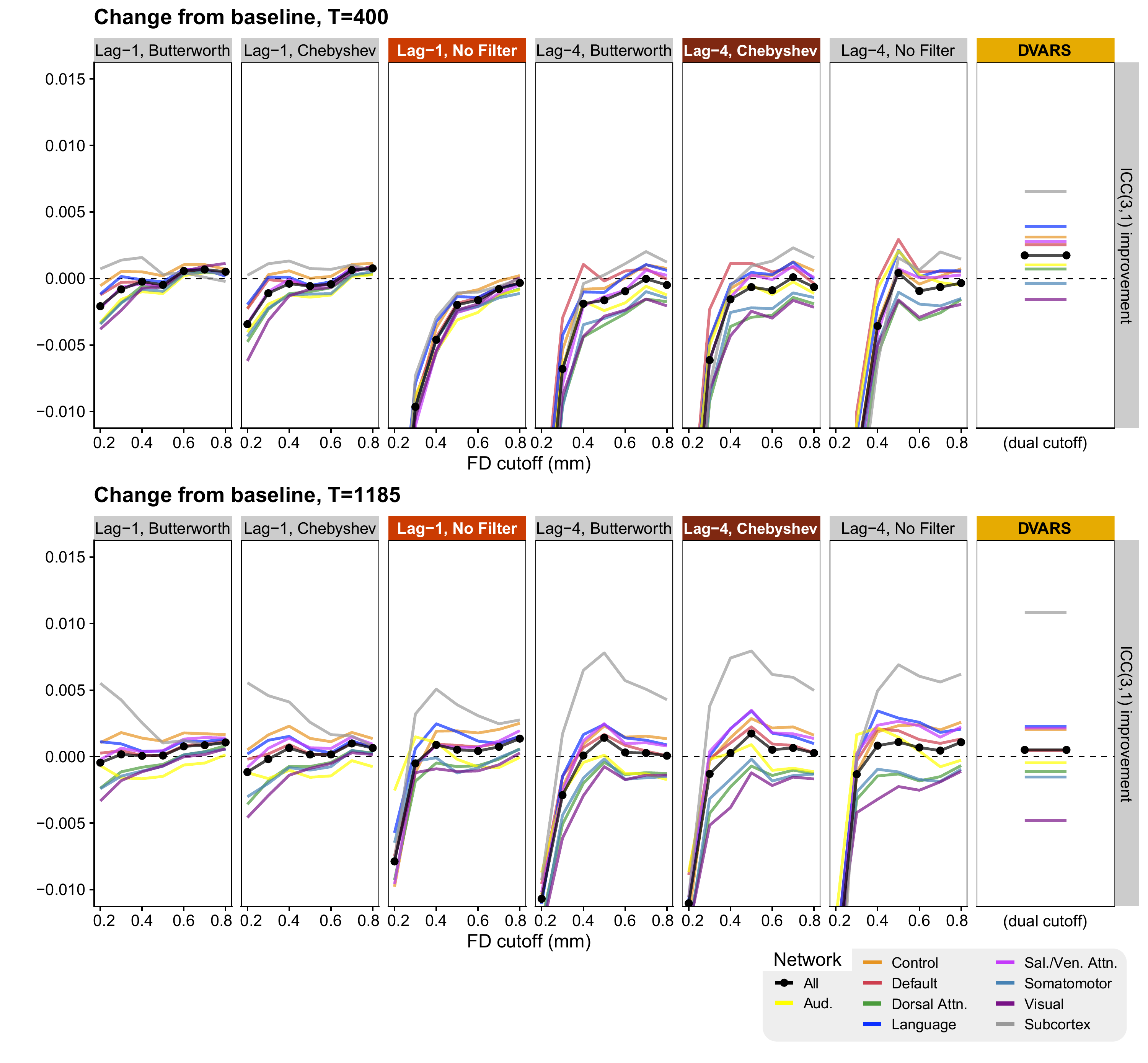}
    \caption{\small \textbf{Expanded comparison of motion scrubbing methods.} We initially evaluated six forms of motion scrubbing: two different notch filters at the respiratory frequency or no filtering applied to the RP timeseries, and either lag-1 differences or lag-4 differences for computing FD. The mean ICC improvement from the CC2+MP6 baseline is shown for connections involving each network (or for all connections), for each method, and for cutoffs between 0.2 mm and 0.8 mm. The bottom row shows results using the full 14.4 minute scan, while the top row shows results using the middle third (almost five minutes). DVARS is included in the last panel for comparison. In the main text we proceeded with the original formulation of FD (lag-1, no filter) and the version which appeared to consistently perform the best at most cutoffs (lag-4, chebyshev), referred to as "modFD".}
    \label{app:fig:MotionExpanded}
\end{figure}

In multiband data, the RPs from which FD is calculated may contain head movements which accompany normal respiration, as well as artifactual variance in the phase encode direction at the respiratory frequency. \cite{power2019distinctions} and \cite{fairCorrectionRespiratoryArtifacts2020} both proposed applying a notch filter to the RPs at the respiratory frequency. Specifically, \cite{power2019distinctions} used a 10th-order Butterworth filter between 0.2 and 0.5 Hz, while \cite{fairCorrectionRespiratoryArtifacts2020} used a second-order IIR notch filter between 0.31 and 0.43 Hz, both in MATLAB. \cite{williams2022advancing} compared two IIR notch filters, one at 0.2 to 0.5 Hz and another at 0.31 to 0.43 Hz, and found evidence in favor of the narrower filter. Here we consider all choices of RP lags (lag-1 or lag-4) and temporal filtering (a notch filter similar to that in \cite{power2019distinctions}, a notch filter similar to that in \cite{fairCorrectionRespiratoryArtifacts2020}, or none at all). We do not consider the lowpass filter used by \cite{williams2022advancing} because we are not lowpass filtering the fMRI data. To approximate the MATLAB notch filters in R, we used the \texttt{gsignal} package: a .2 Hz - .5 Hz 10th order Butterworth filter (\texttt{butter}) and a .31 - .43 Hz 20dB Chebyshev Type 2 filter (\texttt{cheby2}).  

\autoref{app:fig:MotionExpanded} shows the change in reliability compared to the CC2+MP6 baseline for each variant of motion scrubbing. The bottom row of plots shows that when using the full scans, FC estimates are most reliable with FD calculated with lag-4 differences and the Chebyshev filter. The two other versions of FD calculated with lag-4 differences also improve upon those which use lag-1 differences. Based on these results, we adopted lag-4 with Chebyshev filtering as \textit{modFD} in our main analysis.

For the top panel of plots, FC was calculated using the middle third of each fMRI scan instead of the full duration. In the case of less data, we observe that the benefit of motion scrubbing is even less clear.


\newpage
\section{Simultaneous nuisance regression and alternatives}
\label{app:nreg_framework}

In this work we perform scrubbing, noise covariate removal, and high-pass filtering in a simultaneous nuisance regression framework. Using a single ``omnibus" regression instead of multiple sequential regressions avoids the problem of later regressors removing orthogonality between the cleaned data and prior regressors \citep{lindquistModularPreprocessingPipelines2019}. For example, consider performing 24 realignment parameter denoising followed by global signal regression. The cleaned data will not necessarily be orthogonal to the RPs after the second regression, meaning that covariance with motion can be reintroduced. Instead, if the RPs and global signal regressors are combined in the design matrix for a simultaneous regression, the cleaned data will be orthogonal to both the RPs and the global signal.

Even if perfect orthogonality with the nuisance regressors is not a primary concern, simultaneous regression may be preferred because it integrates information from all the identified nuisance signals to estimate the nuisance regression fit. For example, \cite{carpOptimizingOrderOperations2013} has warned that applying a temporal filter before denoising risks blurring temporally-constrained burst noise into adjacent timepoints. These ringing artifacts would create a lower noise floor for scrubbing. However, if filtering and denoising are performed together with simultaneous regression, denoising regressors that load onto large peaks in the data can remove them or at least attenuate their impact on the model fit for the temporal filter, thereby avoiding ringing artifacts. In the same way, spike regressors for scrubbing completely remove the influence of artifact-contaminated volumes from estimation of the DCT highpass filter and denoising. In fact, scrubbing with spike regressors is equivalent to scrubbing by dropping corresponding volumes from the design matrix and the BOLD data \citep{power2015recent}.
Lastly, a simultaneous nuisance regression framework has been recommended and used before, such as in \cite{caballero-gaudesMethodsCleaningBOLD2017}, AFNI (\url{http://afni.nimh.nih.gov}; NIMH Scientific and Statistical Computing Core, Bethesda, Maryland), and \cite{parkesEvaluationEfficacyReliability2018}. We note that it is possible to modify sequential regression to obtain the same results as from simultaneous regression: if one set of nuisance regressors is removed from both the data and a different set of regressors, then when that second set of regressors is removed from the data, the cleaned data will be the same as if both sets of regressors were removed simultaneously \citep{lindquistModularPreprocessingPipelines2019}. To use the previous example, if both the data and the global signal undergo 24 RP denoising, then regressing the modified global signal from the modified data yields identical results as a simultaneous nuisance regression that includes both the 24 RPs and the global signal. For the same reason, applying a linear temporal filter to both the data and the design matrix before nuisance regression is identical to adding nuisance regressors for that filter to the design matrix. There are a few cases where it may in fact be preferable to execute simultaneous regression as a modified sequential regression: sometimes it is more computationally efficient \citep{power2015recent}, other times; such as when a temporal filter is non-linear or otherwise cannot be formulated as a collection of nuisance regressors, then filtering both the data and the design matrix prior to nuisance regression is the closest one can get to simultaneous regression. 

Still, some people may prefer to apply temporal filtering after nuisance regression rather than simultaneously. But this strategy would face an additional dilemma if the data were censored: if the temporal filter expects constant time differences, one may need to add back the censored volumes by interpolating from the newly cleaned data. Interpolation could introduce new artifacts, especially to autocorrelation structure, which may negatively impact the fit of the filter as well as latter analyses. Simultaneous regression, where possible, avoids the need for interpolation. 

\newpage
\section{Additional results}
\label{app:moreresults}

\begin{figure}
    \centering
    \includegraphics[width=1\textwidth]{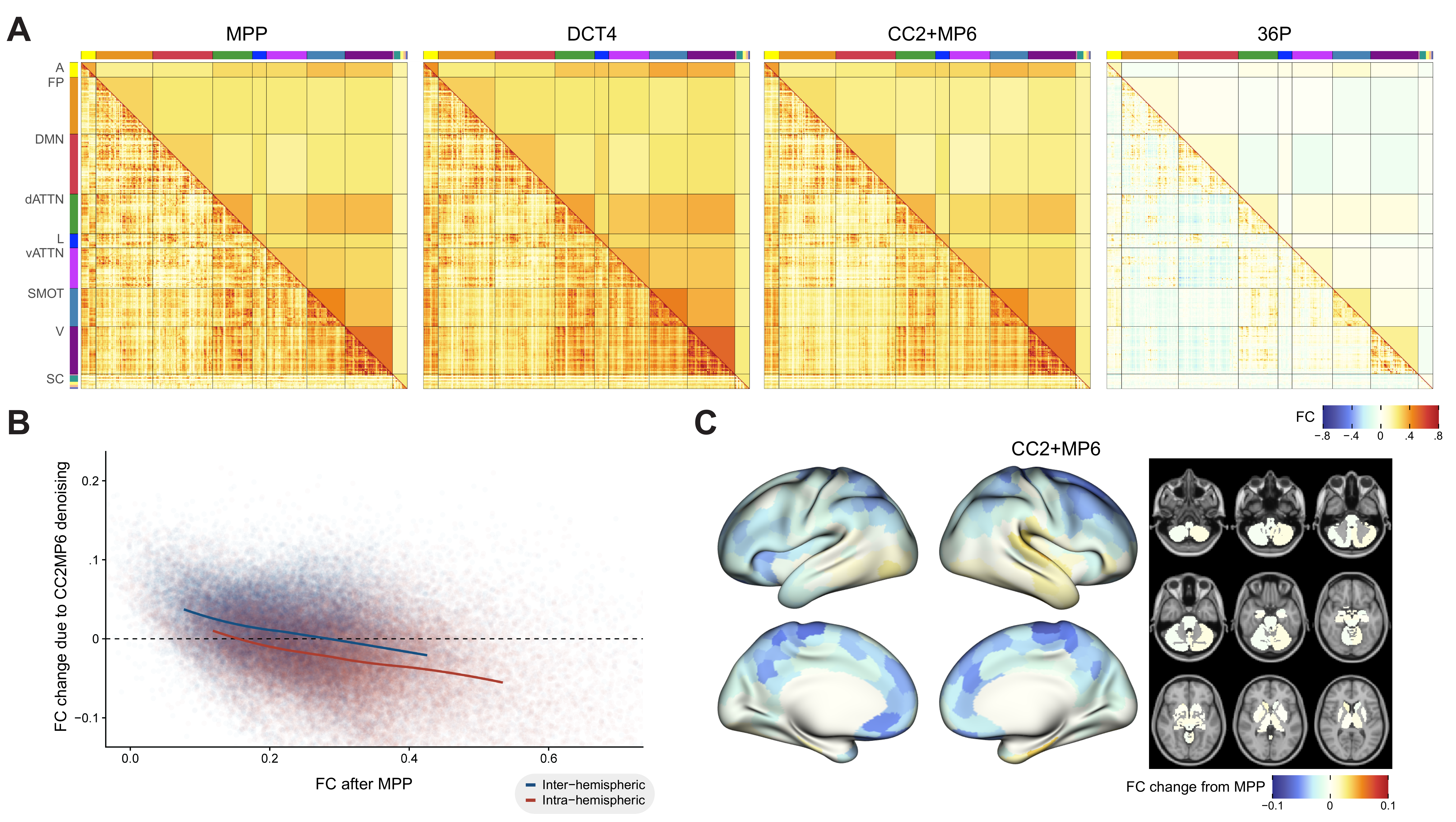}
    \caption{\small \textbf{Denoising tends to lower FC strength of stronger connections and proximal connections.} \textbf{(A)} Average FC estimates across subjects and sessions for different denoising strategies, from left to right: minimally preprocessed (MPP), four DCT bases (DCT4), aCompCor with two components per noise ROI (CC2) plus six RPs (CC2+6MP), and the 36 parameter model (36P). CC2+MP6 and 36P additionally include the DCT4 regressors. 
    The top halves of the matrices represent the mean FC strength for each network pair, i.e. for each corresponding region in the lower triangles. \textbf{(B)} Effect of baseline denoising (CC2+MP6) on FC strength for inter- and intra-hemispheric cortical connections. Each point represents a pair of parcels, and lines represent lowess smoothers. Stronger connections and intra-hemispheric connections tend to be weakened more by denoising, possibly reflecting the removal of an artifactual upward bias in FC strength for proximal connections. \textbf{(C)} Average change in FC estimates across subjects and sessions, and across all connections involving a given parcel, for the CC2+MP6 denoising strategy compared to MPP FC. The values are equivalent to the row/column means of the matrix for CC2+MP6 in Panel A minus those for MPP.}
    \label{app:fig:baselineFC}
\end{figure}

\autoref{app:fig:baselineFC} displays the effect of different baseline denoising strategies on the strength of FC. Panel A shows average FC estimates across all subjects and sessions for the four denoising strategies also presented in \autoref{fig:baselineICC}. We observe two primary trends: first, we see clear global differences in FC strength between CC2+MP6, which slightly decreases the magnitude of FC values, and 36P, which considerably reduces FC strength to the point where negative FC values are induced, especially for off-diagonal connections (between regions belonging to different functional networks). While the large global reduction in FC values for 36P is due to the inclusion of the global signal \citep{murphyImpactGlobalSignal2009}, smaller reduction in FC values for CC2+MP6 and 36P denoising may be attributable to the attenuation of upwardly biased FC strength for proximal connections caused by head motion and previously reported in past studies \citep{vandijkInfluenceHeadMotion2012, powerSpuriousSystematicCorrelations2012, satterthwaiteImpactInScannerHead2012}.
Second, we see a checkerboard pattern of FC strength within network pairs (that is, within each constituent rectangle in the matrices of FC values), particularly for the MPP data, suggesting differences in FC strength for intra- and inter-hemispheric connections. This pattern is attenuated by CC2+MP6 denoising, explored further in Panel B. This supports the idea that inflated FC strength for proximal connections is being reduced. Panel C shows the mean change in FC after CC2+MP6 denoising for each cortical parcel and subcortical structure, across each of their 418 connections. The mean changes vary smoothly across space, with greater reductions in the parietal and frontal regions.

\begin{figure}
    \centering
    \includegraphics[width=.45\textwidth]{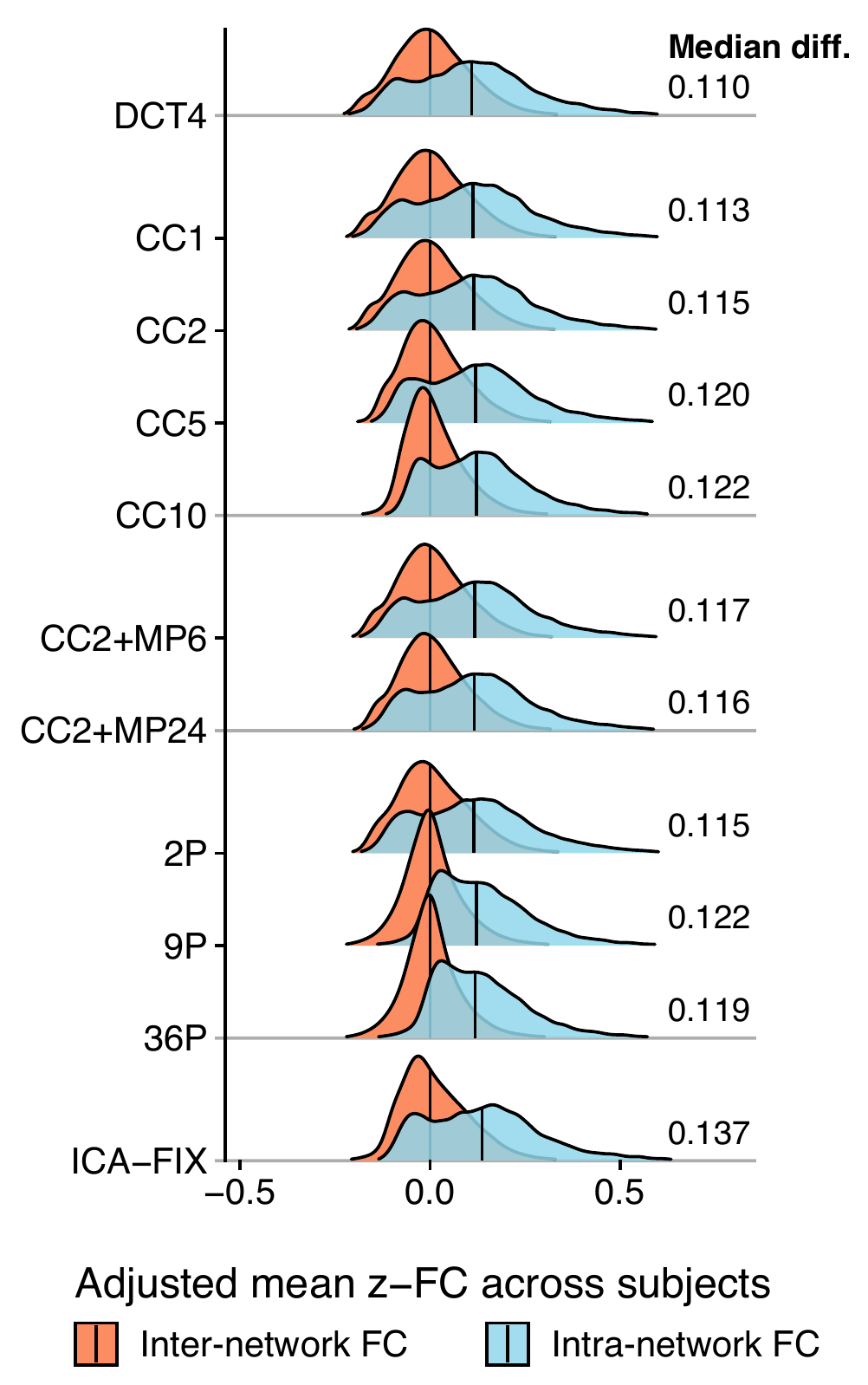}
    \caption{\small \textbf{Comparison of the difference between intra- and inter-network FC strength, for the denoising methods.} Histograms for z-transformed FC values for each denoising method are shown, with separate histograms for inter-network FC pairs and intra-network FC pairs. For each denoising method, the median inter-network FC value was subtracted from all FC values, shifting both histograms along the x-axis to align the medians of the inter-network FC values across methods. This allows a clearer visual comparison between the magnitude of the difference between inter- and intra-network FC. Strengthened within-network connectivity, relative to between-network connectivity, may reflect the uncovering of expected neural signals from noise contamination. But none of the denoising methods show much clearer separation compared to the others. The difference between the medians increases consistently with greater CompCor order, yet 9P has a higher median difference than 36P.}
    \label{app:fig:baselineFC_diff}
\end{figure}

\autoref{app:fig:baselineFC_diff} explores how the difference between within-network and between-network FC strength differs across the denoising methods. Stronger within-network FC, as opposed to broader connectivity patterns, is sometimes thought to indicate more correct FC estimates. But the difference between within- and between-network FC strength is not clearly greater for any one denoising method. 

\begin{figure}
    \centering
    \includegraphics[width=1\textwidth]{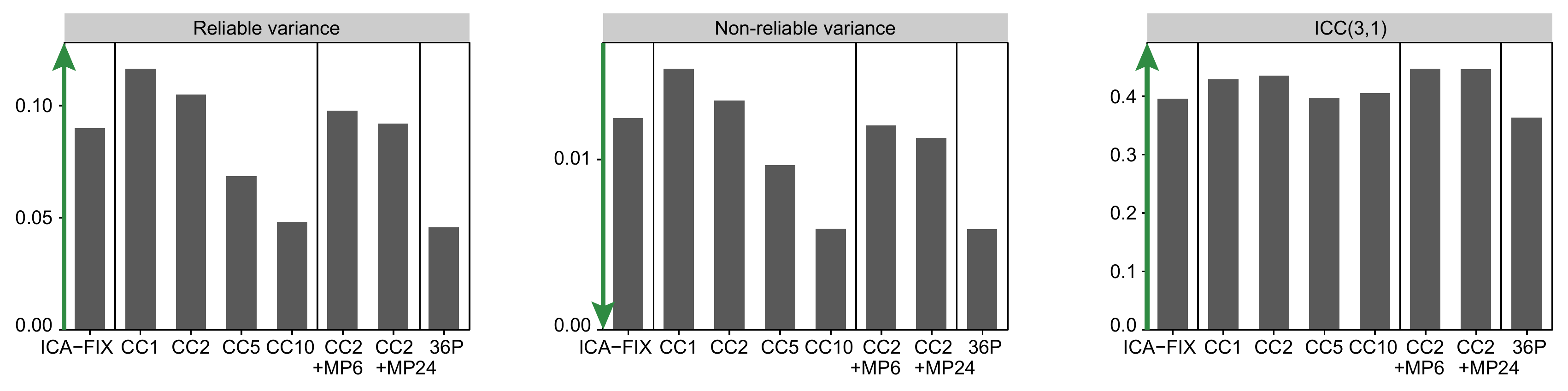}
    \caption{\small \textbf{CC2+MP6 represents the best trade-off between preserving between-subject signal and minimizing within-subject noise.} For select denoising methods, we show the two variance components from which ICC(3,1) is calculated: MSB, a measure of inter-subject variation (signal, in this context), and MSR, a measure of within-subject variation (noise). As a measure of reliability, ICC increases with greater inter-subject variation (MSB) or lesser within-subject variation (MSR). Arrows along the y-axis of each panel point in the direction corresponding to greater ICC. All denoising methods reduce both MSB and MSR. Results for the CC$x$ methods suggest that using more nuisance regressors leads to stronger decreases in both MSB and MSR. For aCompCor, the trade-off is best for CC2, and even better after adding the six RPs to the set of nuisance regressors.}
    \label{app:fig:baselineVarDecomp}
\end{figure}

\autoref{app:fig:baselineVarDecomp} decomposes the ICC for baseline denoising into reliable variance (MSB) and non-reliable variance (MSR). ICC increases with the former, and decreases with the latter. This figure shows that methods with relatively high MSB tend to have relatively high MSR, and vice versa, meaning that superior ICC is achieved by balancing the trade-off between noise reduction and signal retention. 

\begin{figure}
    \centering
    \includegraphics[page=1, width=.98\textwidth]{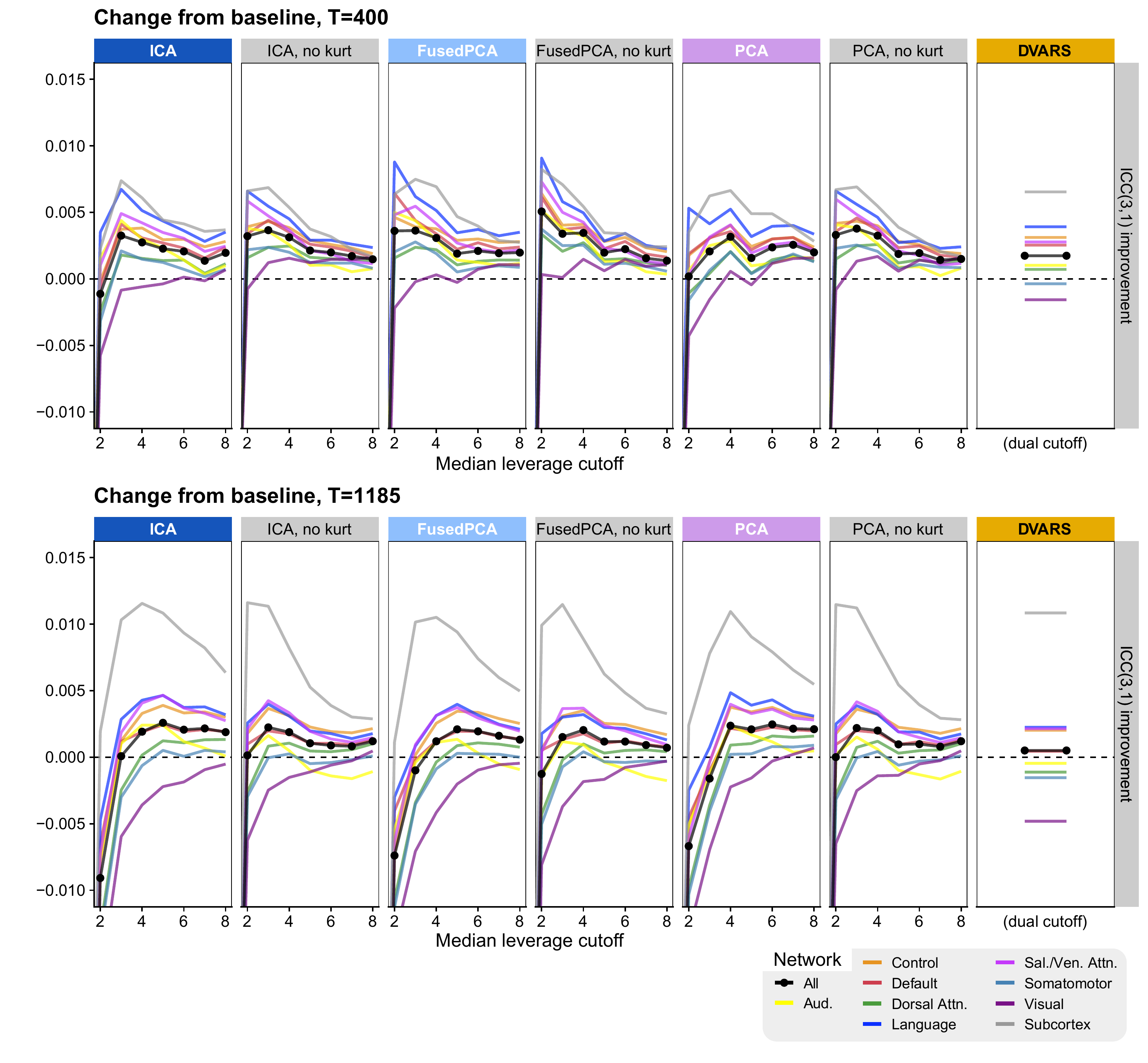}
    \caption{\small \textbf{Expanded comparison of projection scrubbing methods.} This plot shows the three projection scrubbing methods discussed in the main text, along with their corresponding alternatives in which kurtosis is not used to select the noise ICs. For these alternative methods, all highest-variance components up to the number selected by PESEL are used to compute leverage. The mean ICC improvement from the CC2+MP6 baseline is shown for connections involving each network (or for all connections), for each method, and for cutoffs between 2 and 8 times the median. The bottom row shows results using the full 14.4 minute scan, while the top row shows results using the middle third (almost five minutes). DVARS is included in the last panel for comparison. The three methods which use kurtosis to inform noise IC selection yield greater reliability improvements at the higher cutoff values, suggesting both better separation of artifact-contaminated volumes in the leverage timecourse, and that volumes with more egregious artifacts correctly have greater leverage values. For all projection scrubbing methods, stricter cutoffs (up to a point) further benefit subcortical connections, which are known to have lower SNR, and also benefit most connections when less data is used to calculate FC. Meanwhile, more lenient cutoffs alleviate the negative impact of scrubbing on the visual cortex. Together these observations suggest that scrubbing may have greater benefit when less neural signal is available: the noise reduction effected by scrubbing is more impactful than any loss of true neural signal caused by censoring volumes.}
    \label{app:fig:ProjectionExpanded}
\end{figure}

\autoref{app:fig:ProjectionExpanded} examines the effect of removing kurtosis-based noise IC selection, and of using shorter scan duration, on the impact of projection scrubbing on FC reliability. 

\begin{figure}
    \centering
    \includegraphics[width=.7\textwidth]{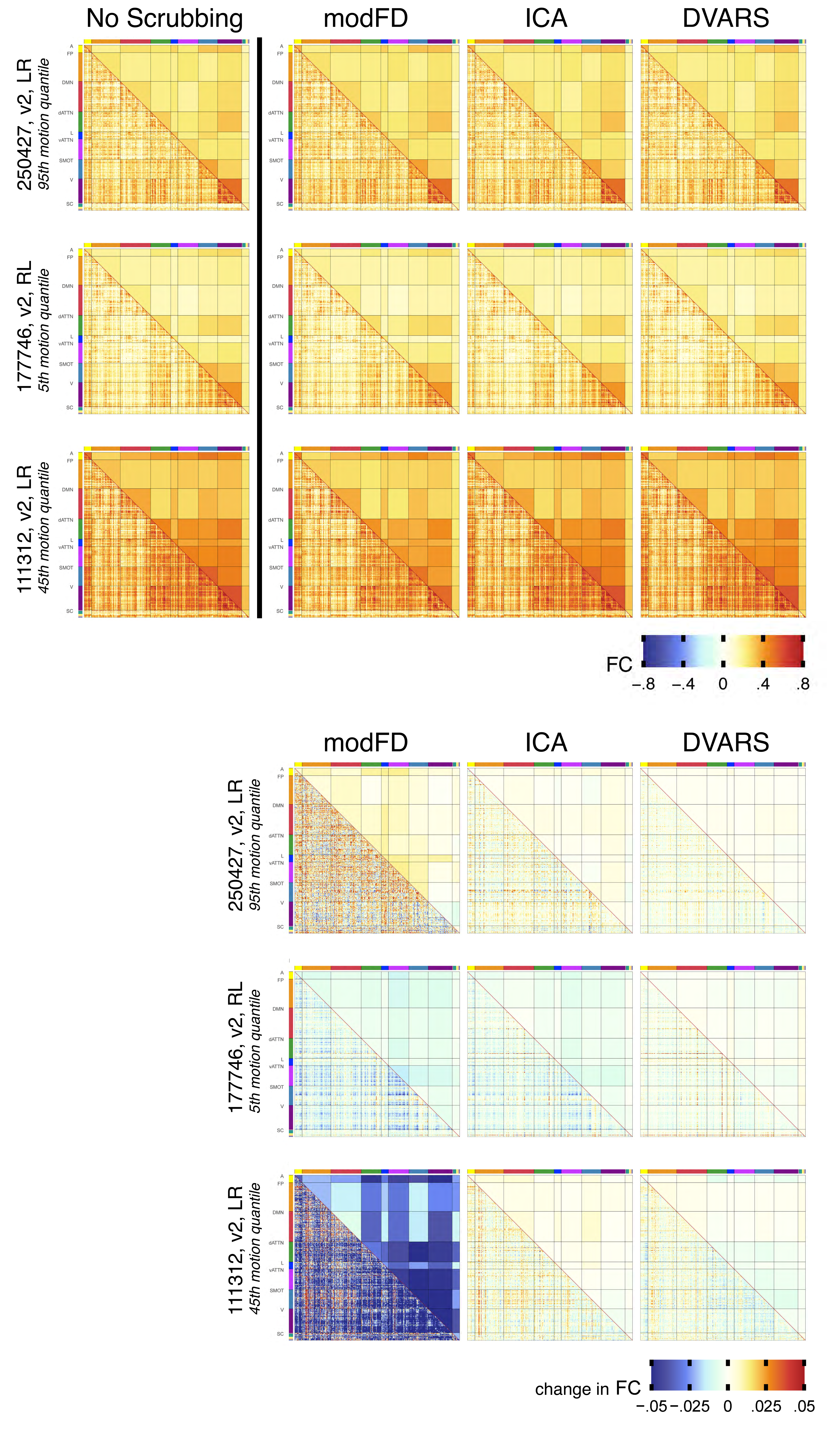}
    \caption{\small \textbf{Effect of scrubbing on FC magnitude for three selected sessions.} All scrubbing methods generally do not yield stark changes to FC estimates, compared to denoising alone. The subjects are the same subjects shown in \autoref{fig:Example}, \autoref{app:fig:Example2}, and \autoref{app:fig:Example3}.}
    \label{app:fig:scrubFCgrid}
\end{figure}

\autoref{app:fig:scrubFCgrid} provides matrices showing all FC pairs, before and after scrubbing, for three selected sessions. 

\begin{figure}
    \centering
    \includegraphics[page=1, width=1\textwidth]{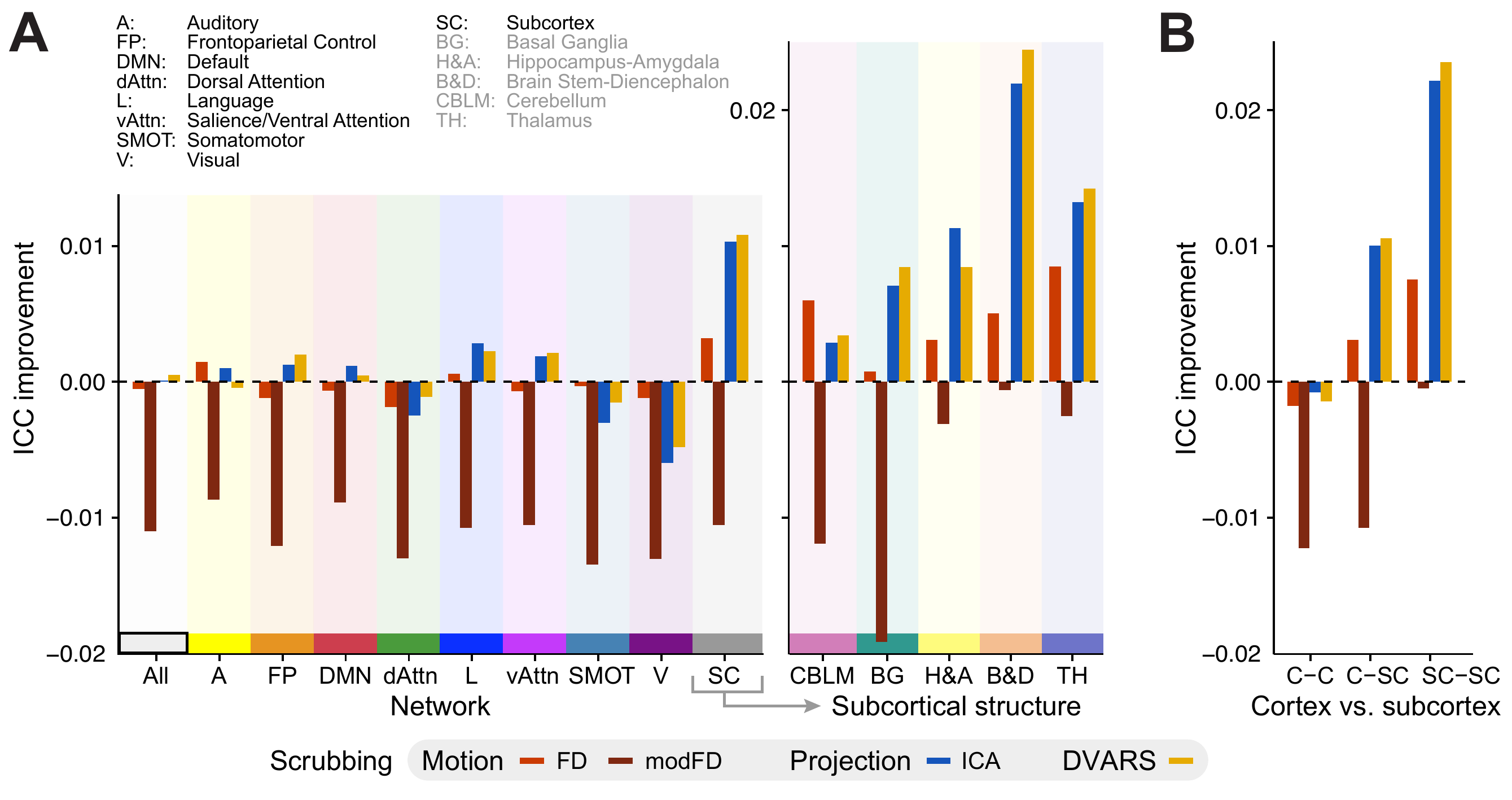}
    \caption{\small \textbf{Projection scrubbing results in the greatest improvement in FC reliability across most types of functional connections.}
      \textbf{(A)} The average change in ICC (over baseline CC2+MP6 denoising)
      across all connections involving a given cortical network or subcortical
      group. Data-driven scrubbing produces greater improvement to FC reliability
      than either form of motion scrubbing across nearly all networks
      and subcortical groups. \textbf{(B)} Effect of scrubbing on reliability of
      FC for cortical-cortical (C-C) connections, cortical-subcortical (C-SC)
      connections, and subcortical-subcortical (SC-SC) connections. Connections
      involving subcortical regions (SC-SC and C-SC) show the greatest
      improvement in reliability due to scrubbing, especially with data-driven
      methods.}
    \label{app:fig:ICCexpanded}
\end{figure}

\autoref{app:fig:ICCexpanded} explores the change in ICC for different types of functional connections due to scrubbing, relative to the CC2+MP6 baseline regression-based denoising. Panel A shows the average change in ICC across all connections involving a particular cortical network (left) or subcortical group (right). Each bar summarizes a group of rows (or colummns) of the FC matrix. For all cortical networks, modified motion scrubbing yields the greatest decreases in reliability. Projection scrubbing and DVARS have overall similar improvement for subcortical connections. 

Panel B groups connections differently, comparing connections within the cortex to those within the subcortex and those between cortical and subcortical regions. For connections involving the subcortex, data-driven scrubbing methods improve reliability much more than motion scrubbing.

\begin{figure}
    \centering
    \includegraphics[page=1, width=1\textwidth]{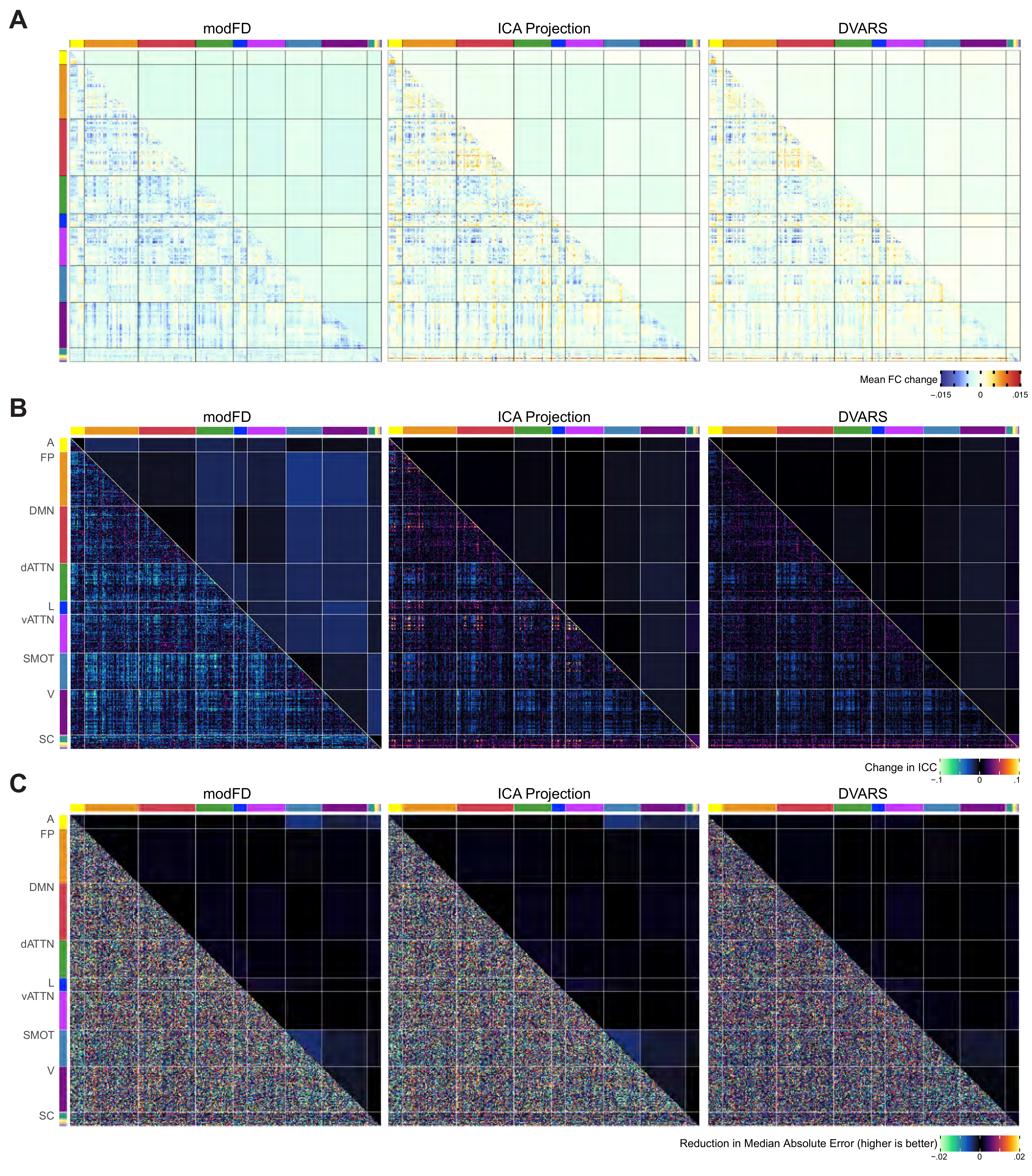}
    \caption{\small \textbf{Connection-level results for the effects of scrubbing on FC strength, reliability, and validity.} \textbf{(A)} Matrix of mean change in FC values for modFD, ICA projection scrubbing, and DVARS. On average, scrubbing lowers FC strength, especially for intra-cortical connections. (Within each network, connections are listed left cortex first and right cortex second, such that the darker rectangles of blue occur in blocks of left-left and right-right connections.) This is consistent with the idea that scrubbing reduces the effects of motion, one of them being inflated short-distance (within-hemisphere) connectivity, compared to long-distance (betweeen-hemisphere) connectivity. \textbf{(B)} Matrix of change in ICC values. Patterns in the matrix for DVARS are similar to those in that of projection scrubbing except without the marked improvements for the intra-left hemisphere connections, while matrices for the other projection scrubbing methods look very similar to the matrix for ICA projection scrubbing including this quirk. \textbf{(C)} Matrix of reduction in median absolute error (MAE) across subjects for the validity analysis. Note that warmer colors indicate a reduction in MAE, meaning higher validity, so warm colors represent an improvement and cool colors represent a worsening.} 
    \label{app:fig:scrubEffectGrid}
\end{figure}

\autoref{app:fig:scrubEffectGrid} shows the mean change in FC, the change in ICC, and the median absolute difference between the FC estimate and the ground truth in the validity study, for each functional connection and the three scrubbing methods. Some individual connections exhibit much greater improvement (or worsening) on reliability or validity than the average connection. 

\begin{figure}
    \centering
    \includegraphics[width=.8\textwidth]{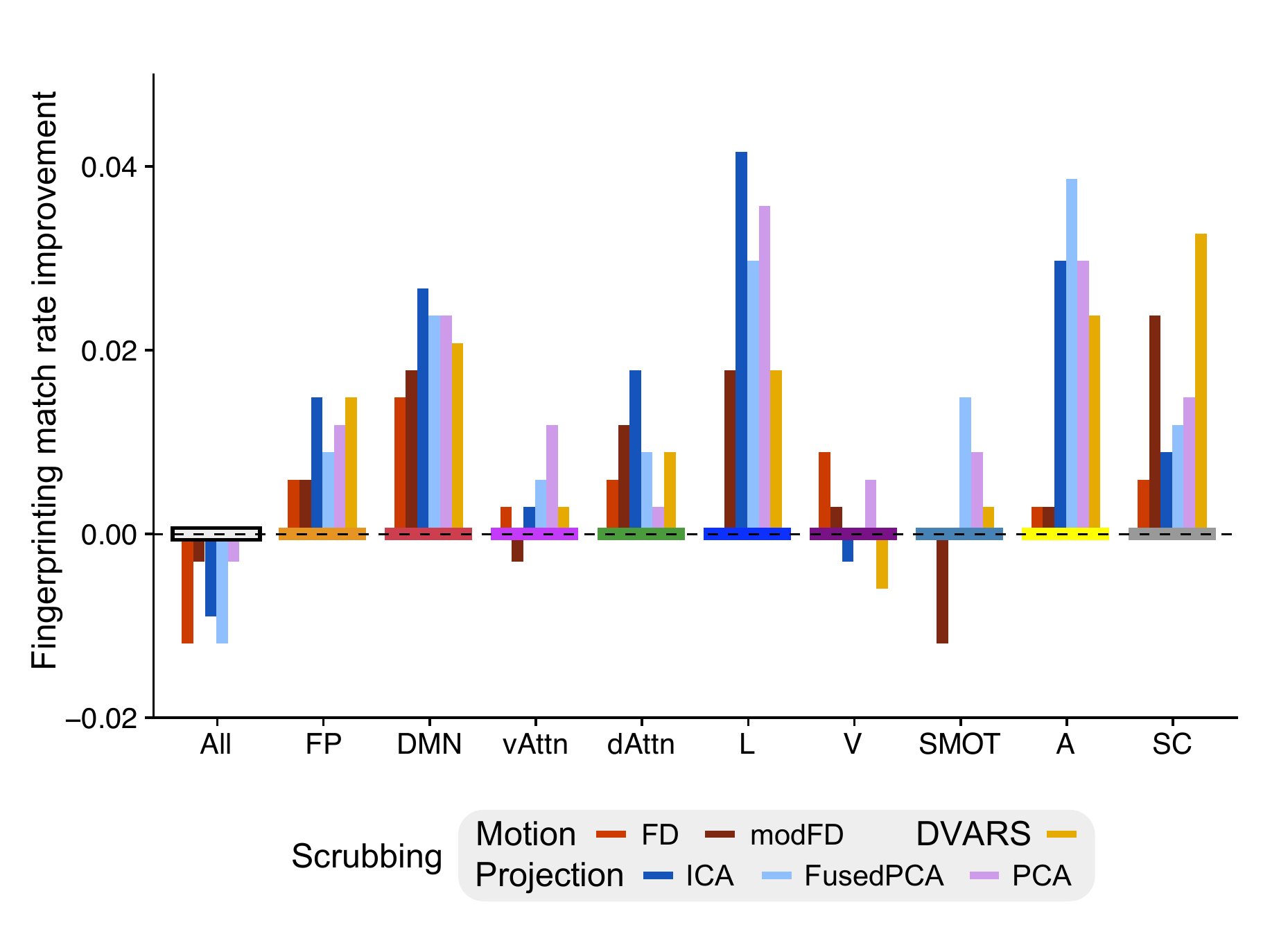}
    \caption{\small \textbf{Expanded comparison of the effect of scrubbing on fingerprinting success.} This plot includes PCA and FusedPCA projection scrubbing, which perform similarly to ICA projection scrubbing. An exception is that PCA and FusedPCA projection scrubbing appear to benefit fingerprinting using the somatomotor network, while ICA projection scrubbing appears to have no effect.}
    \label{app:fig:scrubFingerprintExpanded}
\end{figure}

\autoref{app:fig:scrubFingerprintExpanded} expands on \autoref{fig:Fingerprint} by showing PCA and FusedPCA projection scrubbing.


\newpage
\section{Alternative baselines}
\label{app:baselines}

\begin{figure}[H]
    \centering
    \includegraphics[width=.65\textwidth, clip]{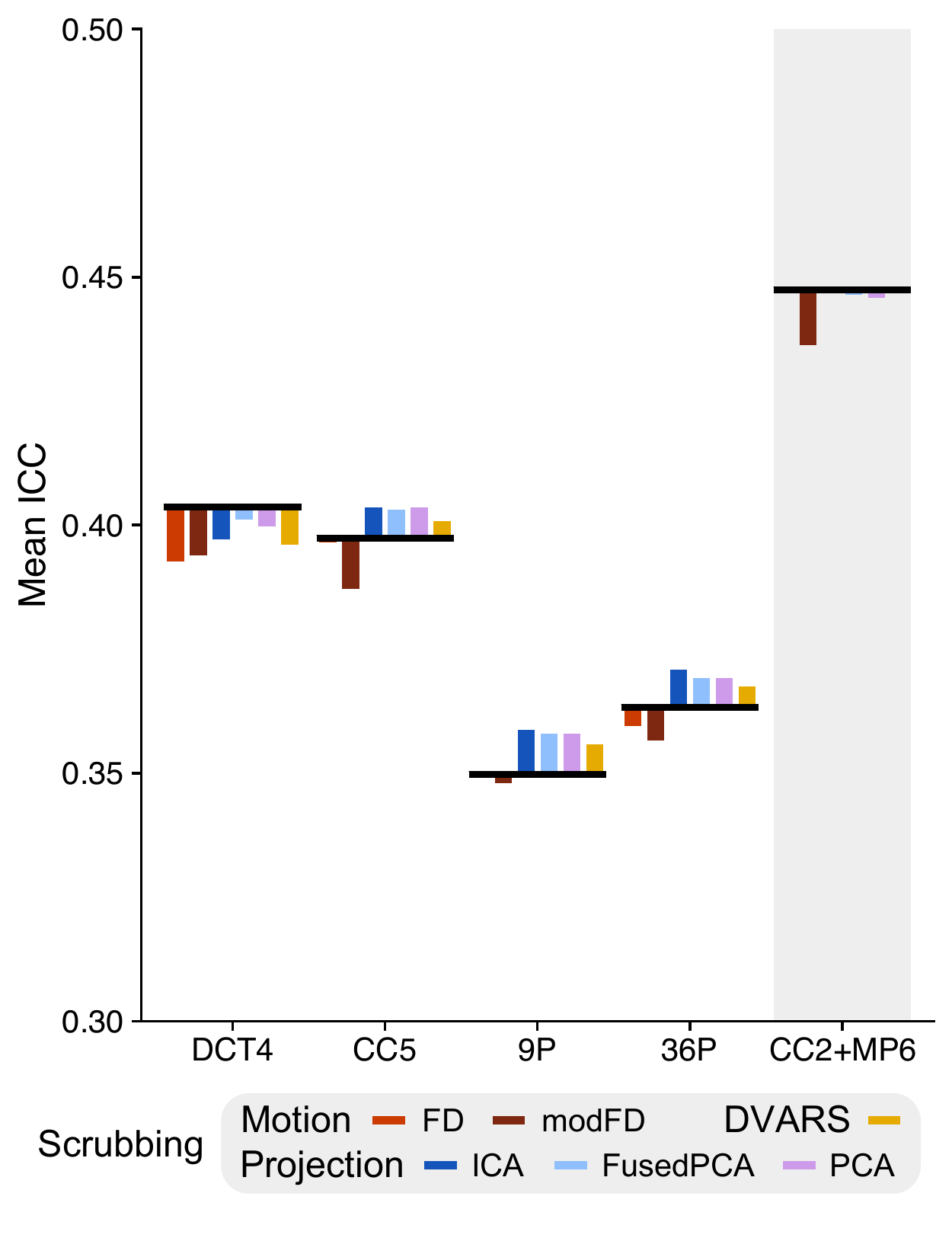}
    \caption{\small \textbf{The benefit of ICA projection scrubbing for reliability is consistent across a variety of denoising baselines.} We estimate FC after applying five different denoising strategies: detrending only (DCT4), aCompCor with five components per ROI (CC5), aCompCor with two components plus six RPs (CC2+MP6, discussed in the main text), the 9 parameter model (9P), and the 36 parameter model (36P). All denoising strategies are regression-based, and the latter four also include the four DCT bases for detrending. The mean ICCs across connections for each denoising method are indicated by black horizontal lines. Then, FC estimation is repeated while combining each scrubbing method with each denoising method. We use the same cutoffs as in our primary analysis: 3 times the median for projection scrubbing, the dual cutoff for DVARS, 0.3 mm for FD, and 0.2 mm for modFD. The colored vertical lines indicate the change in mean ICC attributable to scrubbing by connecting the mean ICC with scrubbing to the mean ICC without scrubbing, for each baseline and scrubbing method. For example, colored lines that extend upward from a black line indicate an improvement to FC reliability from the respective baseline.}
    \label{app:fig:AltBases}
\end{figure}

In our primary analysis we compared scrubbing methods using aCompCor with two components per ROI plus six RPs (CC2+MP6) as our baseline. CC2+MP6 was selected because it yielded the highest mean ICC. However, other denoising strategies we considered are also commonly used for fMRI data cleaning. In \autoref{app:fig:AltBases}, we explore the improvement in reliability of FC estimates using scrubbing in conjunction with a few of these other baselines. With the DCT4 baseline, scrubbing is not beneficial, but projection scrubbing has the least negative impact. Projection scrubbing is beneficial with all other baselines, and ICA projection scrubbing consistently yields the greatest improvements to reliability while modified motion scrubbing yields the least. This finding generalizes the reliability comparison between the different scrubbing methods presented in the main text. Another notable pattern is that scrubbing improves reliability more for baselines having lower mean ICC, suggesting that scrubbing can compensate for less effective denoising strategies. Thus, choosing an effective scrubbing method can be even more impactful when less optimal denoising is used. 

\end{document}